\documentclass[11pt]{article}
\pdfoutput=1
\usepackage{geometry}
\usepackage{graphicx}
\usepackage{amssymb}
\usepackage[square,authoryear]{natbib}
\newcommand{\ra}[1]{\renewcommand{\arraystretch}{#1}}
\newcommand{\captionfonts}{\small}
\makeatletter  
\long\def\@makecaption#1#2{%
  \vskip\abovecaptionskip
  \sbox\@tempboxa{{\captionfonts #1: #2}}%
  \ifdim \wd\@tempboxa >\hsize
    {\captionfonts #1: #2\par}
  \else
    \hbox to\hsize{\hfil\box\@tempboxa\hfil}%
  \fi
  \vskip\belowcaptionskip}
\makeatother   

\title{Spatial patterns of tidal heating}

\author{Mikael Beuthe\\
\it Royal Observatory of Belgium,\\
\it Avenue Circulaire 3, 1180 Brussels, Belgium\\
\it E-mail: mikael.beuthe@observatoire.be}
\date{December 19, 2012} 

\begin{document}
\maketitle

\begin{abstract}

In a body periodically strained by tides, heating produced by viscous friction is far from homogeneous.
The spatial distribution of tidal heating depends in a complicated way on the tidal potential and on the internal structure of the body.
I show here that the distribution of the dissipated power within a spherically stratified body is a linear combination of three angular functions.
These angular functions depend only on the tidal potential whereas the radial weights are specified by the internal structure of the body.
The 3D problem of predicting spatial patterns of dissipation at all radii is thus reduced to the 1D problem of computing weight functions.
I compute spatial patterns in various toy models without assuming a specific rheology: a viscoelastic thin shell stratified in conductive and convective layers, an incompressible homogeneous body and a two-layer model of uniform density with a liquid or rigid core.
For a body in synchronous rotation undergoing eccentricity tides, dissipation in a mantle surrounding a liquid core is highest at the poles.
Within a soft layer (or asthenosphere) in contact with a more rigid layer, the same tides generate maximum heating in the equatorial region with a significant degree-four structure if the soft layer is thin.
The asthenosphere can be a layer of partial melting in the upper mantle or, very differently, an icy layer in contact with a silicate mantle or solid core.
Tidal heating patterns are thus of three main types: mantle dissipation (with the icy shell above an ocean as a particular case), dissipation in a thin soft layer and dissipation in a thick soft layer.
Finally, I show that the toy models predict well patterns of dissipation in Europa, Titan and Io.
The formalism described in this paper applies to dissipation within solid layers of planets and satellites for which internal spherical symmetry and viscoelastic linear rheology are good approximations.

\end{abstract}

{\small
Keywords:
Tides, Solid body - Planetary dynamics - Europa - Io -Titan}
\\

{\it
\noindent
The final version of this preprint is published in Icarus (www.elsevier.com/locate/icarus)\\
doi:10.1016/j.icarus.2012.11.020}

\section{Introduction}
\label{Introduction}
\markright{1 INTRODUCTION\hfill}

Fifty years ago, William Kaula found that the heat dissipated by tidal friction within the Moon is extremely nonuniform both radially and laterally \citep{kaula1963,kaula1964}.
At that time, nonuniform tidal heating was a rather academic subject as these variations were not observable: tidal heating in the Moon is indeed much smaller than radiogenic heating.
Spatial variations of tidal heating were actually a byproduct of computing the total power dissipated in the body, a key factor in modeling orbital evolution.
Kaula's calculations were based on the microscopic (or micro) approach to tidal dissipation, which starts with the computation of viscoelastic tidal strains.
A bit earlier, \citet{munk1960} had given an approximate formula for the total power dissipated by tides with a macroscopic (or macro) approach.
This alternative approach does not require the computation of tidal strains: the tidal bulge lags the tidal forcing by an angle parameterizing the viscous response.
Their formula, however, was limited to an incompressible homogeneous body, in contrast with the micro approach which is applicable to any model of internal structure.

New theoretical developments had to wait until 1978, when two papers significantly improved tidal heating computations.
First, \citet{peale1978} reconsidered both macro and micro approaches to tidal heating, correcting several errors in the various formulas.
They also mapped tidal heating variations for eccentricity tides and obliquity tides, showing that tidal dissipation in a homogeneous Moon is maximum at the poles in the former case and at the equator in the latter.
Furthermore they showed that the presence of a large liquid core enhances dissipation in the mantle.
Second, \citet{zschau1978} derived a simple formula for the total dissipated power in a spherically stratified compressible body, in which the influence of the body's internal structure appears through $\mbox{\it Im}(k_2)$, the imaginary part of the gravity tidal Love number $k_2$.
Zschau's formula was the first step toward reconciling the macro and micro approaches, since $k_2$ can be computed if the internal structure of the body is known.
\citet{tobie2005} later applied the variational method to relate $\mbox{\it Im}(k_2)$ to the imaginary part of the volume-integrated strain power.
They also computed the radial distribution of the dissipated power in terms of deformation functions that can be evaluated with standard methods for any spherically stratified model.
I will show that the formulas of \citet{zschau1978} and \citet{tobie2005} are recovered by spatially averaging the local power obtained in the micro approach, thus bridging the last gap between the macro and micro approaches.
My paper, however, is primarily about the angular distribution of the dissipated power.

Spatial variations in tidal heating became relevant when spacecraft sent back incredible surface data showing that the Galilean satellites Io and Europa undergo strong tidal deformations and heating.
Tidal heating was the only explanation for Io's volcanism \citep{peale1978} but what was going on beneath the surface was a mystery.
Maybe the distribution of volcanoes could be used in order to constrain the internal structure of the satellite?
\citet{segatz1988} suggested that dissipation occurred either in the whole mantle or mostly in a thin asthenosphere close to the surface.
These two models predict completely different patterns of surface heat flux with maximum dissipation at the poles in the former case and at the equator in the latter (a mix of the two is of course possible).
Galileo data have often been interpreted as favoring the asthenospheric dissipation model though the issue remains controversial \citep{lopes1999,tackley2001b,kirchoff2011,veeder2012,hamilton2012}.
Another idea consists in using long wavelength topography as an indirect measure of the heat flux (valid if the topography is isostatically compensated), but the promising analysis of Voyager data \citep{ross1990} was not confirmed by Galileo measurements \citep{thomas1998}.
Lateral variations of surface heat flux however also depend on the heat transport mechanism which could be determinant.

On icy satellites like Europa, Titan and Enceladus, spatial variations of tidal heating are interesting for other reasons.
In these satellites, tidal heating can melt the ice at depth and create a global subsurface ocean.
The covering icy shell varies in thickness because of spatial variations in tidal heating and solar insolation.
\citet{ojakangas1989a,ojakangas1989b} computed icy shell thickness variations on Europa with the aim of predicting nonsynchronous rotation and polar wander.
\citet{nimmo2007} and \citet{nimmo2010} used the same method to predict long wavelength topography on Europa and on Titan, respectively.
Variations of the thickness of the icy shell (or more generally the lithosphere) also influence surface tectonics \citep{beuthe2010a}.

Finally, spatial variations of tidal heating are important because of their strong coupling to convection.
Several convection models have used as input tidal heating predicted by spherically stratified models \citep{tackley2001b,tobie2003,roberts2008}.
An important limitation of this approach is the neglect of lateral viscosity variations on tidal heat production, which can be taken into account by giving up the assumption of spherical symmetry and solving simultaneously for convection and tidal dissipation \citep{behounkova2010,han2010}.
Chaos terrain on Europa could be a visible result of spatially varying tidal heating enhanced by a local drop in viscosity.

Until now, predicting spatial patterns of tidal dissipation meant computing the dissipated power at every point within the body.
This laborious procedure obscures the link between the internal structure and the resulting pattern, and makes it difficult to look for all possible patterns generated by realistic internal structures.
In this paper, I show that the dissipated power at a given radius  within a spherically stratified body is the linear combination of three basic patterns that depend only on the tidal potential.
The coefficients weighting the patterns depend are radial functions which can be computed with standard methods developed for tidal deformation problems once the internal structure of the body has been specified.

I study the influence of the internal structure on dissipation patterns by computing the dissipation weight functions in various toy models without assuming a specific rheology.
The toy models are the thin icy shell above an ocean, the homogeneous body (relevant to a completely solid body with little stratification or to a solid core) and the incompressible two-layer body, either with a liquid core or with rigid core, the latter case being relevant to dissipation in a soft layer (such as an asthenosphere) above a more rigid layer.
It is well-known that dissipation patterns are completely different if dissipation occurs in the deep mantle and in a thin asthenosphere.
Besides these two classes of patterns, I show that dissipation in a thick asthenosphere leads to a third type of dissipation pattern, with maximum heating at the equator as in asthenospheric dissipation but with a lower content in harmonic degree four.
Toy models predict well dissipation patterns within real bodies though not their magnitude.
I will give three examples of this by computing dissipation weight functions for realistic internal structures of Europa, Titan and Io.

\section{Dissipated power}
\label{Section2}
\markright{2 DISSIPATED POWER\hfill}

\subsection{Power, strains and tidal potential}

In the micro approach to dissipation, the dissipated power within the planet or satellite is expressed in terms of tidal strains (this formula is derived in Appendix~A and compared with other expressions found in the literature).
If tides operate at only one angular frequency $\omega$, the dissipated power per unit volume averaged over one orbital period is given by
\begin{equation}
P =  \omega \mbox{\it Im}(\tilde\mu) \left( \tilde\epsilon_{ij}^{} \, \tilde\epsilon_{ij}^{\,*} - \frac{1}{3} \left| \tilde\epsilon \right|^2 \right) + \frac{\omega}{2} \, \mbox{\it Im}(\tilde{K}) \left| \tilde\epsilon \right|^2 \, ,
\label{PowerStrainMain}
\end{equation}
where $\tilde\mu$ (resp. $\tilde{K}$) is the complex shear (resp. bulk) modulus, $\tilde\epsilon_{ij}^{}$ is the Fourier transform of the strain tensor and $\tilde\epsilon$ is the trace of $\tilde\epsilon_{ij}^{}$.
All quantities implicitly depend on the frequency $\omega$ and on the point ${\bf x}$ within the planet where the power is evaluated.
In this paper, the tilde on viscoelastic parameters indicates that they are complex and frequency-dependent (the tilde is dropped if the parameters are purely elastic).

If there are several tidal frequencies, the total power averaged over time is a sum over these frequencies (interferences vanish, see Eq.~(\ref{AveragedPower})) so that it becomes essential to know the frequency dependence of the viscoelastic parameters (i.e.\ the rheology).
Unfortunately the rheology of planetary bodies is poorly constrained \citep{jackson2007,karato2008}.
Earth's mantle has been mainly studied at frequencies that are much higher (laboratory experiments), moderately higher (seismic attenuation and seismic anisotropy) or much lower (Chandler wobble and postglacial rebound) than tidal frequencies \citep{karato2010}.
It is indeed difficult to determine Earth's viscous response at tidal frequencies \citep[e.g.][]{benjamin2006,nakada2012} and even more so for other bodies.
Maxwell rheology is often used in studies of tidal dissipation because it is the simplest model in which the response changes from elastic to viscous as the frequency decreases.
It is however not clear how the Maxwell viscosity is related to the true viscosity of the material \citep{ross1986,bills2005b,sotin2009}.
Rheological models depending on more parameters such as the Andrade model \citep{castillo2011} or the extended Burgers model \citep{nimmo2012} could be more realistic.
Moreover there has been a long-standing debate on whether viscous deformations in Earth's mantle are mainly due to diffusion creep or to dislocation creep, corresponding to a linear or nonlinear rheology, respectively \citep{karato1993}.
In this paper, I assume that the rheology is linear without being more specific about it except in applications to real bodies for which I use the Maxwell model.
Besides I consider only the dominant tidal frequency; this restriction is appropriate for Solar System bodies, but it should be lifted for exoplanets with short orbital periods.

Dissipation within the Earth is generally much larger in shear than in uniform compression ($\mbox{\it Im}(\tilde\mu)\gg{\mbox{\it Im}(\tilde K)}$), at least in the seismic frequency band \citep{anderson1989}.
For this reason, most models of planetary dissipation assume that the bulk modulus is real and independent of frequency \citep[p.~142]{ranalli1995}, though it is not necessarily true in presence of a high fraction of partial melt \citep{schmeling1985}.
In Earth's asthenosphere, the ratio of bulk to shear dissipation could possibly reach 30\% \citep{durek1995}.
Bulk dissipation should thus be kept in mind when studying bodies (such as Io) where asthenospheric dissipation could be the dominant mechanism.

The strains appearing in Eq.~(\ref{PowerStrainMain}) are induced by tidal deformations which are in turn related to the tidal potential.
The tidal potential at the surface of the deformed body can be expanded in spherical harmonics of degree $\ell$ and order $m$ \citep{kaula1964}, each component being the superposition of several terms of angular frequencies $\omega_{\ell m j}$,
\begin{equation}
U(t,\theta,\phi) = Re \left( \sum_{\ell,m,j} \Phi_{\ell m j}(\theta,\phi) \, \exp \left( i \omega_{\ell m j} t \right) \right) \, ,
\end{equation}
where $Re(x)$ means real part of $x$, $\theta$ is the colatitude and $\phi$ is the longitude in a frame attached to the body.
The coefficients $\Phi_{\ell m j}(\theta,\phi)$ are defined by this equation if you know the tidal potential from the formulas in \citet{kaula1964}; they are not the same as the complex coefficients of the Fourier series of $U(t,\theta,\phi)$ since the frequencies $\omega_{\ell m j}$ are not all different.
If the orbital eccentricity and the obliquity of the body are not zero, an infinite number of terms (indexed by $j$) contribute to the potential at a given degree $\ell$ and order $m$.
I neglect all tidal perturbations except the dominant contribution of degree $\ell=2$ and I sum over the order $m$.
The component of the tidal potential at frequency $\omega$ can be written as
\begin{eqnarray}
U_\omega(t,\theta,\phi) &=& Re \left( \Phi_\omega(\theta,\phi) \, \exp \left( i \omega t \right) \right)
\nonumber \\
&=& \frac{1}{2} \, \Phi_\omega(\theta,\phi) \, \exp(i \omega t) + c.c. \, ,
\label{TidalFourier}
\end{eqnarray}
where $c.c.$ means complex conjugate.

Strains are related to the tidal potential through displacements.
If the internal structure of the body is spherically symmetric, displacements due to tides can be written as \citep{alterman1959,takeuchi1972,saito1974}
\begin{eqnarray}
\tilde u_r &=& y^{}_1(r) \, \Phi_\omega \, ,
\nonumber \\
\tilde u_\theta &=& y^{}_3(r) \, \frac{\partial \Phi_\omega}{\partial \theta} \, ,
\nonumber \\
\tilde u_\phi &=& y^{}_3(r) \, \frac{1}{\sin\theta} \, \frac{\partial \Phi_\omega}{\partial \phi} \, ,
\label{spheroidal}
\end{eqnarray}
where $(\tilde u_r,\tilde u_\theta,\tilde u_\phi$) are the Fourier components of the displacement defined as in Eqs.~(\ref{epsilonTF}).

The two radial functions $y_1(r)$ and $y_3(r)$ depend on the internal structure of the body and have the dimension of inverse acceleration.
These functions are part of a set of six radial functions $(y_1,...,y_6)$ that solve six coupled linear differential equations of first order (I drop from now on the explicit dependence on $r$).
Strains depend on $y_1$, $y_3$ and their derivatives.
If derivatives are eliminated with the help of the differential equations, strains also depend on the functions $y_2$ and $y_4$ respectively associated with the stresses $\sigma_{rr}$ and $\sigma_{r\theta}$ (or $\sigma_{r\phi}$).
The remaining functions $y_5$ and $y_6$ are related to the gravity potential.
The derivatives $(y_1',y_3')$ are related to $(y^{}_1,...,y^{}_4)$ by Eqs.~(82) of \citet{takeuchi1972}:
\begin{eqnarray}
r y_1' &=& \frac{1}{\tilde{K}+4\tilde\mu/3} \left(  r y^{}_2 - (\tilde{K}-2\tilde\mu/3) \left( 2 y^{}_1 - 6 y^{}_3 \right) \right) \, ,
\nonumber
\\
\frac{r y^{}_4}{\tilde\mu} &=& r y_3' - y^{}_3 + y^{}_1 \, .
\label{y4def}
\end{eqnarray}
The quantity $y^{}_4/\tilde\mu$ actually measures the shear strain (see Eqs.~(\ref{strainsRAD}) below).
In an incompressible medium, the first equation reduces to
\begin{equation}
r y_1' = - 2 y^{}_1 + 6 y^{}_3 \, .
\label{y1primebis}
\end{equation}
Since stresses with a radial component vanish at the surface $r=R$, the boundary conditions for $y^{}_2$ and $y^{}_4$ are
\begin{equation}
y^{}_2(R)=y^{}_4(R)=0 \, .
\label{bc}
\end{equation}
Beware that the definitions of $y_i$ vary between authors (I follow here \citet{takeuchi1972}).
In particular, the functions $(y_1,y_2,y_3,y_4,y_5,y_6)$ of \citet{sabadini2004} are equivalent to the functions $(y_1,y_3,y_2,y_4,-y_5,-y_6)$ of \citet{takeuchi1972}.

\citet{kaula1963,kaula1964} already used the $y^{}_i$ functions in his pioneering papers on tidal dissipation; his example was recently followed by \citet{tobie2005} and \citet{roberts2008}.
This formalism has the advantage that it is widely used in geophysics to compute the deformation of Earth's surface for complicated internal structures \citep{spada2011}.
\citet{peale1978} unfortunately returned to the method of \citet{love1911} which is difficult to apply correctly to structures more complicated than two layers of the same density.

In spherical coordinates, I divide strain components in three classes:
(1) the purely radial component $\epsilon_{rr}$ is the {\it radial strain}, (2) the purely angular components $(\epsilon_{\theta\theta},\epsilon_{\phi\phi},\epsilon_{\theta\phi})$ are the {\it tangential strains}, and (3) the mixed components $(\epsilon_{r\theta},\epsilon_{r\phi})$ are the {\it radial-tangential shear strains}.
I now insert Eqs.~(\ref{spheroidal}) into the strain-displacement equations in spherical coordinates \citep[][Eqs.~(20)]{takeuchi1972}.
Note that the non-diagonal strains $(e_{\theta\phi},e_{\phi{r}},e_{r\theta})$ in \citet{takeuchi1972} must be multiplied by $1/2$ if they are to be the components of the strain tensor.
The radial and shear strains read
\begin{eqnarray}
\tilde \epsilon_{rr} &=& y_1' \, \Phi_\omega \, ,
\nonumber \\
\tilde \epsilon_{r\theta} &=& \frac{1}{2} \frac{y^{}_4 }{\tilde\mu} \, \frac{\partial \Phi_\omega}{\partial \theta} \, ,
\nonumber\\
\tilde \epsilon_{r\phi} &=& \frac{1}{2} \, \frac{y^{}_4}{\tilde\mu} \, \frac{1}{\sin\theta} \, \frac{\partial \Phi_\omega}{\partial \phi} \, .
\label{strainsRAD}
\end{eqnarray}
Doing the same for the tangential strains, I get
\begin{eqnarray}
\tilde \epsilon_{\theta\theta} &=& \frac{1}{r} \left( y^{}_3 \, \bar{\cal O}_1 + y^{}_1 \right) \Phi_\omega \, ,
\nonumber\\
\tilde \epsilon_{\phi\phi} &=& \frac{1}{r} \left( y^{}_3 \,  \bar{\cal O}_2 + y^{}_1 \right) \Phi_\omega \, ,
\nonumber\\
\tilde \epsilon_{\theta\phi} &=& \frac{1}{r} \, y^{}_3 \, \bar{\cal O}_3 \, \Phi_\omega \, ,
\label{strainsTAN}
\end{eqnarray}
 where $\bar{\cal O}_{1,2,3}$ are differential operators of degree two on the sphere defined by Eqs.~(\ref{opODEF1})-(\ref{opODEF2}).

Though Eqs.~(\ref{strainsRAD})-(\ref{strainsTAN}) are well-known, their tensorial structure with respect to rotations is always ignored in geophysics:
$\tilde \epsilon_{rr}$ is a scalar, $\tilde \epsilon_{r\alpha}$ are the components of a vector and $\tilde \epsilon_{\alpha\beta}$ are the components of a second-order symmetric tensor (Greek indices denote $\theta$ or $\phi$).
The tensorial character of strains under rotations is the key to simplifying the formula for the dissipated power.

\subsection{Strain invariants}
\label{StrainInvariants}

From a technical point of view, this is the key section of the paper.
In order to evaluate the dissipated power in terms of the tidal potential, I need to compute in Eq.~(\ref{PowerStrainMain}) the quantities $\left| \tilde\epsilon \right|^2$ and $\tilde\epsilon^{}_{ij} \, \tilde\epsilon^{\,*}_{ij}$  which are invariant under arbitrary coordinate transformations \citep{malvern1969}.
This general invariance is not obvious once the strains have been expressed in spherical coordinates.
Nevertheless invariance under rotations should remain obvious in that basis.
It is indeed easy to construct rotationally invariant terms: take the product of two scalars (such as $\tilde \epsilon_{rr}$), take the scalar product of two vectors (such as $\tilde \epsilon_{r\alpha}$), contract two tensors (such as $\tilde \epsilon_{\alpha\beta}$), or take the trace of a tensor.

Let us define
\begin{eqnarray}
E_{dilat} &=& r^2 \left| \tilde\epsilon \right|^2 \, ,
\nonumber \\
E_2 &=& r^2 \, \tilde\epsilon^{}_{ij} \, \tilde\epsilon^{\,*}_{ij} \, ,
\end{eqnarray}
with the subscript {\it dilat} denoting {\it dilatation}, since $\tilde\epsilon$ is the infinitesimal change of volume.

In spherical coordinates, $E_{dilat}$ reads
\begin{equation}
E_{dilat} = r^2 \left| \tilde\epsilon_{rr} + \tilde\epsilon_{\theta\theta} + \tilde\epsilon_{\phi\phi} \right|^2 \, .
\label{E1def}
\end{equation}
The terms $\tilde\epsilon_{rr}$ (scalar) and $\tilde\epsilon_{\theta\theta} + \tilde\epsilon_{\phi\phi}$ (trace of a tensor) are each invariant under rotations but it is easier to keep them together in Eq.~(\ref{E1def}).
$E_2$ is the sum of three terms that are each invariant under rotations:
\begin{equation}
E_2 = E_{radial} + E_{shear} + E_{tangent} \, ,
\label{E2}
\end{equation}
where
\begin{eqnarray}
E_{radial} &=&  r^2 \, \tilde\epsilon^{}_{rr} \, \tilde\epsilon^{\,*}_{rr} \, ,
\nonumber \\
E_{shear} &=& 2 r^2 \left(   \tilde\epsilon^{}_{r\theta} \, \tilde\epsilon^{\,*}_{r\theta} +  \, \tilde\epsilon^{}_{r\phi} \, \tilde\epsilon^{\,*}_{r\phi}  \right) \, ,
\nonumber \\
E_{tangent} &=& r^2 \left( \tilde\epsilon^{}_{\theta\theta} \, \tilde\epsilon^{\,*}_{\theta\theta} + \tilde\epsilon^{}_{\phi\phi} \, \tilde\epsilon^{\,*}_{\phi\phi} + 2 \, \tilde\epsilon^{}_{\theta\phi} \, \tilde\epsilon^{\,*}_{\theta\phi} \right) \, .
\label{E2subinv}
\end{eqnarray}
$E_{radial}$, $E_{shear}$ and $E_{tangent}$ are invariant because they are respectively the product of two scalars,  the scalar product of two vectors and the contraction of two tensors.

Inserting Eqs.~(\ref{strainsRAD})-(\ref{strainsTAN}) and the first Eqs.~(\ref{opD4der}) into Eq.~(\ref{E1def}), I write the first invariant as
\begin{equation}
E_{dilat} = \left| \left( r y_1' + 2 y^{}_1 +  y^{}_3 \, \Delta \right) \Phi_\omega \right|^2  \, ,
\label{E1}
\end{equation}
where $\Delta$ is the spherical Laplacian defined by Eq.~(\ref{laplacianDEF1}).
Inserting Eqs.~(\ref{strainsRAD}) and (\ref{opD2der}) into the first and second Eqs.~(\ref{E2subinv}), I write the radial and shear invariants as
\begin{eqnarray}
E_{radial} &=& \left| r y_1' \right|^2 \left| \Phi_\omega \right|^2 \, ,
\nonumber
\\
E_{shear} &=& \frac{1}{2} \left| \frac{r y_4}{\tilde\mu} \right|^2 {\cal D}_2(\Phi_\omega,\Phi^*_\omega) \, .
\label{Eshear}
\end{eqnarray}
Inserting Eqs.~(\ref{strainsTAN}) and (\ref{opD4der}) into the third Eqs.~(\ref{E2subinv}), I write the tangential invariant as the sum of two terms invariant under rotations:
\begin{equation}
E_{tangent} = E_{tan1} + E_{tan2} \, ,
\end{equation}
where
\begin{eqnarray}
E_{tan1} &=& \left| y^{}_3 \right|^2 \left( {\cal D}_4 \left(\Phi_\omega ,  \Phi_\omega^* \right) - \left| \Delta \Phi_\omega \right|^2 /2 \right) \, ,
\nonumber
 \\
E_{tan2} &=& 2 \left| \left( y^{}_1 + y^{}_3 \, \Delta/2 \right) \Phi_\omega \right|^2 \, .
\label{Etan2}
\end{eqnarray}

Noting that $\Phi_\omega$ is a spherical harmonic degree two, I insert Eqs.~(\ref{eigen}) and (\ref{opD4DEF3}) into Eqs.~(\ref{E1}) to (\ref{Etan2}):
\begin{eqnarray}
E_{dilat} &=&  \left| r y_1' + 2 y^{}_1 + \delta_2 \, y^{}_3 \right|^2 \left| \Phi_\omega \right|^2 \, ,
\nonumber \\
E_{radial} &=& \left| r y_1' \right|^2 \left| \Phi_\omega \right|^2 \, ,
\nonumber \\
E_{shear} &=& (1/4) \left| r y^{}_4/\tilde\mu \right|^2 \left( \Delta - 2\delta_2 \right) \left| \Phi_\omega \right|^2 \, ,
\nonumber \\
E_{tan1} &=& (1/4)
\left| y_3 \right|^2 \left( \Delta \Delta - 2 \left( 2 \delta_2 + 1 \right) \Delta + 2 \delta_2 \left( \delta_2 + 2 \right) \right)
\left| \Phi_\omega \right|^2 \, ,
\nonumber\\
E_{tan2} &=& 2 \left| y^{}_1+ (\delta_2/2) \, y^{}_3 \right|^2
\left| \Phi_\omega \right|^2  \, ,
\label{InvFact}
\end{eqnarray}
where $\delta_2=-6$ is the degree-two eigenvalue of the spherical Laplacian $\Delta$ (see Eq.~(\ref{eigen})).
Strain invariants are now factorized into radial and angular parts.

\subsection{Factorized power}
\label{FactorizedPower}

The last step consists in rewriting the dissipated power given by Eq.~(\ref{PowerStrainMain}) in terms of the various strain invariants of Section~\ref{StrainInvariants}:
\begin{equation}
P = \frac{\omega}{r^2} \, \mbox{\it Im}(\tilde\mu) \left(E_{radial} + E_{shear} + E_{tan1} + E_{tan2} - \frac{1}{3} \, E_{dilat} \right) + \frac{\omega}{2r^2} \, \mbox{\it Im}(\tilde{K}) \, E_{dilat} \, .
\label{PowerStrainInv}
\end{equation}
Examining the factorized strain invariants (Eqs.~(\ref{InvFact})), we see that the power at a given radius $r$ depends on the angles $(\theta,\phi)$ through a linear combination of three factors,
\begin{equation}
\left| \Phi_\omega \right|^2, \, \Delta\left| \Phi_\omega \right|^2, \, \Delta\Delta\left| \Phi_\omega \right|^2 \, ,
\label{basis1}
\end{equation}
showing that there are at most three independent spatial patterns.

The three invariants without derivatives of $\left| \Phi_\omega \right|^2$ can be combined as
\begin{equation}
E_{radial} + E_{tan2} - E_{dilat}/3
= (2/3) \left| r y_1' - y^{}_1 - (\delta_2/2) \,  y^{}_3 \right|^2 \left| \Phi_\omega \right|^2 \, .
\label{identity}
\end{equation}

The three functions (\ref{basis1}) are not the best choice of basic angular functions because of compensations between the three terms.
Instead I will maintain as much as possible the distinction between the radial, radial-tangential shear and tangential dissipative terms.
These contributions have indeed very different magnitudes in very different interior models (see Section~\ref{TwoLayer}).
Besides $\left| \Phi_\omega \right|^2$, I thus propose to use as basic angular functions the combinations appearing in $E_{shear}$ and $E_{tan1}$ (see Eqs.~(\ref{InvFact})):
\begin{eqnarray}
\Psi^{}_A &=& (nR)^{-4} \left| \Phi_\omega \right|^2 \, ,
\nonumber \\
\Psi^{}_B &=& \frac{1}{12} \left( \Delta + 12 \right) \Psi^{}_A \, ,
\nonumber \\
\Psi^{}_C &=& \frac{1}{48} \left( \Delta\Delta + 22 \Delta + 48\right) \Psi^{}_A \, .
\label{PsiABC} 
\end{eqnarray}
These functions are dimensionless since the factor $(nR)^{-4}$ absorbs the dimension of the squared tidal potential ($n$ is the mean motion and $R$ is the surface radius, see Eq.~(\ref{TidalPot2})).
The numbers 12, 22 and 48 result from setting $\delta_2=-6$ in Eqs.~(\ref{InvFact}).
The patterns $\Psi^{}_B$ and $\Psi^{}_C$ are normalized so that the terms without derivatives are the same in the three functions.
In the following, the index $J$ collectively refers to $(A,B,C)$.

Inserting Eqs.~(\ref{InvFact}), (\ref{identity}) and (\ref{PsiABC}) into Eq.~(\ref{PowerStrainInv}), I write the dissipated power as
\begin{equation}
P =  \frac{\omega(nR)^4}{2r^2} \left( \mbox{\it Im}(\tilde\mu) \left( f^{}_A \, \Psi^{}_A + f^{}_B \, \Psi^{}_B + f^{}_C \, \Psi^{}_C \right) + \mbox{\it Im}(\tilde K) \, H^{}_K \, \Psi^{}_A \right)  ,
\label{fullpower}
\end{equation}
where the {\it weight functions} $f^{}_J$ and $H^{}_K$ are positive and depend on the internal radial functions $y^{}_i$:
\begin{eqnarray}
f^{}_A &=& \frac{4}{3} \left| r y_1' - y^{}_1 + 3 y^{}_3 \right|^2 \, ,
\nonumber \\
f^{}_B &=& 6 \left| \frac{ r y^{}_4}{\tilde\mu}  \right|^2  \, ,
\nonumber \\
f^{}_C &=& 24 \left| y^{}_3 \right|^2 \, ,
\nonumber \\
H^{}_K &=& \left| r y_1' + 2 y^{}_1 - 6 y^{}_3 \right|^2 \, .
\label{weights}
\end{eqnarray}
The weights functions have the dimension of squared inverse acceleration.
When averaging over angles, I will need the sum of the weight functions $f^{}_J$ denoted as
\begin{equation}
H_\mu =  f^{}_A  + f^{}_B  + f^{}_C \, .
\label{defHmu}
\end{equation}
In practice, $y_1'$ is evaluated from $y^{}_1$, $y^{}_2$ and $y^{}_3$ (see Eqs.~(\ref{y4def})).
The terms $B$, $C$ and $K$ are associated with radial-tangential shear, tangential and bulk dissipation, respectively, whereas the term $A$ is a combination of radial and tangential dissipation.
The terms $A$ and $K$ depend on the same angular function and can be grouped into one term.
Formulas (\ref{PsiABC}), (\ref{fullpower}) and (\ref{weights}) are the central results of this paper.

At the surface, weight functions can be related to the displacement Love numbers which characterize how the body responds to tidal forcing.
The displacements $y_1(R)$ and $y_3(R)$ are indeed proportional to the tidal Love numbers $h_2$ and $l_2$, while $y_2(R)$ and $y_4(R)$ vanish because of Eqs.~(\ref{bc}):
\begin{equation}
 \left( y^{}_1, y^{}_2 , y^{}_3, y^{}_4 \right) \Big|_{r=R} = \left(  \frac{h_2}{g} , 0,  \frac{l_2}{g}, 0 \right) \, ,
\label{LoveDispl}
\end{equation}
where $g$ is the surface gravity.
Moreover, the first of Eqs.~(\ref{y4def}) evaluated at the surface gives
\begin{equation}
R \, y_1'(R) = - \frac{2\tilde\nu}{1-\tilde\nu} \frac{h_2 - 3 l_2}{g} \, ,
\label{y1primesurf}
\end{equation}
where I replaced $\tilde \mu$ and $\tilde K$ by Young's modulus $\tilde E$ and Poisson's ratio $\tilde\nu$:
\begin{eqnarray}
\tilde K &=& \frac{\tilde E}{3(1-2\tilde\nu)} \, ,
\nonumber \\ 
\tilde\mu &=& \frac{\tilde E}{2(1+\tilde\nu)} \, .
\label{ModuliRelation1}
\end{eqnarray}
I compute the weight functions at the surface with Eqs.~(\ref{weights}) to (\ref{y1primesurf}):
\begin{eqnarray}
\left( f^{}_A \, , f^{}_B \, , f^{}_C \,  \right) \Big|_{r=R} &=&  \frac{|h_2|^2}{g^2}
\left(
\frac{4}{3} \left| \frac{1+\tilde\nu}{1-\tilde\nu} \right|^2 \left| 1- 3 \frac{l_2}{h_2} \right|^2
, \, 0 \, ,
24 \left| \frac{l_2}{h_2} \right|^2
\right) \, ,
\nonumber \\
H^{}_K \Big|_{r=R} &=& 4 \, \frac{|h_2|^2}{g^2} \left| \frac{1-2\tilde\nu}{1-\tilde\nu} \right|^2 \left| 1 -3 \frac{l_2}{h_2} \right|^2 \, .
\label{WeightFunSurface}
\end{eqnarray}
The amplitude of tidal deformation (quantified by $|h_2|$) varies by orders of magnitude depending on the global structure of the body.
By contrast, the ratio $|l_2/h_2|$ typically ranges from 0.2 to 0.3 \citep[][Fig.~10]{beuthe2010a}.
Eqs.~(\ref{WeightFunSurface}) show that ratios of weight functions at the surface are insensitive to the magnitude of the viscoelastic response $|h_2|$.
In good approximation, this property also holds inside the body.
It is thus useful to define dimensionless {\it rescaled weight functions} by
\begin{equation}
\left( \bar f^{}_J , \bar H^{}_\mu, \bar H^{}_K \right) = \left( \frac{|h_2|^2}{g^2} \right)^{-1} \left( f^{}_J , H^{}_\mu, H^{}_K \right) \, .
\label{rescaled1}
\end{equation}
For example, rescaled weight functions do not depend on the rheology if the body is incompressible and homogeneous (Section~\ref{HomogeneousBody}) or if it is consists of an incompressible viscoelastic mantle above a liquid core of the same density (Section~\ref{TwoLayer}).

Weight functions have a few properties that do not depend on a specific internal structure. 
First, $f^{}_B$ always vanishes at the surface (see Eqs.~(\ref{WeightFunSurface})) and at internal fluid/solid interfaces (where a condition similar to Eq.~(\ref{bc}) holds).
Pattern~B can thus be disregarded close to such boundaries, for example in a thin icy shell above an ocean.
Second, $f^{}_A$ and $f^{}_C$ vanish in the incompressible limit at the boundary of an infinitely rigid layer because $y^{}_1=y^{}_3=0$ there.
Dissipation thus approximately follows Pattern~B close to rigid/viscoelastic boundaries, for example at the bottom of an icy shell in contact with a silicate mantle.
Third, I can relate $H^{}_K$ to $f^{}_A$ at the surface with Eqs.~(\ref{WeightFunSurface}):
\begin{equation}
\left.\frac{H^{}_K}{f^{}_A}\right|_{r=R} = \frac{4}{3} \, \left| \frac{\tilde \mu}{\tilde K} \right|^2 = 3 \left| \frac{1-2\tilde\nu}{1+\tilde\nu} \right|^2 \, .
\label{bulksurf}
\end{equation}
This ratio is less than one-half for silicates ($\nu\approx1/4$) and less than one-fifth for ice ($\nu\approx1/3$).
If the material is incompressible, it is possible to go further by applying Eq.~(\ref{y1primebis}):
\begin{eqnarray}
f^{}_A &=& 12 \left|  y^{}_1 - 3 y^{}_3 \right|^2 \, ,
\nonumber \\
H^{}_K &=& 0 \, ,
\label{weightsIncomp}
\end{eqnarray}
so that $H^{}_K$ is expected to be much smaller than $f^{}_A$ at all radii if incompressibility is a good approximation.
Since dissipation also receives contributions from terms {\it B} and {\it C}, the contribution of bulk dissipation to total dissipation remains in general minor even if $\mbox{\it Im}(\tilde K)\approx \mbox{\it Im}(\tilde\mu)$ (see discussion in Section~\ref{Section2}).

\section{Spatial patterns, global power and surface flux}
\label{Section3}
\markright{3 SPATIAL PATTERNS, GLOBAL POWER AND SURFACE FLUX\hfill}

\subsection{Harmonic content of angular functions}

The angular functions $\Psi^{}_J$ defined by Eqs.~(\ref{PsiABC}) can be computed from the expansion of the squared norm of the tidal potential in spherical harmonics.
Since the tidal potential is of degree two, the expansion of its squared norm only includes functions of degrees zero, two and four:
\begin{equation}
\left| \Phi_\omega \right|^2 = (nR)^4 \left( \Psi^{}_0 + \Psi^{}_2 + \Psi^{}_4 \right) \, ,
\label{expansion1}
\end{equation}
where
\begin{equation}
\Psi^{}_\ell = \sum_{m=0}^\ell  P^{}_{\ell{m}}(\cos\theta)  \left( a^{}_{\ell{m}} \cos m\phi + b^{}_{\ell{m}} \sin m\phi \right) \, .
\label{expansion2}
\end{equation}
The constant $\Psi^{}_0=a^{}_{00}$ is the spatial average of the squared norm of the nondimensional tidal potential,
\begin{equation}
\Psi^{}_0 = \frac{(nR)^{-4}}{4\pi} \int_S \left| \Phi_\omega \right|^2 \sin\theta \, d\theta d\phi \, ,
\label{Psi0}
\end{equation}
while $\Psi^{}_2$ and $\Psi^{}_4$ represent spatial variations with zero average.
Table~\ref{TableSqCoeff} gives the numerical values of the coefficients for a satellite in synchronous rotation with non-zero eccentricity and obliquity.

\begin{table}\centering
\ra{1.3}
\small
\caption{Squared norm of the nondimensional tidal potential for synchronous rotation: non-zero spherical harmonic coefficients $(a^{}_{\ell{m}},b^{}_{\ell{m}})$ of Eq.~(\ref{expansion2}) for eccentricity tides ($I=0$) and for obliquity tides ($e=0$).
The coefficients are computed with Eqs.~(\ref{TidalPot2}) and (\ref{CoeffEven})-(\ref{CoeffOdd}).
The symbols $s_p$ and $c_p$ denote $\sin\omega_p$ and $\cos\omega_p$, respectively, where $\omega_p$ is the argument of pericentre.}
\vspace{1.5mm}
\begin{tabular}{@{}lcccccc@{}}
\hline
\vspace{0.3mm}
  & $a^{}_{00}$ & $a^{}_{20}$ & $a^{}_{22}$ & $a^{}_{40}$ & $a^{}_{42}$ & $a^{}_{44}$ \\
\cline{2-7}
eccentricity tides ($\times \, e^2$) & $21/5$ & $- 33/7$ & $9/14$ & $387/140$ & $-27/140$ & $-3/160$ \\
obliquity tides ($\times \sin^2 \! I$) & $3/5$ & $3/7$ & $3/14$ & $-36/35$ & $3/35$ & 0 \\
\hline
\vspace{0.3mm}
  & $a^{}_{21}$ & $a^{}_{41}$ & $a^{}_{43}$ & $b^{}_{21}$ & $b^{}_{41}$ & $b^{}_{43}$ \\
\cline{2-7}
\vspace{0.3mm}
interference ($\times \, e\sin I$) & $(6/7)s_p$ & $(-81/70)s_p$ & $(9/140)s_p$ & $(12/7)c_p$& $(-18/35)c_p$& $(3/35)c_p$\\
\hline
\end{tabular}
\label{TableSqCoeff}
\end{table}%

\begin{figure}
   \centering
   \includegraphics[width=14cm]{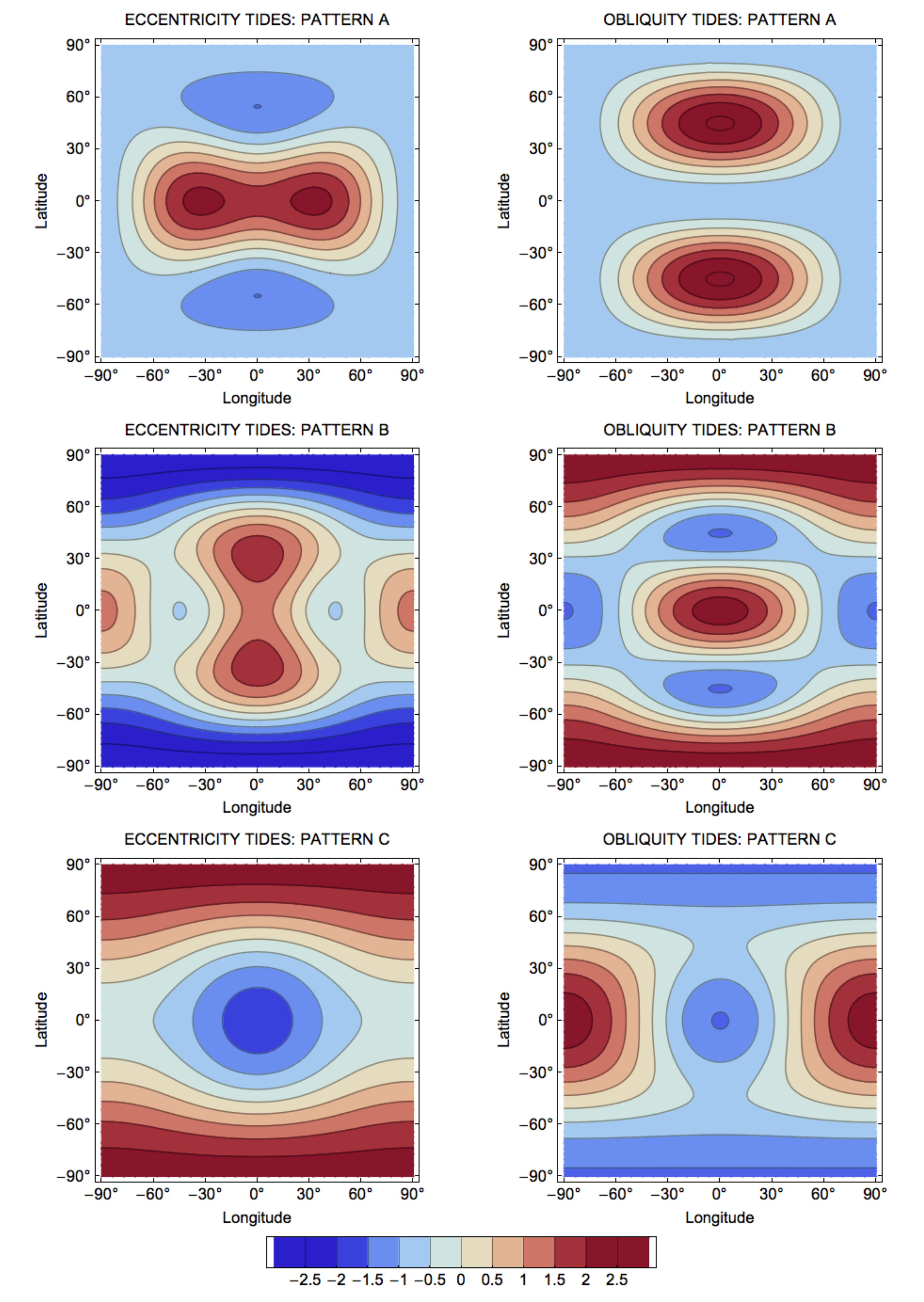}
   \caption{Basic patterns of tidal heating  for eccentricity tides (left panels) and obliquity tides (right panels) for a body in synchronous rotation.
   In each case, the mean value has been subtracted from the angular function $\Psi^{}_J$ before normalizing it by its standard deviation.
   The origin of coordinates coincides with the sub-primary point when the primary is at pericentre (eccentricity tides) or at the ascending node (obliquity tides).
   Patterns are repeated from $90^\circ$ to $-90^\circ$. }
   \label{FigPatterns}
\end{figure}

In terms of the harmonic functions $\Psi_\ell$, the angular functions $\Psi^{}_J$ read
\begin{eqnarray}
\Psi^{}_A &=& \Psi^{}_0 + \Psi^{}_2 + \Psi^{}_4 \, ,
\nonumber \\
\Psi^{}_B &=& \Psi^{}_0 + \frac{1}{2} \, \Psi^{}_2 - \frac{2}{3} \,  \Psi^{}_4  \, ,
\nonumber \\
\Psi^{}_C &=& \Psi^{}_0 - \Psi^{}_2 + \frac{1}{6} \, \Psi^{}_4 \, .
\label{PsiABCsh}
\end{eqnarray}
The angular dependence cancels in the following combination:
\begin{equation}
\Psi^{}_A + 2 \, \Psi^{}_B + 2 \, \Psi^{}_C = 5 \, \Psi^{}_0 \, .
\label{combili}
\end{equation}
Eqs.~(\ref{PsiABCsh}) give us a qualitative understanding of the spatial patterns.
$\Psi^{}_A$ and $\Psi^{}_B$ mainly differ by the sign of the degree-four term.
By contrast, the term of degree two in $\Psi^{}_C$ is of opposite sign with respect to $\Psi^{}_A$ (or $\Psi^{}_B$) and the degree-four contribution to $\Psi^{}_C$ is much smaller, even though $\Psi^{}_C$ contains derivatives of the fourth order.
The harmonic content of the angular functions can be measured with the variance (see Eq.~(\ref{VarianceDef})) if one knows the tidal potential.
Table~\ref{TableDegreeContent} shows that the content of the variance in degree four is much larger for obliquity tides than for eccentricity tides though $\Psi^{}_C$ is still mainly of degree two.
Fig.~\ref{FigPatterns} shows the basic patterns of tidal heating (angular functions $\Psi^{}_J$) for a body in synchronous rotation with either eccentricity tides or obliquity tides, drawn from Eqs.~(\ref{PsiABCsh}) in which I substitute Eq.~(\ref{expansion2}) and the values of Table~\ref{TableSqCoeff}.

\begin{table}\centering
\ra{1.3}
\small
\caption{Harmonic content of the basic spatial patterns for synchronous rotation: contribution of degree four (in \%) to the variance of $\Psi^{}_J$ for eccentricity tides ($I=0$) and for obliquity tides ($e=0$). The remainder of the variance is due to degree two.}
\vspace{1.5mm}
\begin{tabular}{@{}lcccc@{}}
\hline
\vspace{0.3mm}
& $\Psi^{}_A$ & $\Psi^{}_B$ & $\Psi^{}_C$ & unit \\
\hline
\vspace{0.3mm}
eccentricity tides & 30 & 44 & 1 & \%\\
obliquity tides & 64 & 76 & 5 & \% \\
\hline
\end{tabular}
\label{TableDegreeContent}
\end{table}%

The relative importance of the three patterns can be determined from the comparison of the weight functions but one should keep in mind that the $\Psi^{}_J$ do not have the same variance.
I thus define equal-variance weight functions $\hat f^{}_J$ through
\begin{equation}
\hat f^{}_J = \frac{sd(\Psi^{}_J)}{sd(\Psi^{}_A)} \, \bar f^{}_J \hspace{1cm} (J=A,B,C) \, .
\label{fABCnorm}
\end{equation}
The standard deviations $sd(\Psi_J)$ are tabulated in Table~\ref{TableVariance} for synchronous rotation (eccentricity tides only or obliquity tides only).
For eccentricity tides, $\hat f^{}_B/\bar f^{}_B= 0.56$ and $\hat f^{}_C/\bar f^{}_C= 0.84$.
For obliquity tides, $\hat f^{}_B/\bar f^{}_B= \hat f^{}_C/\bar f^{}_C= 0.61$.

\begin{table}\centering
\ra{1.7}
\small
\caption{Standard deviation of the basic spatial patterns for synchronous rotation.}
\vspace{1.5mm}
\begin{tabular}{@{}lccc@{}}
\hline
& $sd(\Psi^{}_A)$ & $sd(\Psi^{}_B)$ & $sd(\Psi^{}_C)$  \\
\hline
\vspace{0.3mm}
eccentricity tides ($\times \, e^2$) & $\frac{6}{5} \, \sqrt{\frac{38}{7}} = 2.80$ & $\frac{3}{5} \, \sqrt{\frac{47}{7}}=1.55$ & $\frac{3}{5} \, \sqrt{\frac{107}{7}}=2.35$ \\
\vspace{0.7mm}
obliquity tides ($\times \sin^2 \! I$) & $\frac{6}{5} \, \sqrt{\frac{2}{7}} = 0.64$ & $\frac{3}{5} \, \sqrt{\frac{3}{7}} = 0.39$ & $\frac{3}{5} \, \sqrt{\frac{3}{7}} =0.39$ \\
\hline
\end{tabular}
\label{TableVariance}
\end{table}%

Instead of $\Psi^{}_J$, I could use $\Psi^{}_\ell$ as basic angular functions and express the spatially varying part of the power as a linear combination of only two functions ($\Psi_2$ and $\Psi_4$). 
The radial weights associated with $\Psi_2$ and $\Psi_4$, however, are not always positive.
For example, the degree-two component of the power has a minimum either at the poles or at the equator, depending on the sign of the radial function.
Physically, these two cases are interpreted as different spatial patterns.

\subsection{Global power}
\label{GlobalPower}

I show here that the dissipated power integrated over the volume of the body is equal to the global power computed with the macro approach to tidal heating discussed in the Introduction.
The angular average of the local power at radius $r$ can be read from Eqs.~(\ref{fullpower}), (\ref{defHmu}) and (\ref{PsiABCsh}):
\begin{equation}
P^{}_0 = \frac{\omega{(nR)^4}}{2r^2} \left( \mbox{\it Im}(\tilde\mu) \, H^{}_\mu+ \mbox{\it Im}(\tilde K) \, H^{}_K \right) \Psi_0 \, .
\label{defP0}
\end{equation}
$H^{}_\mu$ and $H^{}_K$ have the same dependence on the functions $y^{}_i$ as the radial sensitivity functions $H^{}_\mu$ and $H^{}_K$ in \citet{tobie2005} (my formulas are more compact but strictly equivalent).
For a satellite in synchronous rotation undergoing eccentricity tides, $\Psi_0=21e^2/5$ (see Table~\ref{TableSqCoeff}) and $n=\omega$, in which case $P^{}_0$  is identical to the function $h_{tide}(r)$ of \citet{tobie2005} (their Eq.~(37)) except for a difference in sign explained below.
After having shown that the global power is equal to $4\pi\int{h_{tide}}(r)r^2dr$, these authors correctly interpret $h_{tide}(r)$ as the radial distribution of the dissipation rate per unit volume averaged over angles.

Using variational principles, \citet{tobie2005} prove that conservation of energy relates the sum of the strain and kinetic energies to the potential energy, the former two being integrated throughout the body whereas the latter is computable at the surface.
The dissipated part of this energy balance results in
\begin{equation}
\int_0^R \left( \mbox{\it Im}(\tilde\mu) \, H^{}_\mu + \mbox{\it Im}(\tilde K) \, H^{}_K \right) dr = - \frac{5 R}{4 \pi G} \, \mbox{\it Im}(k_2)\, ,
\label{ConsEnergy}
\end{equation}
where $R$ is the radius of the tidally perturbed body, $k_2$ is the tidal gravity Love number of degree two and $G$ is the gravitational constant.
I obtain the correct sign in the right-hand side of Eq.~(\ref{ConsEnergy}) by correcting Eqs.~(34)-(35) of \citet{tobie2005} as follows: $y_5(R_S)$ must be replaced by its complex conjugate in their Eq.~(34) so that the left-hand side of their Eq.~(35) has an additional minus sign.

The local power integrated over the volume of the body is thus given by
\begin{eqnarray}
\dot{E} &=& \int P \, dV 
\nonumber \\
&=&  4 \pi \int P^{}_0 \, r^2 dr
\nonumber \\
&=& - \frac{5\omega{R}}{2G} \,  \mbox{\it Im}(k_2) \, (nR)^4 \, \Psi^{}_0 \, .
\label{Edot1}
\end{eqnarray}
This equation agrees with the formula that \citet{zschau1978} obtained for the energy dissipated during one tidal cycle by considering the dephasing between the tidal potential and the potential induced by tidal deformations (the formulation in terms of $\mbox{\it Im}(k_2)$ was introduced by \citet{platzman1984}).
Therefore Eqs.~(\ref{defP0}), (\ref{ConsEnergy}) and (\ref{Edot1}) prove the equivalence between the micro and macro approaches to tidal heating.

Substituting the values of Table~\ref{TableSqCoeff} into Eq.~(\ref{Edot1}) and setting $n=\omega$, I get the well-known formula for the total dissipated power within a satellite in synchronous rotation \citep{cassen1980,segatz1988,chyba1989},
\begin{equation}
\dot{E} = - \mbox{\it Im}(k_2) \, \frac{(\omega{R})^5}{G} \, \left( \frac{21}{2} \, e^2 + \frac{3}{2} \, \sin^2 I \right) \, ,
\label{Edot2}
\end{equation}
which is valid up to second order in eccentricity $e$ and obliquity $I$ (terms proportional to $e\sin{I}$ vanish when averaged over angles).
In the macro approach to tidal heating, the total power is often expressed in terms of the quantity $1/Q$ which \citet{segatz1988} define through
\begin{equation}
\mbox{\it Im}(k_2) = - \frac{|k_2|}{Q} \, .
\label{defQ}
\end{equation}
I inserted a minus sign in this equation because $Q$ is usually assumed to be positive whereas $\mbox{\it Im}(k_2)$ is negative.
There is however no simple relation between this $1/Q$ and the specific dissipation function of the same name \citep{munk1960}, as discussed by \citet{zschau1978} and \citet{segatz1988}.
Contrary to what is sometimes said in the literature, Eq.~(\ref{Edot2}) is valid when the body is inhomogeneous (but spherically stratified) and compressible.
The macroscopic derivation of this equation does not even require a linear rheology.
Note that existing extensions of Eq.~(\ref{Edot2}) to arbitrary eccentricity and obliquity \citep{wisdom2008,levrard2008} assume that $1/Q$ is proportional to frequency.
While this assumption makes computations easy, it does not correspond to any plausible rheology \citep{efroimsky2007}.
Eq.~(\ref{Edot2})  thus remains the best available formula for the total dissipated power within a satellite in synchronous rotation if the eccentricity and obliquity are not too large.                                                                                

In the Solar System, the obliquity of large satellites other than the Moon has not yet been observed except for Titan \citep{stiles2008,stiles2010} but it is theoretically predicted to be very small \citep{gladman1996,peale1999,bills2005a,baland2012}.
Thus obliquity tides are at present negligible for large satellites except the Moon where they contribute about 40\% of the tidal heating.
Heating patterns are however not observable on the Moon because radiogenic heating is dominant.
Though Moon's obliquity may have been larger in the past, tidal heating due to obliquity tides was always smaller than radiogenic heating \citep{peale1978,wisdom2006}.
Regarding exoplanets, \citet{winn2005} suggested that obliquity tides account for bloated `hot Jupiters'  but this idea has been criticized on the grounds that the corresponding Cassini state is unstable \citep{fabrycky2007,levrard2007,peale2008}.
Tides could also be caused by forced librations in longitude \citep{wisdom2004}, the effect of which can be computed by adding new terms proportional to $P_{22}$ in the tidal potential (Eq.~(\ref{TidalPot1})).

\subsection{Surface flux}
\label{SurfaceFlux}

Observations from space give indications on the distribution of surface heat flux but not on the dissipated power at depth.
Computing angular variations of the surface heat flux thus involves assumptions about heat transport.
Let me first define $\chi^{}_J$ as the fraction of the global power $\dot{E}$ due to the dissipation term $J$,
\begin{equation}
\chi^{}_J = \left( - \frac{5R}{4\pi{G}} \,\mbox{\it Im}(k_2) \right)^{-1} \int_0^R  \mbox{\it Im}(\tilde\mu) \, f^{}_J \, dr \, ,
\label{chiJ}
\end{equation}
with a similar formula for $\chi^{}_K$ ($f^{}_J\rightarrow{f^{}_K}$ and $\tilde\mu\rightarrow{\tilde{K}}$).
By definition, $\sum_J\chi^{}_J+\chi^{}_K=1$.

If the heat is radially transported to the surface, the surface heat flux due to tidal dissipation is given by
\begin{equation}
{\cal F}(\theta,\phi) = {\cal F}^{}_0 \left(  \left(\chi^{}_A + \chi^{}_K \right) \, \frac{\Psi^{}_A}{\Psi^{}_0} + \chi^{}_B \, \frac{\Psi^{}_B}{\Psi^{}_0} + \chi^{}_C \, \frac{\Psi^{}_C}{\Psi^{}_0} \right) \, .
\end{equation}
where ${\cal F}^{}_0$ is the spatially-averaged heat flux at the surface,
\begin{equation}
{\cal F}^{}_0 = \frac{\dot E}{4 \pi R^2} \, .
\end{equation}
Under the assumption of radial heat transport, the coefficients $\chi^{}_J$ are the weights for the three basic angular patterns of the surface flux.
Non-radial transport mechanisms within the planet tend to average lateral variations in heating.
On the other hand, an analysis including a more realistic modeling of heat transport within the planet (for example by convection) should also take into account lateral variations in viscosity due to tidal heating itself.
I further discuss this topic in the Conclusions.
Finally the heat passing through the crust can be channelled into heat pipes (for Io) or fractures (Enceladus' tiger stripes), thus further modifying the distribution of surface heat flux.

\section{Interior structure and spatial patterns: toy models}
\label{Section4}
\markright{4 INTERIOR STRUCTURE AND SPATIAL PATTERNS: TOY MODELS\hfill}

\subsection{Thin shell}
\label{ThinShell}

\subsubsection{Rheology constant with depth}
\label{RheologyConstantWithDepth}

If the body contains an ocean beneath a thin icy shell (or an asthenosphere beneath a lithosphere), displacements do not vary much with depth within the shell and can be approximated by their value at the surface which depend on the displacement Love numbers $h_2$ and $l_2$.
In the membrane limit of thin shell theory, these Love numbers are related by
\begin{equation}
l_2 = \frac{1+\tilde\nu}{5+\tilde\nu} \, h_2 \, ,
\label{l2thinshell}
\end{equation}
where $\tilde\nu$ is the viscoelastic counterpart of Poisson's ratio.
This relation was previously derived for an elastic thin shell \cite[][Eq.~(D.12)]{beuthe2010a}.
Using the correspondence principle, I apply it here to a viscoelastic thin shell.
The membrane limit of thin shell theory is appropriate for loads of large wavelength with respect to the shell thickness $d$: the harmonic degree $\ell$ of the load should satisfy $\ell\ll1.8\sqrt{R/d}$ \citep[][Eq.~(83)]{beuthe2008}, that is $d/R\ll0.8$ for tidal loads of degree two.
This condition is satisfied in thin shell theory, in which one usually assumes that the shell thickness is less than 10\% of the radius.

When does the thin shell assumption break down for Eq.~(\ref{l2thinshell})?
Consider an incompressible ($\tilde\nu=\nu=1/2$) two-layer body of uniform density in the liquid core limit for which the ratio $l_2/h_2$ is given by Eq.~(\ref{l2TwoLayers}) with $\xi=0$.
Deviations from the thin shell limit ($l_2/h_2=3/11$) are smaller than 10\% if $x=1-d/R\gtrsim0.9$.
This constraint coincides with the usual requirement of thin shell theory mentioned above and is satisfied in most models of icy satellites with a subsurface ocean.

Eq.~(\ref{l2thinshell}) is not very sensitive to rheology.
For materials constituting planetary crusts, $\nu$ ranges between 0.25 (silicates) and 0.325 (elastic ice, \citet{gammon1983}).
The value of $\tilde\nu$ depends on the rheological model.
If the rheology is Maxwell, the real part of $\tilde\nu$ varies between the elastic value $\nu$ (high viscosity) and the incompressible value 0.5 (low viscosity), while its imaginary part is always much smaller than its real part.
Neglecting the imaginary part, we see that $(1+\tilde\nu)/(5+\tilde\nu)$ changes from $1/4$ to $3/11$ (less than 10\%) as $\tilde\nu$ changes from $1/3$ to $1/2$.

\begin{figure}
   \centering
     \includegraphics[width=14cm]{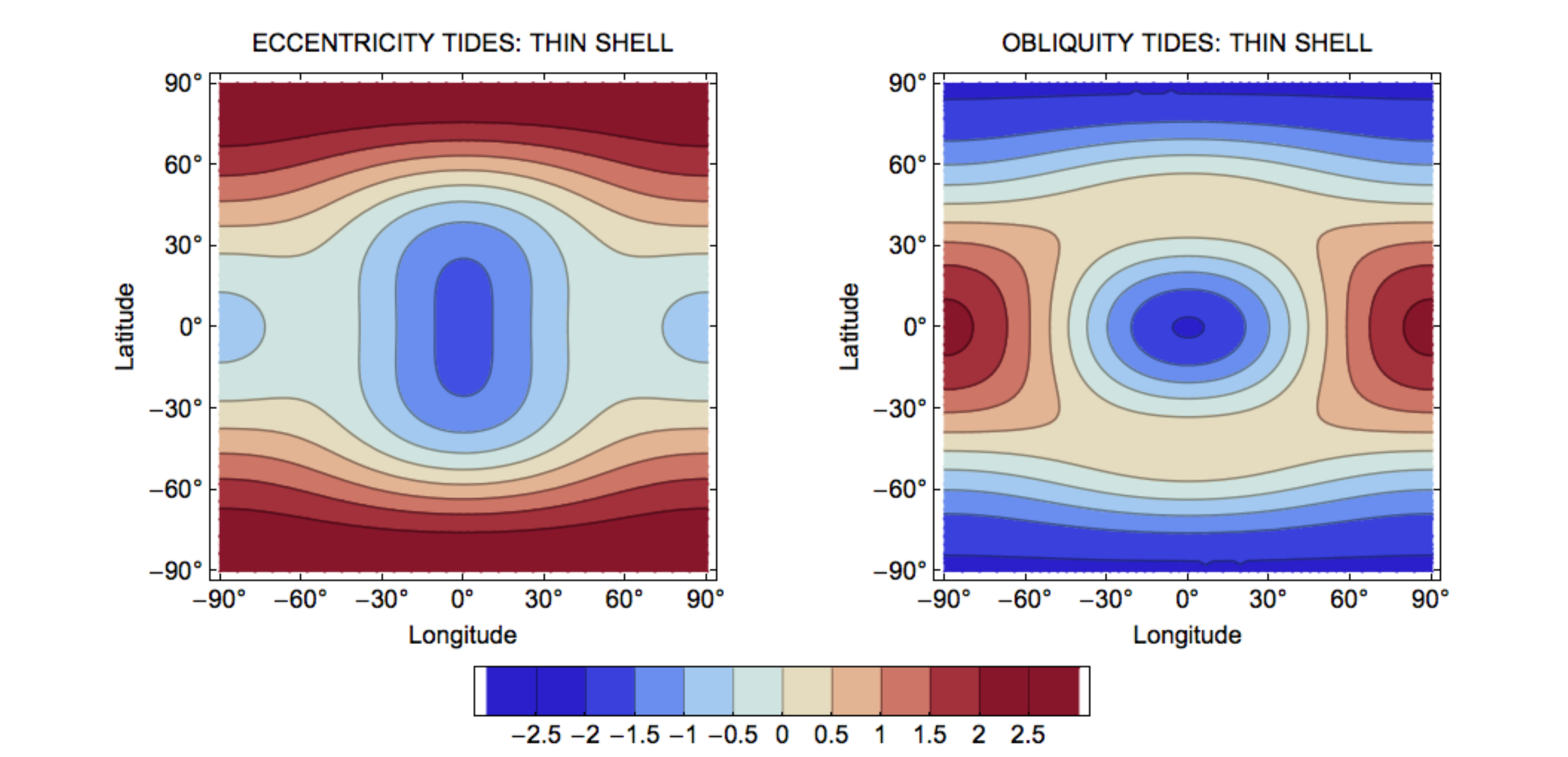}
   \caption{
   Pattern of tidal heating ($\Psi^{}_A + \frac{9}{2}  \Psi^{}_C$) due to eccentricity tides (left panel) or obliquity tides (right panel) within a thin compressible shell underlain by a liquid layer for a satellite in synchronous rotation (other details as in Fig.~\ref{FigPatterns}).
   }
   \label{FigPatternD}
\end{figure}

I compute weight functions in the {\it thin shell approximation} with Eqs.~(\ref{defHmu}), (\ref{WeightFunSurface}) and (\ref{l2thinshell}):
\begin{eqnarray}
\left( f^{}_A \, , f^{}_B \, , f^{}_C \, , H^{}_\mu \right) &=&  \frac{|h_2|^2}{g^2} \left| \frac{1+\tilde\nu}{5+\tilde\nu} \right|^2 \frac{16}{3} \left(  1 \, , 0 \, ,  \frac{9}{2} \, ,  \frac{11}{2} \right) \, ,
\nonumber \\
H^{}_K &=& \frac{6}{11} \left| \frac{1-2\tilde\nu}{1+\tilde\nu} \right|^2 H^{}_\mu \, .
\label{RadialFunThinShell}
\end{eqnarray}
In absence of bulk dissipation, the relative weights of the patterns $\Psi^{}_J$ do not depend on the rheology: Pattern~C dominates in the icy shell ($\hat f^{}_C/\hat f^{}_A=3.8$ for eccentricity tides) whatever the value of $\tilde\nu$.
Of course, the amount of tidal heating depends on the rheology of ice through the common multiplying factor depending on $h_2$ and $\tilde\nu$.
Therefore, the dissipated power in a thin shell above a fluid layer (e.g.\ an icy crust above an ocean)  is well approximated by
\begin{equation}
P =  \frac{8}{3} \, \frac{\omega(nR)^4}{r^2} \, \mbox{\it Im}(\tilde\mu) \, \frac{|h_2|^2}{g^2} \left| \frac{1+\tilde\nu}{5+\tilde\nu} \right|^2 \left(  \left(1+\kappa \right) \Psi^{}_A + \frac{9}{2} \, \Psi^{}_C \right) \, ,
\label{PowerThinShell}
\end{equation}
where $\kappa$ is related to the ratio of bulk dissipation to shear dissipation:
\begin{equation}
\kappa = \frac{11}{2} \, \frac{\mbox{\it Im}(\tilde{K}) H_K}{\mbox{\it Im}(\tilde\mu) H_\mu} \, .
\label{kappa}
\end{equation}
Fig.~\ref{FigPatternD} shows the resulting spatial pattern for eccentricity tides and obliquity tides if there is no bulk dissipation (see Section~\ref{ComparisonLiterature} for a comparison with the literature).

\subsubsection{Love numbers}

In the formula for the dissipated power (Eq.~(\ref{PowerThinShell})), the Love number $h_2$ does not affect the spatial pattern but regulates the total amount of tidal heating.
It can be computed with an interior model, for example the thin shell limit of the incompressible body of uniform density (see Eqs.~(\ref{ThinShellLimit})).
I give here slightly more general formulas (obtained with the same method) valid for an incompressible body of nonuniform density having a fully rigid mantle surrounded by an ocean and a thin icy shell.
The density and elastic structure below the mantle do not matter: a solid or liquid core of higher density may be present.
The ocean and the icy shell have the same density.
In this {\it thin shell/rigid mantle approximation}, the Love numbers are given by
\begin{eqnarray}
h_2 &=& h_2^{(0)} \, \left( 1 + h_2^{(0)} \, \frac{24}{11} \, \frac{\tilde\mu}{\rho{g}R} \, \frac{d}{R} \right)^{-1} \, ,
\nonumber \\
l_2 &=& \frac{3}{11} \, h_2 \, ,
\nonumber \\
k_2 &=& \frac{3\rho}{5\bar\rho} \, h_2 \, .
\label{h2thinshell}
\end{eqnarray}
The parameters $d$, $\rho$ and $\bar\rho$ are the shell thickness, the density of the ice (or ocean) and the average density of the body, respectively.
The factor $h_2^{(0)}$ is the value of $h_2$ in the limit of vanishing shell thickness:
\begin{equation}
h_2^{(0)} = \frac{5 \bar\rho}{5 \bar \rho-3\rho} \, .
\label{defh20}
\end{equation}
Eqs.~(\ref{h2thinshell}) expanded at first order in $d/R$ agree with Eqs.~(3)-(6) of \citet{wahr2006}.
It is however important not to expand Eqs.~(\ref{h2thinshell}) if one wants to compute good estimates of the imaginary part of Love numbers.
Though \citet{wahr2006}'s formulas take into account the density difference between ice and ocean, they are not well-suited to a viscoelastic interpretation because $\mbox{\it Im}(\tilde\mu)$ is not necessarily of the same order as $Re(\tilde\mu)$.
Using Eqs.~(\ref{PowerThinShell}) and (\ref{h2thinshell}), I can check that the integration over the volume of the shell of the dissipated power yields the global power (Eq.~(\ref{Edot2})) as computed with the macro approach (equivalently, one can verify that Eq.~(\ref{ConsEnergy}) is satisfied).

\subsubsection{Depth-dependent rheology}

The above formulas implicitly assumed that the icy shell has a homogeneous rheology characterized by $(\tilde\mu,\tilde\nu)$.
Actually, the assumption of depth-independent rheology is not realistic.
The viscosity of ice indeed depends on the local temperature $T$ of the ice and thus varies a lot between the top ($T\sim100\rm\,K$ for Europa) and bottom ($T\sim273\rm\,K$) of the icy shell, where it is at its melting point.
This variation has a huge effect on the shear modulus $\tilde\mu$, as can be seen for example in a Maxwell model \citep{wahr2009}.
However, Poisson's ratio $\tilde\nu$ varies much less as I already argued in Section~\ref{RheologyConstantWithDepth}.
In practice, one divides the shell into an outer conductive layer and an inner convective layer, the former being mainly elastic and the latter being viscoelastic \citep{mckinnon1999}.
Dissipation mainly occurs in the convective layer whereas the conductive layer influences deformations.

What is the effect of depth-dependent rheology on the thin shell formula for the dissipated power?
There is no problem is letting $\mbox{\it Im}(\tilde\mu)$ be depth-dependent in Eq.~(\ref{PowerThinShell}), but $l_2$ is a constant so that $(\tilde\mu,\tilde\nu)$ cannot vary with depth in the formulas for Love numbers, Eqs.~(\ref{l2thinshell}) and (\ref{h2thinshell}).
In this last equation, we see that $\tilde\nu$ is constant ($\tilde\nu=1/2$) and that $\tilde\mu$ is multiplied by the thickness of the shell $d$.
In thin shell theory, it can be shown that the product $\tilde\mu\,{d}$ in Eqs.~(\ref{h2thinshell}) arises from the integration of the shear modulus over the thickness of the shell \citep{beuthe2008}:
\begin{equation}
\tilde\mu \,d \leftrightarrow \tilde\mu_{av} \,d = \int_{shell} \tilde\mu \, dr \, .
\label{muav}
\end{equation}
In good approximation, $Re(\tilde\mu_{av})$ (resp. $\mbox{\it Im}(\tilde\mu_{av})$) is determined by the conductive (resp. convective) layer where $Re(\tilde\mu)$ (resp. $|\mbox{\it Im}(\tilde\mu)|$) is highest.
Thus the real (resp. imaginary) part of the Love numbers, corresponding to deformation (resp. dissipation), is mainly determined by the conductive (resp. convective) layer.
The substitution $\tilde\mu \rightarrow \tilde\mu_{av}$ preserves energy conservation: one can either check that Eq.~(\ref{ConsEnergy}) is satisfied, or more explicitly that the integration over the volume of the shell of the dissipated power with depth-dependent $\tilde\mu$ yields the global dissipated power (Eq.~(\ref{Edot2})).
Incompressibility ($\tilde\nu=1/2$) is the simplest assumption for $\tilde\nu$ compatible both with depth-dependent rheology and with the absence of bulk dissipation.
In contrast with Eq.~(\ref{muav}), it would not make sense to simply replace $\tilde\nu$ by its average in Eq.~(\ref{l2thinshell}).
Note that the averaging procedure described by Eq.~(\ref{muav}) was heuristically proposed for an elastic icy shell by \citet{wahr2006}.

In Section~\ref{Section5}, I will compare thin shell predictions with the results of interior models of Europa and Titan.

\subsection{Homogeneous body}
\label{HomogeneousBody}

\begin{figure}
   \centering
      \includegraphics[width=7cm]{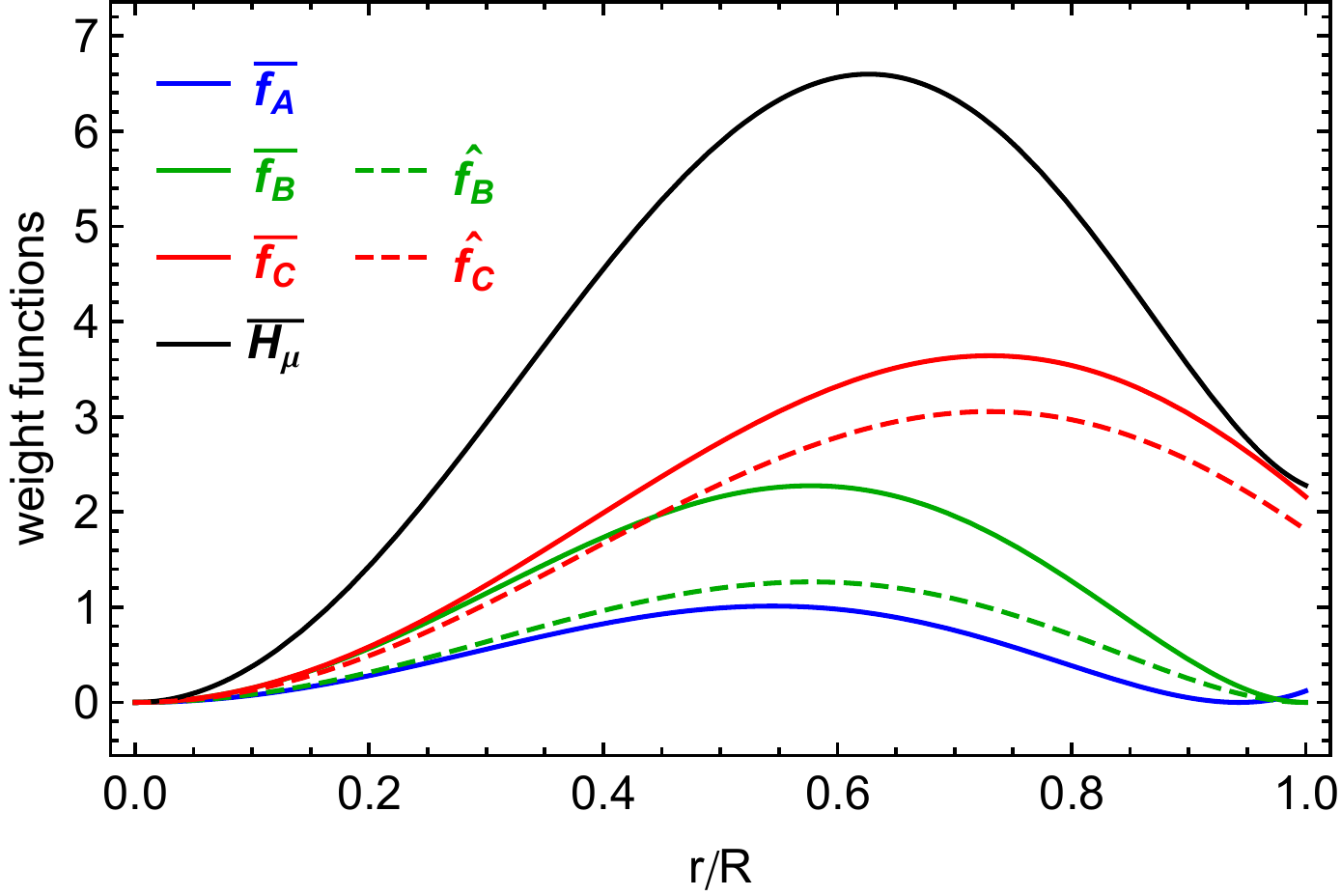}
   \caption{
   Weight functions for the three basic spatial patterns of dissipation in a homogeneous incompressible body.
   The functions $\bar f^{}_J$ are the rescaled weight functions (Eqs.~(\ref{rescaled1})), the sum of which is $\bar H^{}_\mu$.
   The functions $\hat f^{}_B=0.56\,\bar f^{}_B$ and $\hat f^{}_C=0.84\,\bar f^{}_C$ are the equal-variance weight functions for eccentricity tides (see Eq.~(\ref{fABCnorm})), which should be compared with $\hat f^{}_A=\bar f^{}_A$ in order to determine which pattern dominates. }
   \label{FigSensitivityHomog}
\end{figure}

The incompressible homogeneous body is the simplest interior model and serves as a useful approximation either (1) for bodies without liquid layers or (2) for a solid core surrounded by a weak layer (ocean or ice).
The body is characterized by its radius $R$, density $\rho$ and shear modulus $\mu$.
The problem of computing tidal deformations can be formulated in terms of the reduced radius $\bar{r}=r/R$, the surface gravity $g=(4\pi/3)G\rho{R}$ and the reduced shear modulus defined by
\begin{equation}
\mu_{red} = \frac{\tilde\mu}{\rho{g}R} \, .
\end{equation}
The solution to this venerable problem (going back to Lord Kelvin in the 19th century, see \citet{love1911}) is for example given by Eqs.~(7)-(9) of \citet{peale1978} (in which $(a_e/r)^2V_{2mpq}$ is equivalent to $\Phi_\omega$ in my notation).
It is also derived in Appendix~F (see Eq.~(\ref{yiHomog})).
After substitution of this solution, the weight functions given by Eqs.~(\ref{weights}) become
\begin{equation}
\left( f^{}_A, f^{}_B, f^{}_C \right) =   \frac{|h_2|^2}{g^2} \,  \frac{192}{25} \, \bar r^2 \left(  \left( 1 - \frac{9}{8} \bar r^2\right)^2, \, 2 \left( 1 - \bar r^2 \right)^2, \, 2 \left( 1 - \frac{5}{8} \bar r^2 \right)^2 \right) \, ,
\label{WeightsHomogeneous}
\end{equation}
in which $h_2$ is given by the first of Eqs.~(\ref{h2l2OneLayer}).
While $H^{}_K=0$ (incompressible material), the radial sensitivity function is given by
\begin{equation}
H_\mu = \frac{|h_2|^2}{g^2} \, \frac{3}{25} \, \bar r^2 \left( 320 - 560 \, \bar r ^2 + 259 \, \bar r^4 \right) \, ,
\label{HmuHomog}
\end{equation}
with a maximum at $\bar r=0.627$.
The comparison of the equal-variance weight functions $\hat f^{}_{A,B,C}$ (see Fig.~\ref{FigSensitivityHomog}) shows that Pattern $C$ dominates at all depths.

Conservation of energy (Eq.~(\ref{ConsEnergy})) can be analytically checked by radially integrating $H^{}_\mu$ and using the relations between $h_2$, $k_2$ and $\tilde\mu$ (see Eqs.~(\ref{relationh2k2}) and (\ref{h2l2OneLayer})).
I do not prove it here because it is a particular case of the two-layer model (see Eqs.~(\ref{ConsEnergyTwolayers}) with $z(0)=19/2$).
Remarkably, weight functions depend on the rheology through the common factor $|h_2|^2/g^2$, so that the relative weights of heating patterns at a given radius do not depend on any physical parameter.
The homogeneous solution is an example of the complete factorization of the overall deformation described by the tidal Love number $h_2$.

For an Earth-like elastic homogeneous body, allowing for compressibility changes $h_2$ by 10 to 20\% but $k_2$ by only 2\% \citep[][pp.~105-110]{love1911}.
If this result can be extended to a viscoelastic body, we expect sizable compressibility effects on the weight functions (depending on squared strains) but little change in the global power (depending on $\mbox{\it Im}(k_2)$).
Compressibility effects decrease with body size and thus do not affect much the radial sensitivity function $H^{}_\mu$ for satellites of the solar system \citep[][Fig.~4]{tobie2005}.
The effect on weight functions remains however significant in small bodies with a liquid core (see Section~\ref{ComparisonLiterature}). 

\subsection{Incompressible two-layer model}
\label{TwoLayer}

The radial functions $y_i$ have analytical expressions depending on powers of $r$ if the body is approximated as being (1) incompressible and (2) composed of spherically concentric homogeneous layers.
Analytical formulas for the $y_i$ can be determined with the propagator matrix technique \citep[e.g.][]{sabadini2004} and the help of mathematical software (I used Mathematica) though expressions are very lengthy except in the simplest cases.
In this approach, one usually computes tidal deformations in the static limit of infinite tidal period.

\subsubsection{Density effects}

Radial variations in density strongly affect the magnitude of the overall deformation.
The relative weights of dissipation mechanisms, however, depend little on the density structure if the body is far from the fluid limit.
In order to be more quantitative, let us consider a two-layer model with liquid core (radius $R_C$, density $\rho_C$) and viscoelastic mantle (radius $R$, density $\rho_M$, shear modulus $\tilde\mu_M$).
The rescaled weight functions (\ref{rescaled1}) depend on three dimensionless parameters: the reduced core size $x=R_C/R$, the density contrast $\eta=\rho_C/\rho_M$ (supposed to be smaller than 5), and the reduced shear modulus of the mantle equal to $\mu_{red}=\tilde\mu_M/(\bar\rho{g}R)$, with $\bar\rho$ the average density and $g$ the surface gravity.
I will assume that $\mbox{\it Im}(\tilde\mu)\ll{\mbox{\it Re}(\tilde\mu)}$, so that I can approximate $\tilde\mu$ and $\tilde\lambda$ by their real part when computing tidal displacements.
I thus compute the weight functions in the elastic limit: the viscous response appears only in the formula for the power through the global factor $\mbox{\it Im}(\tilde\mu)$.
Density effects have a negligible effect on the rescaled weight functions if the body is far from the fluid limit, that is if
\begin{equation}
| \mu_{red} | \gtrsim 0.1 \left( \eta-1 \right) \, .
\label{constraint}
\end{equation}

If the density is uniform, the fluid limit is not relevant because the rescaled weight functions are independent of $\mu_{red}$ (see below).
Fig.~\ref{FigDensity} shows how rescaled weight functions change as the density contrast between the liquid core ($R_C=R/2$) and mantle increases from one to three.
Density effects are small as long as $|\mu_{red}|\gtrsim0.2$ (see Eq.~(\ref{constraint})), but are large if the body deforms like a fluid ($\mu_{red}\rightarrow0$).
The condition (\ref{constraint}) is generally satisfied in the dissipative layers of terrestrial planets and satellites of the solar system, though it could be violated for Io if the mantle is extremely soft with a shear modulus less than $(\eta-1)\rm\,GPa$ \citep{fischer1990,spohn1997}.
It is true that the asthenosphere (if present) is characteristically much softer but its density contrast with the rest of the mantle is small.

\begin{figure}
   \centering
      \includegraphics[width=7.5cm]{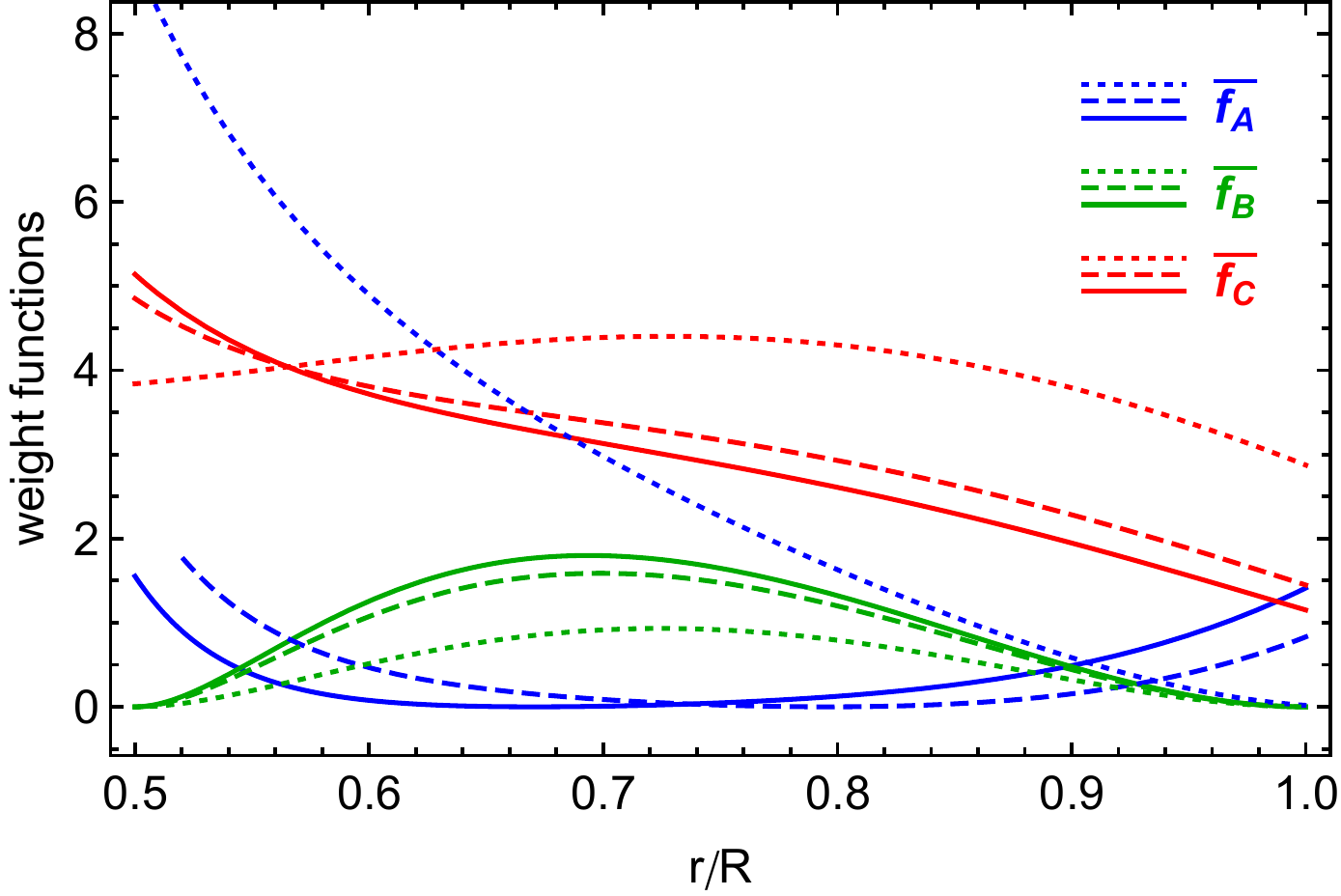}
   \caption{
   Influence of the density contrast $\eta=\rho_C/\rho_M$ between core and mantle on the rescaled weight functions (\ref{rescaled1}) for a two-layer incompressible body having a liquid core with $R_C=0.5R$: $\eta=1$ (continuous curves), $\eta=3$ and $\mu_{red}=0.2$ (dashed curves), $\eta=3$ and $\mu_{red}\rightarrow0$ (dotted curves).
   The densities $\rho_C$ and $\rho_M$ are adjusted so as to yield the same average density $\bar\rho$ and same surface gravity $g$; the relation between $\mu_{red}$ and $\tilde\mu_M$ is thus not affected by the density contrast.
   }
   \label{FigDensity}
\end{figure}

\subsubsection{Uniform density and rheology}

Since the density structure has in most cases a small effect on the rescaled weights functions, I will from now on assume that the body is of uniform density.
Rescaled weight functions in such models depend only on rheological contrasts between layers: the overall magnitude of the deformation factorizes (see Appendix~F).
In Appendix~G, I give the analytical solution for tidal displacements within an incompressible body of uniform density $\rho$ composed of a viscoelastic core (radius $R_C$, shear modulus $\tilde\mu_C$) and a viscoelastic mantle (radius $R$, shear modulus $\tilde\mu_M$).
The model can be characterized by three dimensionless ratios (see Eqs.~(\ref{defxi})): $R_C/R$, $\tilde\mu_C/\tilde\mu_M$ and $\mu_{red}=\tilde\mu_M/(\rho{g}R)$.
The weight functions $(f^{}_A, f^{}_B, f^{}_C)$ and $H^{}_\mu$ can be computed by substituting Eqs.~(\ref{yiFun4}), (\ref{defa}), (\ref{CoeffMantlecd})-(\ref{CoeffMantleab}) into Eqs.~(\ref{weights}) and (\ref{weightsIncomp}).

I will examine two interesting limits of the two-layer model, being (1) the dissipation within the mantle surrounding a liquid core, and (2) the dissipation within a soft layer above a rigid layer.
These two special cases already display the different possible types of dissipation patterns.
Besides they have the advantage that the rheology appears in the functions $y_i$ through the common factor $h_2$ describing the overall deformation.
As a result, the relative weights of the three basic patterns depend only on the core radius.
Dissipation patterns can thus be studied in these two models without specifying the rheology of the mantle.
For an arbitrary core radius, the coefficients characterizing the functions $y^{}_i$ in the mantle are given by Eqs.~(\ref{defa}) and (\ref{CoeffMantlecd})-(\ref{CoeffMantleab}) in which the ratio $\xi$ of shear moduli is either set to zero (liquid core) or tends to infinity (infinitely rigid core).

\subsubsection{Liquid core}
\label{LiquidCore}

In the first limit case, the body consists of a viscoelastic mantle surrounding a liquid core of the same density.
This model is equivalent to the one that \citet{peale1978} used to compute tidal dissipation in the Moon but my equations are simpler.
If the core radius vanishes, the model is equivalent to the homogeneous case (see Eq.~(\ref{yiHomog})).
If the core radius tends to the surface radius, the model is a special case of the thin shell model discussed in Section~\ref{ThinShell} (compare Eqs.~(\ref{ThinShellLimit}) with Eqs.~(\ref{l2thinshell}) and (\ref{h2thinshell})).

\begin{figure}
   \centering
      \includegraphics[width=15cm]{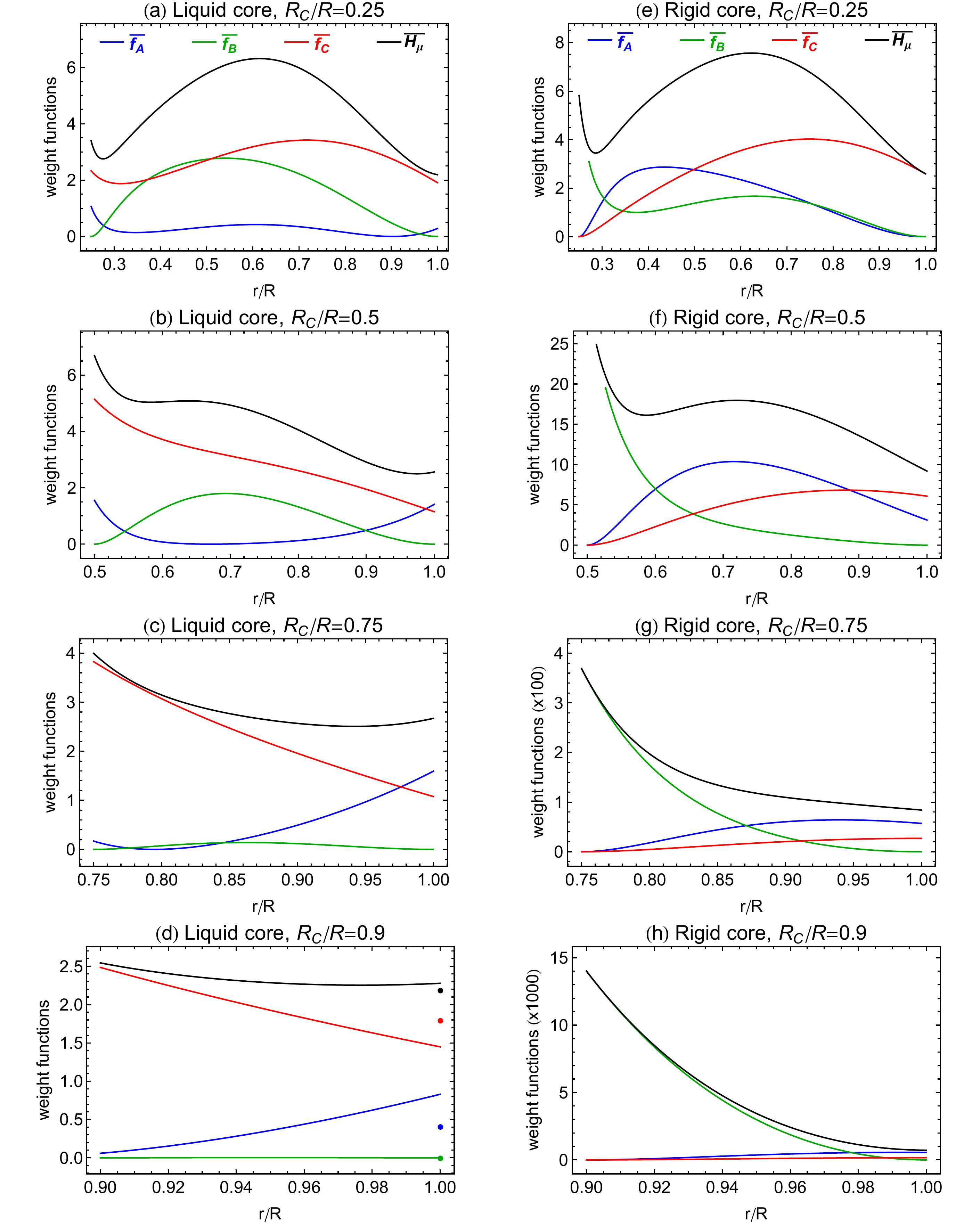}
   \caption{
   Rescaled weight functions (\ref{rescaled1}) for the three basic spatial patterns of dissipation in the mantle of a two-layer incompressible body with either a liquid core (left panels) or an infinitely rigid core (right panels).
   Each row corresponds to a different core radius with $R_C/R$ taking the values $0.25$, $0.5$, $0.75$, $0.9$ from top to bottom.
   The big dots on the right-hand side of Panel~(d) are the values of the functions in the thin shell limit ($R_C\rightarrow{R}$).
   Note the much larger scale in Panels (g) and (h).
   Equal-variance weight functions are not shown for clarity; they could be drawn as in Fig.~\ref{FigSensitivityHomog} from the relations valid for eccentricity tides: $\hat f^{}_B=0.56\,\bar f^{}_B$ and $\hat f^{}_C=0.84\,\bar f^{}_C$.
   }
   \label{FigSensitivityTwoLayer}
\end{figure}

Figs.~\ref{FigSensitivityTwoLayer}(a,b,c,d) shows the rescaled weight functions in the mantle surrounding a liquid core for different values of the core radius.
The location of maximum average dissipation (maximum $H^{}_\mu$) shifts from the value $r/R=0.627$ (homogeneous model) to the core-mantle boundary as the core size increases.
The comparison of the weight functions shows that Pattern~C is generally dominant at all depths.
At the surface, Pattern~A slightly dominates Pattern~C if $R_C\approx0.5-0.75R$ (Figs.~\ref{FigSensitivityTwoLayer}(b,c)), in which case the resulting pattern is mostly of degree four (the component of harmonic degree two approximately cancels, see Eqs.~(\ref{PsiABCsh})) but with a pattern reversed with respect to Pattern~B, as in Fig.~6(f) of \citet{tobie2005}.
Pattern~A is however negligible at greater depth so that the radially-integrated power is dominated in all cases by Pattern~C.
The contribution of Pattern~B is significant if the core is small ($f^{}_B/f^{}_C$ maximum for $R_C\approx{R/4}$).
The influence of core size is summarized in Fig.~\ref{FigPartialPower}(a) showing the relative contribution of each dissipation term to the total power.

\subsubsection{Rigid core}
\label{RigidCore}

In the second limit case, the body consists of a viscoelastic mantle surrounding an infinitely rigid core of the same density, though `mantle' and `core' are only names for the layers and do not necessarily refer to their physical counterparts.
This toy model is relevant to at least two realistic configurations.
First, the model is a good approximation for dissipation within an icy layer just above a rocky mantle, as ice is much softer than rocks.
Second, the model describes the essentials of asthenospheric tidal heating in which tidal dissipation mainly occurs within a soft layer at the top of the mantle.

Figs.~\ref{FigSensitivityTwoLayer}(e,f,g,h) show the rescaled weight functions in the mantle surrounding an infinitely rigid core for different values of the core radius.
The comparison of the weight functions shows that Pattern~B dominates near the core-mantle boundary where shear is maximum and decreases to zero near the surface.

\begin{figure}
   \centering
   \includegraphics[width=15cm]{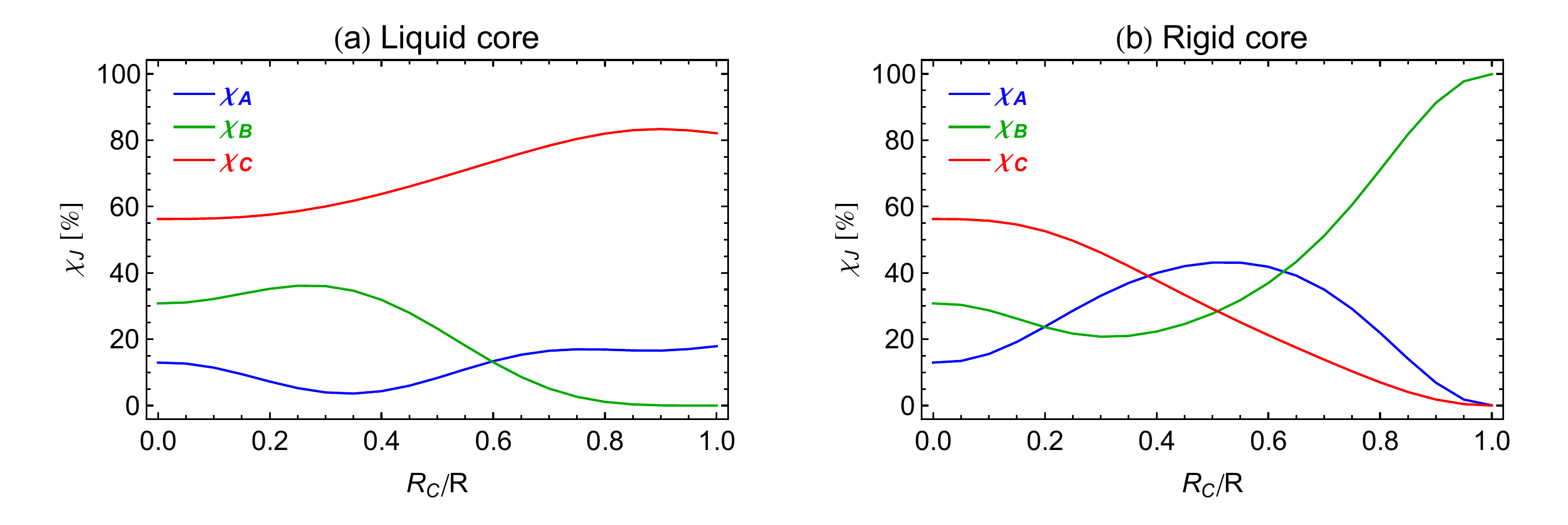}
   \caption{Contribution of Pattern $\Psi^{}_J$ (see Eq.~(\ref{chiJ})) to the global dissipated power for a two-layer incompressible body of uniform density as a function of the core radius: (a) liquid core, (b) infinitely rigid core.}
   \label{FigPartialPower}
\end{figure}

If the soft layer is thick, the weights of the different patterns are of comparable magnitude, with Pattern~C, A or B successively dominating as the thickness of the rigid layer increases from zero to the surface radius (see Fig.~\ref{FigPartialPower}(b)).
For example, a soft layer with thickness $R/2$ leads to weights satisfying on average $f^{}_B\approx{f^{}_C}<f^{}_A$, so that Pattern~A dominates by virtue of Eq.~(\ref{combili}).
As the soft layer becomes thinner (say $R_C\gtrsim0.75R$), Pattern~B dominates Pattern~C in the lower part of the layer with the reverse situation near the surface.
Therefore, the dissipation pattern within an icy layer in contact with the mantle is given by Pattern~B at the bottom of the layer and by a mix of Patterns~A and C at the top of the layer, as in Figs.~10(a,b,e,f) of \citet{tobie2005}.

When the soft layer is very thin (say $R_C\gtrsim0.9R$), Patterns~A and C become negligible because dissipation is mainly caused by shear heating due to the coupling of the soft layer with the rigid layer.
This mechanism accounts for the dominance of Pattern~B in models of asthenospheric dissipation in Io \citep{segatz1988}.
Such models include a rigid lithosphere above the asthenosphere so that shear dissipation also occurs at the asthenosphere-lithosphere boundary, yielding a U-shape for the curve $f_B$ (see application to Io in Section~\ref{Section5}).
As Io shows more volcanic activity in equatorial regions not far from the sub-Jovian and anti-Jovian points \citep{lopes1999,tackley2001b,kirchoff2011,veeder2012,hamilton2012}, models of asthenospheric dissipation (less heating at the poles) have been more popular than models of mantle dissipation (more heating at the poles).
There is however no evidence for the degree-four structure exhibited by Pattern~B.
Another possibility is provided by a model having a rigid lower mantle and a soft upper mantle of large thickness, in which case Pattern~A gives a significant contribution partly cancelling the degree-four term due to Pattern~B and leading to a degree-two structure reverse of Pattern~C (for example $\Psi^{}_A+2\Psi^{}_B=5\Psi^{}_0-2\Psi^{}_C$ from Eq.~(\ref{combili})).
The influence of asthenosphere thickness is summarized in Fig.~\ref{FigPartialPower}(b) showing the relative contribution of each dissipation term to the total power.

\subsubsection{Conservation of energy}

In the two limit cases (liquid or rigid core of reduced radius $x$), dissipation occurs only in the mantle.
It is thus not difficult to analytically check conservation of energy (Eq.~(\ref{ConsEnergy})) by noting that
\begin{eqnarray}
\int_x^1 H^{}_\mu \, d\bar r &=& \frac{2}{5} \, \frac{|h_2(x)|^2}{g^2} \, z(x) \, ,
\nonumber
\\
\mbox{\it Im} (\tilde\mu) &=& - \frac{5}{2} \, \frac{\rho g R}{z(x)} \, \frac{\mbox{\it Im}(h_2(x))}{|h_2(x)|^2} \, ,
\label{ConsEnergyTwolayers}
\end{eqnarray}
where $h_2(x)$ and $z(x)$ are defined by Eqs.~(\ref{h2x})-(\ref{zrigid}) and $h_2=(5/3)k_2$ (see Eq.~(\ref{relationh2k2})).
Considering still the two limit cases, I can use either one of Eqs.~(\ref{ConsEnergyTwolayers}) to write a compact formula for the global power dissipated by eccentricity tides in terms of core size and mantle rheology:
\begin{equation}
\dot E = \frac{63}{4} \,  e^2 \, \frac{(\omega{R})^5}{G} \, \mbox{\it Im}(\mu_{red}) \, \frac{z(x)}{\left| 1 + z(x) \, \mu_{red} \right|^2} \, .
\end{equation}
This equation serves to estimate the core size and the rheology yielding the required global power output (e.g.\ Fig.~1 in \citet{peale1979} and Fig.~1 in \citet{cassen1980}).

\section{Real bodies: Europa, Titan and Io}
\label{Section5}
\markright{5 REAL BODIES: EUROPA, TITAN AND IO\hfill}

\subsection{General features of interior models}

I will show here that the toy models of Section~\ref{Section4} predict well the spatial patterns of tidal heating within each layer of a real body that can be approximated by concentric homogeneous layers.
Toy models however do not predict the magnitude of dissipation in the layer, this factor depending on the complete structure of the body.
Thus comparisons cannot be done between layers unless the tidal response of the real body has been computed.

Choosing as illustrations the satellites Europa, Titan and Io, I approximate their internal structure with four or five layers that are homogeneous and incompressible.
In its general features, the interior model for Titan is also relevant to Ganymede and Callisto, which probably have a subsurface ocean sandwiched between thick ice layers \citep{spohn2003};
Triton is yet another candidate \citep{hussmann2006,gaeman2012}.
Global physical parameters (see Table~\ref{TableGlobal}) are taken from \citet{schubert2009} and \citet{sotin2009} for Europa, from \citet{iess2010} for Titan and from \citet{moore2007} for Io.
Solid layers are viscoelastic with Maxwell rheology (see Eq.~(\ref{MaxwellRheology})).
Each layer is characterized by its thickness $\delta R$, density $\rho$, elastic shear modulus $\mu$, and  viscosity $\eta$.
As in Section~\ref{TwoLayer}, I compute the functions $y^{}_i$ with the propagator matrix technique for an incompressible body in the static limit.

\begin{table}\centering
\ra{1.3}
\small
\caption{Global parameters for Europa, Titan and Io: surface radius $R$, average density $\bar\rho$, normalized axial moment of inertia MoI=$C/MR^2$, surface gravity $g$ and tidal angular frequency $\omega$.}
\vspace{1mm}
\begin{tabular}{@{}llllll@{}}
\hline
& $R$ (km) & $\bar\rho$ ($\rm{kg/m^3}$) & MoI & $g$ ($\rm{m/s^2}$) & $\omega$ ($\rm{rad/s}$)  \\
\hline
Europa & 1561 & 3013 & 0.346 & 1.31 & $2.05\times10^{-5}$ \\
Titan & 2575 & 1881 & 0.341 & 1.35 & $4.56\times10^{-6}$  \\
Io & 1822 & 3527 & 0.378 & 1.80 & $4.11\times10^{-5}$ \\
\hline
\end{tabular}
\label{TableGlobal}
\end{table}%

\subsection{Interior models of Europa, Titan and Io}

I model Europa with five layers: a liquid iron core, a silicate mantle, an ocean and an icy crust stratified into convective and conductive layers.
Table~\ref{TableEuropa} gives the values of the internal physical parameters; they are based on those of \citet[][Table~1 and Fig.~5]{tobie2003}, except that I adjusted the core and mantle radii so as to obtain the correct mass and moment of inertia.
Constraints on density structures and on rheology are reviewed by \citet{schubert2009} and \citet{sotin2009}, respectively.
Estimates of the ice thickness range from 100~m for the upper elastic layer to 30-50~km for the total ice thickness (see reviews by \citet{billings2005} and \citet{nimmo2009}).

\begin{table}\centering
\ra{1.3}
\small
\caption{Interior model of Europa}
\vspace{1mm}
\begin{tabular}{@{}lllll@{}}
\hline
Layer & $\delta R$ (km) & $\rho$ ($\rm{kg/m^3}$) & $\mu$ (GPa) & $\eta$ (Pa.s) \\
\hline
Iron core & 587 & 5500 & 0 & 0\\
Silicate mantle & 840 & 3500 & 70 & $10^{20}$ \\
Ocean & 114 & 1000 & 0 & 0 \\
Convective ice & 12 & 920 & 3.3 & $10^{14}$ \\
Conductive ice & 8 & 920 & 3.3 & $10^{20}$ \\
\hline
\end{tabular}
\label{TableEuropa}
\end{table}%

I model Titan with five layers: a rocky core, a mantle of high-pressure ice, an ocean and an ice layer stratified into convective and conductive layers.
\citet{fortes2012} proposed several internal structures, among which I choose his `dense-ocean' model (see Table~\ref{TableTitan}), which yields the correct moment of inertia and is compatible with the value of tidal $k_2$ measured with Cassini \citep{iess2010,iess2012}.
For the rocky core, I assume the same rheology as in Europa's mantle.
The elastic shear modulus of ice is determined from $\rho{V_S^2}$, with the S-wave velocity given by $V_S=1880\rm\,m\,s^{-1}$ \citep{tobie2005b}, while the ice viscosity is chosen to be the same as in Europa's crust.
It is not clear whether the icy crust convects \citep[e.g.][]{sohl2003,mitri2008,nimmo2010}.
I thus assume that convection is limited to the bottom 10\% of the icy shell in order to illustrate the effect of a thick conductive ice layer above an ocean.
This interior model yields $Re(k_2)=0.57$.

\begin{table}\centering
\ra{1.3}
\small
\caption{Interior model of Titan}
\vspace{1.5mm}
\begin{tabular}{@{}lllll@{}}
\hline
Layer & $\delta R$ (km) & $\rho$ ($\rm{kg/m^3}$) & $\mu$ (GPa) & $\eta$ (Pa.s) \\
\hline
Rocky core & 1984 & 2650 & 70 & $10^{20}$\\
Icy mantle & 241 & 1351 & 4.8 & $10^{14}$ \\
Ocean & 130 & 1000 & 0 & 0 \\
Convective ice & 10 & 931 & 3.3 & $10^{14}$ \\
Conductive ice & 90 & 931 & 3.3 & $10^{20}$ \\
\hline
\end{tabular}
\label{TableTitan}
\end{table}%

I model Io with four layers: a liquid iron core, a silicate mantle of uniform density but rheologically divided into a deep mantle and an upper asthenosphere, and a crust (or lithosphere).
Table~\ref{TableIo} gives the values of the internal physical parameters.
Given plausible values for the crust thickness and density, I constrain the density structure with the average mass and moment of inertia \citep{moore2007}.
As there is still one free parameter, I set to 0.41 the ratio of core radius to deep mantle radius, the same value as in Europa's interior model.
With this choice, I intend to illustrate the predictive power of the toy models of Section~\ref{Section4}: apart from their overall magnitude, weight functions will look the same in Europa's mantle and Io's deep mantle.

I set the asthenosphere thickness to 50~km which is the lower bound for the conductive layer detected with Galileo \citep{khurana2011}.
Since Io's internal temperature distribution and degree of partial melting are unknown \citep{keszthelyi2007,moore2007,kohlstedt2009}, the viscoelastic parameters $(\mu,\eta)$ are not determined from known properties of rocks but are chosen so as to satisfy the global constraint of $\mbox{\it Im}(k_2)=-|k_2|/Q=-0.015\pm0.003$ \citep{lainey2009}.
This constraint comes from astrometric observations of the secular accelerations of Io, Europa and Ganymede.
The substitution of $\mbox{\it Im}(k_2)$ in Eq.~(\ref{Edot2}) with $e=0.0041$ and $I=0$ yields a total power of $\dot{E}=(9\pm2)\times10^{13}\rm\,W$ compatible with the estimates $6-16\times10^{13}\rm\,W$ from heat flux measurements \citep{moore2007}.
For the deep mantle, I adopt $(\mu,\eta)$ values suggested by \citet{spohn1997}.
For the asthenosphere, I choose $(\mu,\eta)$ so as to satisfy the global constraint on dissipation.
In this model, tidal heating is nearly equipartitioned between deep mantle (49\%) and asthenosphere (51\%), the dissipation in the lithosphere being always negligible.

\begin{table}\centering
\ra{1.3}
\small
\caption{Interior model of Io}
\vspace{1.5mm}
\begin{tabular}{@{}lllll@{}}
\hline
Layer & $\delta R$ (km) & $\rho$ ($\rm{kg/m^3}$) & $\mu$ (GPa) & $\eta$ (Pa.s) \\
\hline
Iron core & 716 & 6767 & 0 & 0\\
Deep mantle & 1026 & 3349 & 10 & $10^{15}$ \\
Asthenosphere & 50 & 3349 & $5\times10^{-5}$ & $2.5\times10^{10}$ \\
Crust & 30 & 2750 & 70 & $10^{20}$ \\
\hline
\end{tabular}
\label{TableIo}
\end{table}%

\subsection{Dissipation weight functions}

The spatial patterns resulting from weight functions have been discussed at length in Section~\ref{Section4}.
I thus restrict myself here to comparing the weight functions within Europa, Titan and Io with the weight functions of the toy models of Section~\ref{Section4}.

\begin{figure}
   \centering
     \includegraphics[width=15cm]{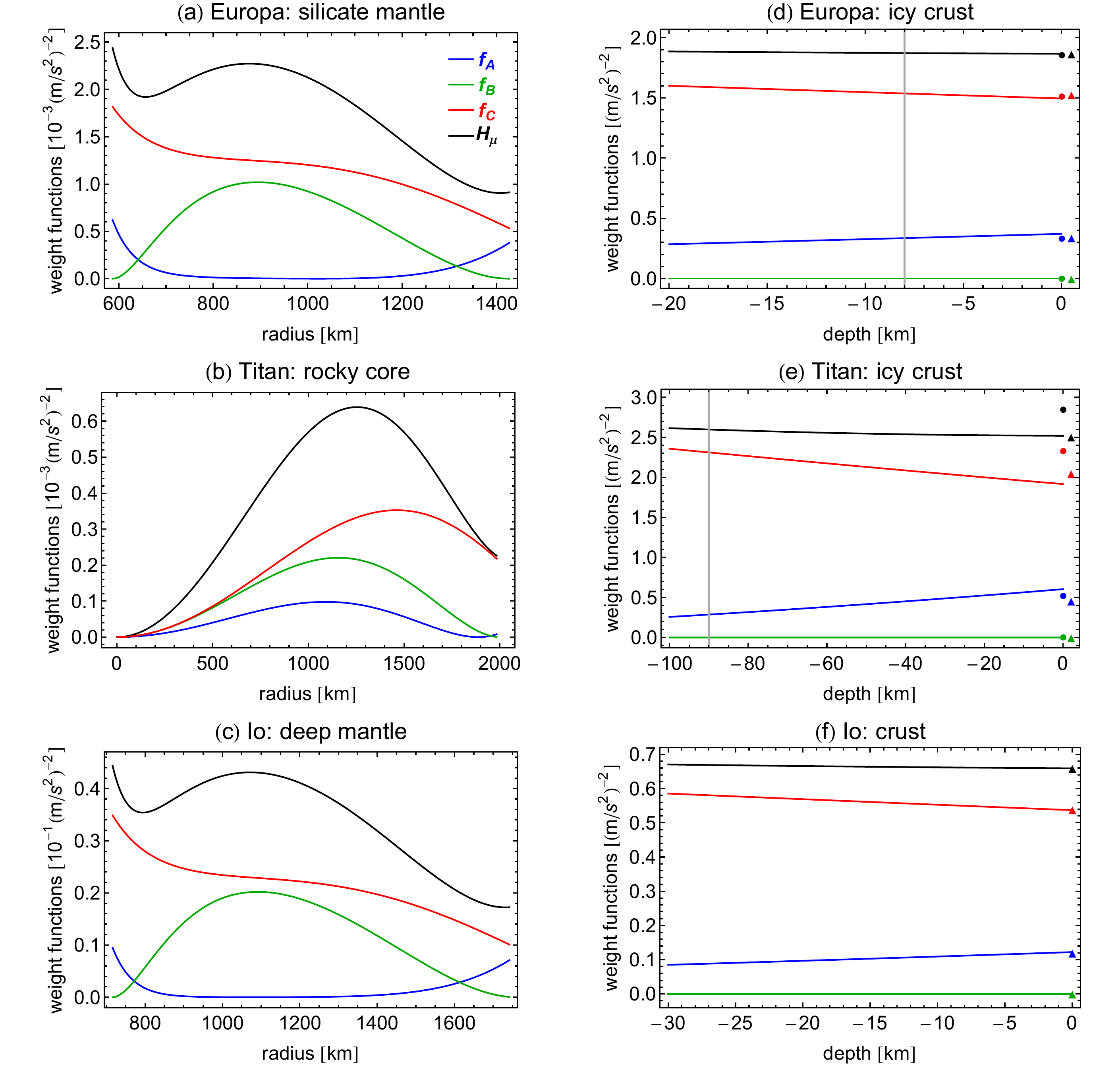}
   \caption{
   Weight functions in the mantle/rocky core/deep mantle (left panels) and in the crust (right panels) of Europa, Titan and Io.
   Interior models are specified in Tables~\ref{TableEuropa}, \ref{TableTitan} and \ref{TableIo}.
   Note the change of scale on y-axis between panels.
   In the right panels, the vertical line separates the convective layer (if present) from the conductive layer.
   The big dots are the values of weight functions in the thin shell/rigid mantle approximation used for floating icy crusts ($|h_2|$=1.21/1.54 for Europa/Titan), whereas the triangles are the values of the weight functions in the thin shell approximation with $|h_2|$ given by the interior model ($|h_2|$=1.21/1.45/0.99 for Europa/Titan/Io).
   }
   \label{FigExamples1}
\end{figure}

Let me start by examining dissipation weight functions in Europa's mantle, Titan's rocky core and Io's deep mantle.
In Europa's mantle (Fig.~\ref{FigExamples1}(a)), weight functions look like those in the mantle of a two-layer body with liquid core (the curves are intermediate between Figs.~\ref{FigSensitivityTwoLayer}(a) and \ref{FigSensitivityTwoLayer}(b) as the core radius is 41\% of the mantle radius).
This similarity could be expected, because the free-slip condition at the mantle/ocean interface mimics one of the boundary conditions at the surface of the body.
The presence of an ocean has also the effect of greatly reducing the amplitude of weight functions in the mantle because deformations of the mantle are due to gravitational coupling: they vanish if the core, mantle and ocean have the same density.
The surrounding ocean actually provides most of the tidal response of the satellite.
In Titan's core (Fig.~\ref{FigExamples1}(b)), weight functions look like those in a homogeneous body (compare to Fig.~\ref{FigSensitivityHomog}), which is not surprising since the icy mantle does not impose a strong shear coupling at the core/mantle boundary.
In Io's deep mantle (Fig.~\ref{FigExamples1}(c)), weight functions look like those in Europa's mantle though their magnitude is one order of magnitude larger.
This resemblance is explained by similar boundary conditions: on the one hand the mantle encloses a liquid core with the same ratio of core radius to mantle radius (41\%), and on the other the mantle is surrounded by a much weaker layer (water for Europa, asthenosphere for Io).

Next, I compare dissipation weight functions in the crusts of Europa, Titan and Io.
In Europa's icy crust (Fig.~\ref{FigExamples1}(d)), weight functions are nearly constant as is expected in a thin shell (compare to Fig.~\ref{FigSensitivityTwoLayer}(d) where the shell is much thicker).
In particular, the stratification of the ice layer has nearly no effect on the weight functions (a zoom on $f^{}_B$ reveals a discontinuity at the convective/conductive transition, but this weight function is close to zero in the icy shell).
The thin shell/rigid mantle approximation (Eqs.~(\ref{RadialFunThinShell}) and (\ref{h2thinshell})-(\ref{muav})) predicts well the values of the weight functions in the icy shell (big dots in Fig.~\ref{FigExamples1}(d)).
One measure of comparison is $\mbox{\it Im}(k_2)$, as it summarizes the effect of internal structure on the global dissipated power (see Eq.~(\ref{Edot2})).
The interior model of Europa specified in Table~\ref{TableEuropa} yields $\mbox{\it Im}(k_2) = - 3.58\times10^{-3}$ whereas the thin shell/rigid mantle approximation gives $\mbox{\it Im}(k_2)=-3.53\times10^{-3}$, a difference of less than 2\% which is mainly due to the elasticity of the mantle and the fluidity of the core (the dissipation in the mantle is negligible).

In Titan's icy crust (Fig.~\ref{FigExamples1}(e)), the dominant weight function $f^{}_C$ decreases by 20\% between the bottom and top of the shell;
this variation is compensated by an increase in $f^{}_B$ so that $H^{}_\mu$ is approximately constant with depth.
Thin shell formulas thus do not yield as good results as for Europa, but are still approximately valid given the large uncertainties in the physical parameters.
The thin shell/rigid mantle approximation for the weight functions (big dots in Fig.~\ref{FigExamples1}(e)) is off by about 10\% mainly because of the large density difference between ocean and icy shell in the `dense ocean' model.
In the thin shell approximation with $h_2$ given by the interior model (triangles in Fig.~\ref{FigExamples1}(e)), estimates of weight functions are off by only a few percents from their average values over the shell thickness.
Spatial patterns in Titan's icy shell are thus well predicted by this type of thin shell approximation.
As for Europa, the stratification of the ice layer has nearly no effect on the weight functions.
Regarding global dissipation, the interior model yields $\mbox{\it Im}(k_2) = - 7.9\times10^{-4}$ whereas the thin shell/rigid mantle approximation gives $\mbox{\it Im}(k_2)=-8.3\times10^{-4}$, a difference in the imaginary part of less than 5\% (the overestimated weight functions are partially compensated by the neglected dissipation in the icy mantle, which amounts to 4\% of the total).

In Io's crust, weight functions are nearly constant and are well predicted by the thin shell approximation with $h_2$ given by the interior model (triangles in Fig.~\ref{FigExamples1}(f)).
The weakness of the asthenosphere underlying the crust results in a nearly free-slip condition for the crust and accounts for the success of this approximation.
The elasticity of the asthenosphere cannot however be neglected when computing the magnitude of the deformation.
If the asthenosphere is absent, the crust is strongly coupled to the mantle and the thin shell approximation cannot be used.
In that case, the values of the weight functions in the crust are well approximated by their values at the top of the mantle, which are sensitive to the size of the liquid core (see Figs.~\ref{FigSensitivityTwoLayer}(a,b,c,d)).
For example, $f^{}_A\approx{f}^{}_C$ if the ratio of core radius to mantle radius is about one-half ($f^{}_B$ is always close to zero in the crust).

\begin{figure}
   \centering
   \includegraphics[width=15cm]{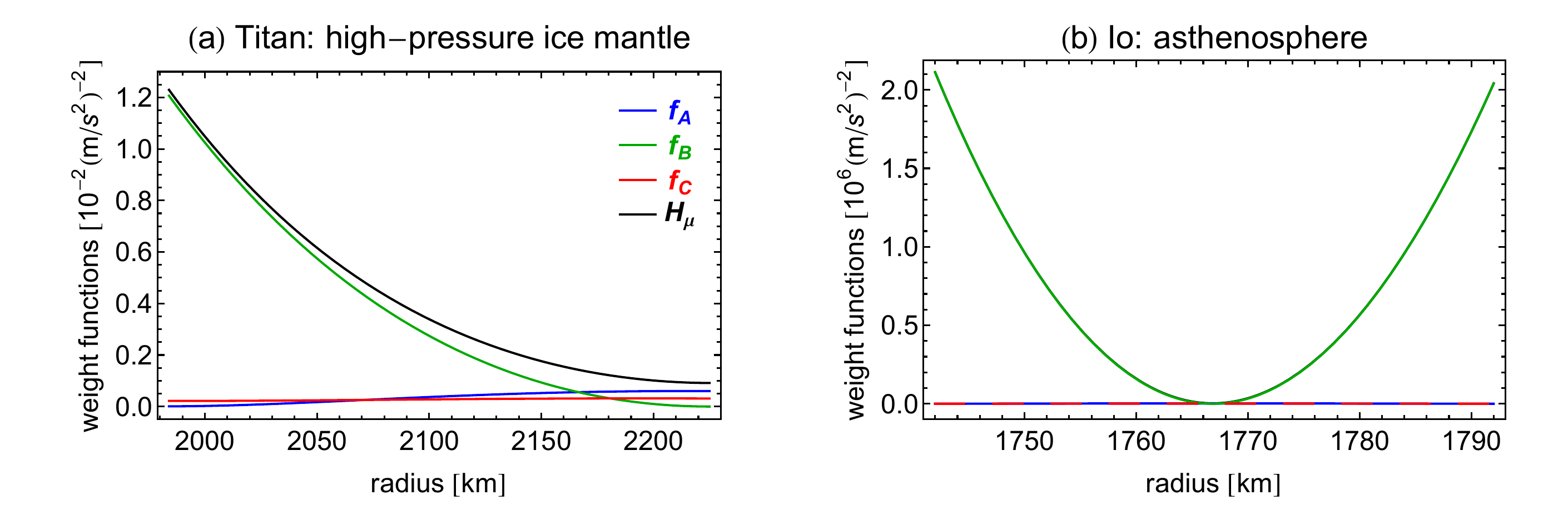}
   \caption{
   Weight functions in (a) the high-pressure ice mantle of Titan and (b) in the asthenosphere of Io.
   Interior models are specified in Tables~\ref{TableTitan} and \ref{TableIo}.
   The scale is not the same in the two panels.
   Curves $f^{}_A$ and $f^{}_C$ cannot be distinguished in Panel~(b) because of the large scale.
   }
   \label{FigExamples2}
\end{figure}

Finally, I compare dissipation weight functions in Titan's icy mantle and Io's asthenosphere.
In Titan's icy mantle (Fig.~\ref{FigExamples2}(a)), weight functions resemble those in the mantle of a two-layer body with an infinitely rigid core of radius $R_C=0.9R$ (see Fig.~\ref{FigSensitivityTwoLayer}(h)).
This result is explained by similar boundary conditions: free-slip condition at the mantle/ocean interface and rocky core much more rigid than the icy mantle.
Besides, the thickness of Titan's icy mantle is about 10\% of the mantle radius as in Fig.~\ref{FigSensitivityTwoLayer}(h).
In the lower half of Io's asthenosphere (Fig.~\ref{FigExamples2}(b)), weight functions look like those in Titan's icy mantle (or equivalently to those in a thin weak layer above a rigid core) but with an even stronger dominance of $f^{}_B$.
The thickness of Io's asthenosphere in the chosen model is indeed less than 3\% of the total radius, which is much less than in Fig.~\ref{FigSensitivityTwoLayer}(h).
In the upper half of Io's asthenosphere, weight functions are nearly the mirror image of the functions in the lower half because the upper boundary is not free but instead in strong shear with the much more rigid crust.

\subsection{Comparison with the literature}
\label{ComparisonLiterature}

The formalism developed in this paper is consistent with previous results found in the literature.
First, the formula for the global power (Eqs.~(\ref{Edot1})-(\ref{Edot2})) obtained by integrating the local power over the volume of the body agrees with the formulas that \citet{zschau1978} obtained using the macro approach to tidal heating.
Second, the formula for the radial distribution of the dissipation rate (Eq.~(\ref{defP0})) is the same as the one derived by \citet{tobie2005}.
In particular, the functions $H^{}_\mu$ and $H^{}_K$ are the same as \citet{tobie2005}'s radial sensitivity functions with the same name.
These equivalences have already been discussed in Section~\ref{GlobalPower}.
Third, the dominance of Pattern~C obtained in Section~\ref{HomogeneousBody} in a homogeneous body agrees with the results of \citet[Figs.~1-2]{peale1978} both for eccentricity tides and for obliquity tides.

Consider now the two maps of Io's surface flux that \citet{segatz1988} computed for mantle dissipation (Model~A) and asthenospheric dissipation (Model~B).
These authors compute tidal strains in the incompressible limit with the propagation matrix method.
It is difficult to reproduce exactly their figures because of some inconsistencies and missing information.
\citet{spohn1997} and \citet{schubert2004} already noted that \citet{segatz1988} forgot a factor $2\pi$ in the angular frequency $\omega$ (the error is limited to the computation of the complex shear modulus and neither affects the factor $\omega$ multiplying $\mbox{\it Im}(\tilde\mu)$ nor $\omega^4$ from the tidal potential).
Thus all viscosity values in \citet{segatz1988} should be divided by $2\pi$ since the Maxwell viscosity is always multiplied by $\omega$.
Besides these authors do not not clearly state the viscosity used for the mantle in Model A and for the asthenosphere in Model B.
As regards Model~A, I duplicate the map of surface flux shown in Fig.~8 of \citet{segatz1988} if the rheology of the mantle is specified by $\mu=10^{10}\rm\,Pa$ and $\eta=3/(2\pi)\times10^{16}\rm\,Pa.s$ (as in their `preferred' model shown in their Fig.~3, though the viscosity mentioned in their text is two-thirds of this value).
My choice leads to a sub-Jovian flux of $0.6\rm\,W{m^{-2}}$ as stated in their paper; the corresponding total power is 10\% smaller than their constraint of $\dot{E}=6\times10^{13}\rm\,W$, with $|k_2|=0.246$ and $Q=28$ (note that the values $|k_2|=0.25$ and $Q=36$ mentioned by \citet{segatz1988} yield a power 28\% lower than their power constraint).
The patterns $(\Psi^{}_A,\Psi^{}_B,\Psi^{}_C)$ have the weights $(7.2,22.3,70.5)\%$ in the surface flux.
As regards Model~B, I duplicate the map of surface flux shown in Fig.~10 of \citet{segatz1988} if the rheology of the asthenosphere is specified by $\mu=10^7\rm\,Pa$ and $\eta=1.5/(2\pi)\times10^8\rm\,Pa.s$, in which case the total power constraint is satisfied ($|k_2|=0.82$ and $Q=84$) and the maximum flux reaches $2.4\rm\,W{m^{-2}}$ as stated in their paper.
If I use the viscosity that \citet{segatz1988} claim to have used for their Fig.~5 ($\eta=1/(2\pi)\times10^8\rm\,Pa.s$), I obtain a power 50\% lower than their total power constraint.
Whatever the precise value of the viscosity, the distribution of surface heat flux is nearly 100\% Pattern~B.

What is the correspondence between the thin shell formulas of Section~\ref{ThinShell} and the results of \citet{ojakangas1989a}?
The substitution of Eq.~(\ref{l2thinshell}) into Eqs.~(\ref{LoveDispl})-(\ref{y1primesurf}) reproduces in a simpler way and with more generality (i.e. allowing for compressibility) the strains computed by these authors (their Eqs.~(B.20)-(B.25)), with their first three coefficients $g_i$ (the coefficient $g_4$ is not needed) corresponding to
\begin{equation}
\left( g_1 , g_2 , g_3 \right) = \frac{3}{4} \, R \omega^2 e  \left( r y_1' , y^{}_3, y^{}_1 \right) \Big|_{r=R} \, .
\end{equation}
In the right-hand side, one must set $\tilde\nu=1/2$ (incompressible limit) and $h_2 = (5/2)C_R$ ($C_R\cong0.5$ corresponds to the choice $\bar\rho=3\rho$ in Eq.~(\ref{defh20}), this density ratio being appropriate for Europa).
Besides, I can duplicate Fig.~1 of \citet{ojakangas1989a} showing the quantity $\overline{ \dot  \epsilon_{ij}^2}$ in units of $(\omega\gamma{e}C_R)^2/2$, in which $e$ is the eccentricity and $\gamma=\omega^2R/g$.
Combining Eq.~(\ref{PowerThinShell}) and Eq.~(\ref{PowerStrainOS89}), I obtain the correspondence
\begin{equation}
\frac{\overline{ \dot  \epsilon_{ij}^2} }{\frac{1}{2} (\omega\gamma{e}C_R)^2} = \frac{8}{3} \left( \frac{5}{2} \right)^2 \left( \frac{3}{11} \right)^2 e^{-2} \left( \Psi^{}_A + \frac{9}{2} \, \Psi^{}_C \right) \, .
\end{equation}

Lastly, consider the spatial patterns of dissipation in icy satellites computed by \citet{tobie2005} who took into account compressibility effects.
First, these authors examine the dissipation within the rocky mantle which is either homogeneous or includes a liquid core, the liquid core being either small or large.
In their Fig.~6, they show the spatial patterns of maximum dissipation rate, surface dissipation rate and surface heat flux (if radial heat transport) for these three models.
As regards the homogeneous model, I can duplicate Figs.~6(a,d,g) of \citet{tobie2005} using my Eqs.~(\ref{WeightsHomogeneous}) valid in the incompressible limit, because compressible effects are very small for a homogeneous body comparable in size to icy satellites.
As regards the two models with liquid core, \citet{tobie2005} already observed in their Fig.~4 that finite compressibility increases (resp.\ decreases) dissipation at the bottom (resp. top) of the mantle, but that the two effects approximately cancel when integrating on the radius.
In particular, compressibility leads to the damping of the weight function $f^{}_A$ at the lower and upper mantle boundaries and to an enhancement of $f^{}_C$ in the lower part of the mantle in comparison with my Fig.~\ref{FigSensitivityTwoLayer}(b) (note that the effect of compressibility on weight functions can vary a lot depending on the value of the reduced shear modulus and the type of boundary conditions).
I can duplicate Figs.~6(b,e,h) and Figs.~6(c,f,i) of \citet{tobie2005} with the parameters given in their paper by computing the functions $y^{}_i$ with a Fortran code allowing for compressibility which was originally developed by Dahlen for the Earth \citep{dahlen1976} (another version of this code is used by \citet{wahr2006,wahr2009}).
The basic differences between the resulting patterns are well explained by the toy model of a viscoelastic mantle above a liquid core.
For example, the component of harmonic degree four is much more visible at the top of the  mantle when the core is large ($R_C\approx{R}/2$, Fig.~6(f) of \citet{tobie2005}) than when the core is small ($R_C\approx{R}/3$, Fig.~6(e) of \citet{tobie2005}), because in the former case $f^{}_A\approx{f}^{}_C$ at the top of the mantle (see discussion in Section~\ref{LiquidCore}).

Next, \citet{tobie2005} examine the dissipation within icy layers for three different models: a thick icy layer in contact with the rocky mantle, a thin icy layer above a subsurface ocean and a thick high-pressure ice layer sandwiched between the rocky mantle and a subsurface ocean.
In their Fig.~10, they show the spatial patterns of dissipation at the top and bottom of these icy layers.
I choose to reproduce their results with the incompressible propagator matrix method because this method works well when there are large rheological contrasts between layers.
I can duplicate nearly all spatial patterns of \citet{tobie2005} with the parameters given in their paper: their Figs.~10(a,e) result from Pattern~B only (bottom of strongly coupled icy shell), their Figs.~10(b,f) result from a mix of Patterns~A and C (top of strongly coupled icy shell) whereas their Fig.~10(d) corresponds to the pattern in a thin icy shell above an ocean.
These results are already well predicted by the toy model of a viscoelastic mantle above a rigid core (see discussion in Section~\ref{RigidCore}) and are also similar to what I obtained for realistic models of Europa and Titan (see my Figs.~\ref{FigExamples1}(d,e) and \ref{FigExamples2}(a)).
The pattern in their Fig.~10(c) is not absolutely correct.
Showing the pattern at the bottom of the thin icy shell in their `Europa-with-ocean' model, it should be the same as the pattern at the top of the shell (their Fig.~10(d)) because the ice shell is extremely thin (1~km is less than 0.1\% of the radius).
In that case, weight functions are nearly constant through the shell (see my Figs.~\ref{FigExamples1}(d,e)).
Compressibility does not play a role since its influence on spatial patterns is negligible in a thin shell.
The difference between Figs.~10(c) and 10(d) of \citet{tobie2005} is however small.

\section{Conclusions}
\label{Conclusions}
\markright{6 CONCLUSIONS\hfill}

In this paper, I showed that three basic spatial patterns are enough to describe the angular dependence of the power dissipated by tides in a spherically stratified body.
This means that the dissipated power at a given radius is the sum of three terms, each term being the product of a radial function (or weight function) depending on the internal structure and an angular function (or spatial pattern) depending only on the tidal potential (Eqs.~(\ref{PsiABC})-(\ref{weights})).
One term results from radial-tangential shear strains, another from tangential strains while the remaining term (usually the subdominant one) receives contributions from radial strains, tangential strains and volume change.
This decomposition presents the advantage that the 3D problem of predicting the spatial pattern of dissipation at all radii is reduced to the 1D problem of computing the tidal displacements.
As a further simplification, the three basic angular functions can be expressed as linear combinations of the harmonic components of degrees zero, two and four of the squared norm of the Fourier-transformed tidal potential (Eqs.~(\ref{PsiABCsh})).
The three basic patterns of tidal dissipation (Fig.~\ref{FigPatterns}) are thus known once the tidal potential is specified, the most important case being a body in synchronous rotation undergoing eccentricity tides.

The angular average of the spatially-dependent dissipated power yields the radial distribution of the dissipated power (already known from conservation of energy), the integration of which gives the classical formula for the global dissipated power.
This last quantity is independently known, being proportional to the surface integral of the product of the tidal potential with the induced potential.
The formula for the dissipated power derived in this paper thus completes the connection between the microscopic approach to tidal dissipation, based on the computation of strains, and the macroscopic approach, based on the phase delay of the induced potential (or on the phase delay of the tidal bulge).

I studied the effect of internal structure on the dissipation pattern with different types of toy models.
In a thin icy shell above an ocean, the dissipation pattern is nearly invariant through the shell thickness while the dissipation amplitude varies with depth as the rheology changes from elastic (conductive layer) to viscoelastic (convective layer).
In an incompressible homogeneous body (also relevant to the solid core), the dissipation pattern depends on depth but not on rheology.
The third toy model is the incompressible body with two viscoelastic layers.
I demonstrated that the density contrast between the two layers weakly affects weight functions ratios.
More generally, assuming uniform density in a multi-layer body leads to the factorization of the overall deformation (described by the Love number $h_2$), so that the spatial pattern in each homogeneous layer depends only on contrasts of shear moduli at the layer boundaries.

An important subset of the two-layer model consists of bodies with a liquid core.
I showed how the weight functions change as the core radius increases from zero (homogeneous model) to the surface (thin shell model).
In all cases, the same pattern is dominant, resulting in more heating at the poles and less heating at the equator, at least if the body is in synchronous rotation and undergoes eccentricity tides (this result was already known).
Adding a liquid layer and an icy shell above the viscoelastic mantle reduces the magnitude of weight functions within the mantle but does not change their ratios, so that the spatial pattern within the mantle remains approximately the same.

Another subset of the two-layer model consists of bodies with a viscoelastic mantle surrounding a core of infinite rigidity.
While this model may seem at first sight unphysical, it describes well the increase in radial-tangential shear dissipation occurring at the interface between a soft and a rigid layer.
This situation could for example  occur in icy satellites if the icy layer is in contact with the rocky mantle.
This behavior is also typical of the asthenospheric dissipation that could explain Io's volcanism, except that radial-tangential shear dissipation is also large at the top of the asthenosphere where it is in contact with the more rigid lithosphere.
If the soft layer is thin, the spatial pattern for eccentricity tides within a body in synchronous rotation corresponds to less heating at the poles and more heating in the equatorial zone, with a significant component of harmonic degree four (the maxima are thus at the north and south of the sub- and anti-primary points).
This pattern was already known from internal models of Io.
If the soft layer is thick, the three basic patterns combine with comparable weights, resulting in a new type of dissipation pattern, with maximum heating at the equator as in asthenospheric dissipation but with a lower content of harmonic degree four.

In summary, there are three classes of spatial patterns associated with 1) mantle dissipation (with the icy shell above an ocean as a particular case), 2) dissipation in a thin soft layer (asthenosphere or icy layer in contact with the mantle or core) and 3) dissipation in a thick soft layer (deep asthenosphere or thick icy layer above a solid core).
I demonstrated the generality of the inferences drawn from toy models by computing weight functions for realistic internal structures of Europa, Titan and Io.
Finally, I duplicated spatial patterns previously obtained in the literature.

The formalism described in this paper rests on two crucial assumptions: linear rheology and spherical symmetry of the internal structure.
Further assumptions on heat transport are necessary in order to compute the surface heat flux.
The easy way out is to assume radial heat transport since it directly leads to the linear superposition of dissipation patterns at all depths whereas lateral heat transport tends to average patterns in complex ways.
For example, \citet{tackley2001b} computed 3D convection within Io using as input tidal heating predicted by spherically stratified models and neglecting lateral variations of viscosity.
At low Rayleigh number, the heat flux follows the input dissipation pattern but small-scale instabilities spread the heat flux as the convective vigor increases.
In these simulations, the Rayleigh number is still much lower than possible values within Io.
In another paper, \citet{tackley2001a} used 2D simulations at a more realistic Rayleigh number in order to show that large-scale lateral flows tend to average the dissipation pattern.
The averaging effect due to convective heat transport is however counteracted by the feedback of nonuniform tidal heating on local viscosity.
Using codes coupling convection and tidal heating, \citet{behounkova2010} and \citet{han2010} showed that tidal heating is significantly higher (resp. lower) in warm (resp. colder) convective plumes.
With their 3D code, \citet{behounkova2010} obtained dissipation patterns in Enceladus' mantle roughly similar to the pattern predicted by the method assuming internal spherical symmetry but with more small-scale structure (\citet{han2010}'s code is 2D and cannot predict global patterns).
Computations coupling convection and tidal heating in 3D are very complex and have neither been applied to all types of dissipation (such as asthenospheric heating) nor been systematically compared to the results obtained with the method assuming internal spherical symmetry.
For these reasons I believe that the analytical approach presented here will remain a useful and practical tool to predict tidal dissipation patterns.


\section*{Appendix A. Dissipated power from strains}
\label{AppendixA}
\markright{APPENDIX A\hfill}

I show here how the dissipated power can be expressed in terms of viscoelastic strains in the Fourier domain.
The formulas are well-known but not always correctly stated in the literature.
Tidal deformations induce stresses $\sigma_{ij}(t,{\bf x})$ and strains $\epsilon_{ij}(t,{\bf x})$ at each point ${\bf x}$ within the planet.
If tides operate at only one angular frequency $\omega$,  stresses and strains can be written in terms of complex Fourier components $\tilde \sigma_{ij}$ and $\tilde \epsilon_{ij}$:
\begin{eqnarray}
\sigma_{ij}(t,{\bf x}) &=&  \frac{1}{2} \, \tilde \sigma_{ij} \, \exp(i \omega t) + c.c. \, ,
\nonumber
\\
\epsilon_{ij}(t,{\bf x}) &=&  \frac{1}{2} \, \tilde \epsilon_{ij} \, \exp(i \omega t) + c.c. \, ,
\label{epsilonTF}
\end{eqnarray}
where $c.c.$ means complex conjugate.
The dependence on ${\bf x}$ of the Fourier components is implicit.
If there are several tidal frequencies, the Fourier decomposition involves a sum over frequencies.
The quantities $\sigma_{ij}$ and $\epsilon_{ij}$ (or $\tilde\sigma_{ij}$ and $\tilde\epsilon_{ij}$) are the physical components of the stress and strain tensors, that is the Cartesian components in a local system of rectangular Cartesian coordinates with axes tangent to the coordinate curves at the point \citep{fung1965,malvern1969}.
In spherical coordinates, this means that the basis vectors are normalized.

In the Fourier domain, the constitutive equations for an elastic solid take the form \citep{fung1965,malvern1969}
\begin{eqnarray}
\tilde \sigma_{ij}' &=& 2 \mu \, \tilde\epsilon_{ij}'  \, ,
\nonumber  \\
\tilde\sigma &=& 3 K \, \tilde\epsilon \, ,
\end{eqnarray}
where the real parameters $\mu$ and $K$ are the shear and bulk moduli, respectively.
Deviatoric stresses and strains (physical components) are defined by
\begin{eqnarray}
\tilde \sigma_{ij}' &=& \tilde \sigma_{ij} - \frac{1}{3} \, \tilde \sigma \, \delta_{ij} \, ,
\nonumber \\
\tilde \epsilon_{ij}' &=& \tilde \epsilon_{ij} - \frac{1}{3} \, \tilde \epsilon \, \delta_{ij} \, ,
\end{eqnarray}
where $\tilde \sigma$ and $\tilde \epsilon$ are the traces of the stress and strain tensors,
\begin{eqnarray}
\tilde \sigma &=& \tilde \sigma_{kk} \, ,
\nonumber \\
\tilde \epsilon &=& \tilde \epsilon_{kk} \, ,
\end{eqnarray}
with implicit summation over repeated indices.

The correspondence principle \citep{fung1965} states that the constitutive equations for a viscoelastic body are similar in form but with complex shear modulus $\tilde\mu(\omega)$ and complex bulk modulus $\tilde{K}(\omega)$ depending on the angular frequency $\omega$:
\begin{eqnarray}
\tilde \sigma_{ij}' &=& 2 \, \tilde\mu(\omega) \, \tilde\epsilon_{ij}'  \, ,
\nonumber
\\
\tilde\sigma &=& 3 \, \tilde{K}(\omega) \, \tilde\epsilon \, .
\label{constitutive2}
\end{eqnarray}
The dependence on frequency of $\tilde\mu(\omega)$ and $\tilde{K}(\omega)$ defines the rheology of the material.

\citet{peale1978} gave the correct formula for the rate at which stresses do work per unit volume:
\begin{equation}
P(t) = \sigma_{ij}(t) \, \dot\epsilon_{ij}(t) \, .
\end{equation}
\citet{kaula1964} incorrectly introduced a factor $1/2$ in this formula (see his Eqs.~(10)-(16)-(19)), arguing that `work is done only with motion, not with just change of force'.
Moreover Kaula thought that a fraction of this energy ($1/2$ for the Moon, $1/27$ for the Earth) went into orbital kinetic energy.

The dissipation power per unit volume averaged over one orbital period $T$ reads
\begin{equation}
P = \frac{1}{T} \int_0^T \sigma_{ij}(t) \, \dot\epsilon_{ij}(t) \, dt \, .
\label{AveragedPower}
\end{equation}
In terms of the Fourier components appearing in Eqs.~(\ref{epsilonTF}), the dissipated power reads
\begin{equation}
P = \frac{\omega}{2} \mbox{\it Im}\left( \tilde \sigma_{ij}^{} \, \tilde \epsilon_{ij}^{\,*} \right) \, ,
\label{PowerImag}
\end{equation}
where the asterisk denotes the complex conjugate.
If stresses and strains are in phase, $\tilde \sigma_{ij}^{}\,\tilde \epsilon_{ij}^{\,*}$ is real so that the dissipated power vanishes.
Eq.~(\ref{PowerImag}) agrees with Eq.~(24) of \citet{tobie2005}.

In terms of deviatoric tensors, the dissipated power reads
\begin{eqnarray}
P &=&  \frac{\omega}{2} \mbox{\it Im}\left( \tilde \sigma_{ij}' \, \tilde \epsilon_{ij}'^{\,*} + \frac{1}{3} \, \tilde\sigma \, \tilde\epsilon^* \right)
\nonumber \\
&=& \omega \mbox{\it Im}(\tilde\mu) \, \tilde\epsilon_{ij}' \, \tilde\epsilon_{ij}'^{\,*} + \frac{\omega}{2} \mbox{\it Im}(\tilde{K}) \left| \tilde\epsilon \right|^2 \, .
\label{PowerStressStrain}
\end{eqnarray}
The first term is due to the change of body shape whereas the second term is due to the change of volume.
Note that Eq.~(10) of \citet{kaula1964} has an incorrect factor $1/6$ instead of $1/2$ in the term proportional to $\tilde{K}$.

In terms of non-deviatoric strains, the dissipated power reads
\begin{equation}
P =  \omega \mbox{\it Im}(\tilde\mu) \left( \tilde\epsilon_{ij}^{} \, \tilde\epsilon_{ij}^{\,*} - \frac{1}{3} \left| \tilde\epsilon \right|^2 \right) + \frac{\omega}{2} \mbox{\it Im}(\tilde{K}) \left| \tilde\epsilon \right|^2 \, .
\label{PowerStrain}
\end{equation}
Note that Eq.~(8) of \citet{segatz1988} is wrong by having a factor $\omega/2$ instead of $\omega$ in front of $\mbox{\it Im}(\tilde\mu)$.

Following \citet{ojakangas1989a}, many authors write the dissipated power in terms of the averaged `squared' strain rate defined as
\begin{eqnarray}
\overline{ \dot  \epsilon_{ij}^2}
&=& \frac{1}{T} \int_0^T \dot\epsilon_{ij}(t)  \, \dot\epsilon_{ij}(t) \, dt
\nonumber \\
&=& \frac{\omega^2}{2} \, \tilde\epsilon_{ij}^{} \, \tilde\epsilon_{ij}^{\,*} \, .
\end{eqnarray}
For an incompressible body with Maxwell rheology, Eq.~(\ref{PowerStrain}) becomes
\begin{equation}
P =  \frac{2\mu \, \overline{ \dot  \epsilon_{ij}^2}}{\omega} \, \frac{\omega \tau_M}{1+(\omega\tau_M)^2}  \,  \, ,
\label{PowerStrainOS89}
\end{equation}
where $\tau_M=\eta/\mu$ is the Maxwell time ($\eta$ is the viscosity).
Recall that Maxwell rheology is defined by \citep[e.g.][]{peltier1982,tobie2005,wahr2009}
\begin{equation}
\tilde\mu =  \frac{\mu \, \omega \tau_M}{1+(\omega\tau_M)^2}  \left( \omega \tau_M + i \right)   .
\label{MaxwellRheology}
\end{equation}

\section*{Appendix B. Differential operators on the sphere}
\label{AppendixB}
\markright{APPENDIX B\hfill}

The tidal potential at the surface of a spherical planet is a scalar function which means that it is invariant under rotations of the coordinate system.
The dissipated power at a given radius is also a scalar function which depends on derivatives of the Fourier-transformed tidal potential.
My purpose is to construct a scalar function from derivatives of another scalar function.
Let $f(\theta,\phi)$ be a complex scalar function defined on the sphere.
How can we combine derivatives of $f$ into a scalar?
The standard method consists in (1) constructing tensors with covariant derivatives $\nabla_\alpha$ and (2) contracting all indices with the inverse metric on the unit sphere,
\begin{equation}
g^{\alpha\beta} = diag(1,\sin^{-2}\theta) \, ,
\end{equation}
with Greek indices denoting angular coordinates $(\theta,\phi)$.
Covariant derivatives and other operators on the sphere are discussed in detail in Appendices~B and D of \citet{beuthe2008}.

A well-known scalar expression of order two in derivatives is the contracted double covariant derivative, called the spherical Laplacian:
\begin{equation}
\Delta \, f= g^{\alpha\beta} \, \nabla_\alpha \nabla_\beta \, f \, .
\label{laplacianDEF1}
\end{equation}
The eigenfunctions of $\Delta$ are spherical surface harmonics of degree $\ell$ with eigenvalues
\begin{equation}
\delta_\ell = -\ell (\ell+1) \, .
\label{eigen}
\end{equation}
Contracting covariant derivatives of $f$ and $f^*$ yields scalars of order two and four:
\begin{eqnarray}
{\cal D}_2(f,f^*) &=& g^{\alpha\beta} \left( \nabla_\alpha \, f \right) \left(\nabla_\beta \, f^* \right) \, ,
\nonumber \\
{\cal D}_4(f,f^*) &=& g^{\alpha\gamma} g^{\beta\delta} \left( \nabla^{}_\alpha \nabla^{}_\beta \, f \right) \left( \nabla_\gamma \nabla_\delta \, f^* \right) \, .
\label{opDcov}
\end{eqnarray}
In terms of usual derivatives, ${\cal D}_2$ reads
\begin{equation}
{\cal D}_2(f,f^*) = \left(
\frac{\partial f}{\partial \theta} \frac{\partial f^*}{\partial \theta}
+ \frac{1}{\sin^2\theta} \, \frac{\partial f}{\partial \phi} \, \frac{\partial f^*}{\partial \phi}
\right) \, .
\label{opD2der}
\end{equation}
Since ${\cal D}_4$ has a complicated expression in terms of usual derivatives, I define auxiliary operators as follows.
In the normalized basis \citep[][Section~3.2 and Appendix~A]{beuthe2008}, the operators corresponding to double covariant derivatives are given by
\begin{eqnarray}
\bar {\cal O}_1 \, f &=& \nabla_\theta \nabla_\theta \, f  \, ,
\nonumber \\
\bar {\cal O}_2 \, f &=&  \frac{1}{\sin^2 \theta} \, \nabla_\phi \nabla_\phi \, f \, ,
\nonumber \\
\bar {\cal O}_3 \, f &=& \frac{1}{\sin \theta} \,  \nabla_\theta \nabla_\phi \, f \, .
\label{opODEF1}
\end{eqnarray}
The operators $\bar{\cal O}_i$ are given in terms of usual derivatives by
\begin{eqnarray}
\bar {\cal O}_1 &=& \frac{\partial^2}{\partial \theta^2}  \, ,
\nonumber \\
\bar {\cal O}_2 &=&  \frac{1}{\sin^2 \theta} \left( \frac{\partial^2}{\partial \varphi^2} + \cos \theta \sin \theta \, \frac{\partial}{\partial \theta}  \right) \, ,
\nonumber \\
\bar {\cal O}_3 &=& \frac{1}{\sin \theta}  \left( \frac{\partial^2}{\partial \theta \partial \varphi} - \cot \theta \, \frac{\partial}{\partial \varphi} \right) \, .
\label{opODEF2}
\end{eqnarray}
The operators $\Delta$ and ${\cal D}_4$ can be expressed in terms of usual derivatives through combinations of the operators $\bar{\cal O}_i$:
\begin{eqnarray}
\Delta \, f &=& \left( \bar {\cal O}_1 + \bar {\cal O}_2 \right) f \, ,
\nonumber
\\
{\cal D}_4(f,f^*) &=& \left( \bar {\cal O}_1 f \right) \left( \bar {\cal O}_1f^* \right) + \left( \bar {\cal O}_2 f \right) \left( \bar {\cal O}_2 f^* \right) + 2 \left( \bar {\cal O}_3 f \right) \left( \bar {\cal O}_3 f^* \right) \, .
\label{opD4der}
\end{eqnarray}
Finally,  the operators ${\cal D}_2$ and ${\cal D}_4$ expressed in terms of covariant derivatives (Eqs.~(\ref{opDcov})) can be transformed with Eqs.~(G.2)-(G.3) of \citet{beuthe2010a} so that derivatives appear only through the Laplacian.
If $f$ is a spherical harmonic of degree $\ell$, I get
\begin{eqnarray}
{\cal D}_2(f,f^*) &=& \frac{1}{2} \left( \Delta - 2\delta_\ell \right) |f|^2 \, ,
\nonumber
\\
{\cal D}_4(f,f^*) &=& \frac{1}{4} \Big( \Delta \Delta - 2 \left( 2 \delta_\ell + 1 \right) \Delta + 4 \delta_\ell \left( \delta_\ell +1 \right) \Big) |f|^2 \, ,
\label{opD4DEF3}
\end{eqnarray}
where $\delta_\ell$ results from the action of $\Delta$ on $f$ or $f^*$ (see Eq.~(\ref{eigen})).
However $|f|^2$ is not an eigenfunction of $\Delta$ because it is the sum of harmonic components of even degrees between 0 and $2\ell$.

\section*{Appendix C. Tidal potential for synchronous orbit}
\label{AppendixC}
\markright{APPENDIX C\hfill}

Let $(R,\theta,\phi)$ be the radius, colatitude and longitude of point on the surface of a satellite orbiting with semi-major axis $a$ around a point mass $M$ (the primary).
The time-dependent part of the tidal potential at $(R,\theta,\phi)$ is given at first order in eccentricity $e$ and obliquity $I$ by Eq.~(1) of \citet{kaula1964}:
\begin{eqnarray}
&& U(t,\theta,\phi) =  \frac{GMR^2}{{a}^3}  \left( 
 - \frac{3e}{2} \, P^{}_{20}(\cos\theta) \cos {\cal M}
\right.
\nonumber \\
&& \hspace{28mm}
+ \, \frac{e}{8} \, P^{}_{22}(\cos\theta) \left( -\cos \left( {\cal M}-2\Lambda+2\omega_p \right)
+ 7 \cos \left( 3{\cal M}-2\Lambda+2\omega_p \right) \right)
\nonumber \\
&& \hspace{27.5mm}
\left.
+ \, \frac{1}{2} \sin I \, P^{}_{21}(\cos\theta) \left( \sin \left( 2{\cal M}-\Lambda+2\omega_p\right) + \sin \Lambda \right)
\right) .
\label{TidalPot0}
\end{eqnarray}
$P_{\ell{m}}(x)$ are the associated Legendre functions  of degree $\ell$ and order $m$, ${\cal M}$ is the mean anomaly and $\omega_p$ is the argument of pericentre.
The auxiliary variable $\Lambda$ is defined by
\begin{equation}
\Lambda = \phi+\Theta-\Omega \, ,
\end{equation}
where $\phi$ is the longitude in a frame $({\bf x},{\bf y},{\bf z})$ attached to the satellite (${\bf x}$ and ${\bf y}$ are within the equatorial plane), $\Theta$ is the sidereal time for the reference meridian (the angle between the x-axis and an inertially fixed point $Q$ on the equator) and $\Omega$ is the longitude of the ascending node (measured from $Q$).

If the orbital motion is synchronous with the rotation of the satellite, the sidereal time and the mean anomaly are related by $\Theta={\cal M}+const$.
Note that ${\cal M}$ gives the position of the planet only at pericentre and apocentre, unless eccentricity vanishes, and that the angles $(\Theta,{\cal M})$ are not measured in the same plane, unless obliquity vanishes.
I choose the direction of the x-axis by setting 
\begin{equation}
 \Theta = {\cal M} + \varpi \, .
\label{sidereal}
\end{equation}
where $\varpi$ is the longitude of pericentre:
\begin{equation}
\varpi = \Omega + \omega_p \, .
\end{equation}
If the obliquity vanishes, the x-axis points to the primary when the latter is at pericentre: $\Theta=\varpi$ when ${\cal M}=0$  ({\it mod} $2\pi$).
If the eccentricity vanishes, the x-axis points to the primary when the latter is at the ascending node: $\Theta=\Omega$ when ${\cal M}+\omega_p=0$ ({\it mod} $2\pi$).
In the general case of non-zero obliquity and eccentricity, the x-axis points to the primary neither at pericentre nor at the ascending node.

Substituting Eq.~(\ref{sidereal}) into the tidal potential (\ref{TidalPot0}) and setting ${\cal M}=nt$ (time of pericentre passage set to zero, $n$ being the mean motion), I get
\begin{eqnarray}
&& U(t,\theta,\phi) = (nR)^2 \Big(
-\frac{3}{2}e \, P^{}_{20}(\cos\theta) \cos nt + e \, P^{}_{22}(\cos\theta) \left( \frac{3}{4} \cos nt \cos2\phi + \sin nt \sin2\phi \right) 
\nonumber \\
&& \hspace{33.5mm}
+ \, \sin I \, P^{}_{21}(\cos\theta) \sin \left( nt+ \omega_p \right) \cos \phi
\Big) \, .
\label{TidalPot1}
\end{eqnarray}
The approximation $GM/a^3=n^2$ is valid if the deformed body is much lighter than the primary.

The Fourier coefficient of the tidal potential defined by Eq.~(\ref{TidalFourier}) is thus equal to
\begin{equation}
\Phi_n = (nR)^2 \Big(
-\frac{3}{2} e \, P_{20}(\cos\theta) \, + \, e \, P_{22}(\cos\theta) \left( \frac{3}{4} \cos2\phi - i  \sin2\phi \right) - \, i \, e^{i\omega_p} \sin I \, P_{21}(\cos\theta) \cos \phi
\Big) \, .
\label{TidalPot2}
\end{equation}
The part of the tidal potential due to non-zero obliquity is dephased by the argument of pericentre $\omega_p$ (frame-independent quantity) with respect to the part due to non-zero eccentricity.

\section*{Appendix D. Squared norm of the tidal potential}
\label{AppendixD}
\markright{APPENDIX D\hfill}

Let the Fourier coefficient of an arbitrary tidal potential of degree two be written as
\begin{equation}
\Phi_n = 
(nR)^2 \left( \alpha \, P_{20}(\cos\theta) + \left( \beta \cos2\phi + \gamma \sin2\phi \right) P_{22}(\cos\theta) + \left( \delta \cos \phi + \epsilon \sin \phi \right) P_{21}(\cos\theta) \right) \, ,
\label{TidalPotGen}
\end{equation}
where $(\alpha,\beta,\gamma,\delta,\epsilon)$ are dimensionless complex numbers.
The term in $P_{20}$ appears when the orbit is eccentric, the terms in $P_{22}$ appear either when the rotation is not synchronous with the orbital motion or when the orbit is eccentric or when there are forced librations in longitude, whereas the term depending on $P_{21}$ appears when the obliquity is not zero.

The squared norm of the tidal potential involves products of Legendre functions of degree two.
These can be expressed as linear combinations of Legendre functions with the formulas of \citet[][pp.~358-359]{balmino1994} to which I add
\begin{equation}
P_{21} P_{22} = (3/35) \, P_{43} \, = \, (12/7) \, P_{21} - (18/35) \, P_{41} \, .
\end{equation}

The coefficients $(a^{}_{\ell{m}},b^{}_{\ell{m}})$ of even order in the spherical harmonic expansion (\ref{expansion2}) read
\begin{eqnarray}
a^{}_{00} &=& (1/5) \left( |\alpha|^2 + 12 \left( |\beta|^2 + |\gamma|^2 \right) + 3  \left( |\delta|^2 + |\epsilon|^2 \right) \right) \, ,
\nonumber \\
a^{}_{20} &=& (1/7) \left( 2 |\alpha|^2 - 24 \left( |\beta|^2 + |\gamma|^2 \right) + 3 \left( |\delta|^2 + |\epsilon|^2 \right) \right) \, ,
\nonumber \\
a^{}_{22} &=& (1/14) \left( - 8 Re[\alpha \beta^*] + 3 \left( |\delta|^2 - |\epsilon|^2 \right) \right) \, ,
\nonumber \\
b^{}_{22} &=& (1/7) \, Re[- 4 \alpha \gamma^* + 3 \delta \epsilon^*] \, ,
\nonumber \\
a^{}_{40} &=& (18/35) \left( |\alpha|^2 + 2 \left( |\beta|^2 + |\gamma|^2 -|\delta|^2 - |\epsilon|^2 \right) \right) \, ,
\nonumber \\
a^{}_{42} &=& (3/35) \left( 2 Re[\alpha \beta^*] + |\delta|^2 - |\epsilon|^2 \right) \, ,
\nonumber \\
b^{}_{42} &=&(6/35) \, Re[\alpha \gamma^* + \delta \epsilon^*] \, ,
\nonumber \\
a^{}_{44} &=& (3/70) \left( |\beta|^2 - |\gamma|^2 \right) \, ,
\nonumber \\
b^{}_{44} &=& (3/35) \, Re[\beta \gamma^*] \, ,
\label{CoeffEven}
\end{eqnarray}
while the coefficients of odd order (resulting from interferences between non-zero obliquity terms and other terms) are given by
\begin{eqnarray}
a^{}_{21} &=& (2/7) \, Re[ (\alpha + 6  \beta) \delta^* + 6 \gamma \epsilon^*  ]  \, ,
\nonumber \\
b^{}_{21} &=& (2/7) \, Re[ (\alpha - 6 \beta ) \epsilon^* + 6 \gamma \delta^*  ]  \, ,
\nonumber \\
a^{}_{41} &=& (18/35) \, Re[ (\alpha - \beta) \delta^* - \gamma \epsilon^* ]  \, ,
\nonumber \\
b^{}_{41} &=& (18/35) \, Re[ (\alpha + \beta) \epsilon^* - \gamma \delta^* ]  \, ,
\nonumber \\
a^{}_{43} &=& (3/35) \, Re[ \beta \delta^* - \gamma \epsilon^*] \, ,
\nonumber \\
b^{}_{43} &=& (3/35) \, Re[ \beta \epsilon^* + \gamma \delta^* ] \, .
\label{CoeffOdd}
\end{eqnarray}

\section*{Appendix E. Variance of spatial patterns}
\label{AppendixE}
\markright{APPENDIX E\hfill}

The variance of a function $\Psi$ defined on the unit sphere reads
\begin{equation}
\mbox{var} (\Psi) = \frac{1}{4\pi} \int_S \left( \Psi - \bar \Psi \right)^2 d\sigma \, ,
\label{VarianceDef}
\end{equation}
where $\bar\Psi$ is the average of $\Psi$ on the sphere and $d\sigma$ is the integration measure.
The angular functions $\Psi^{}_{A,B,C}$ take the generic form
\begin{equation}
\Psi^{}_J = x \, \Psi^{}_0 + y \, \Psi^{}_2 + z \, \Psi^{}_4 \hspace{1cm} (J=A,B,C) \, .
\end{equation}
with coefficients $(x,y,z)$ given by Eqs.~(\ref{PsiABCsh}).
The variance of $\Psi^{}_J$ is equal to
\begin{equation}
\mbox{var} (\Psi^{}_J) = y^2 \, \mbox{var} (\Psi^{}_2) + z^2 \, \mbox{var} (\Psi^{}_4) \, ,
\end{equation}
where each term can be evaluated with the expansion (\ref{expansion2}),
\begin{equation}
\mbox{var} \left( \Psi^{}_\ell \right) = \sum_{m=0}^\ell  V^{}_{\ell{m}} \left( (a^{}_{\ell{m}})^2 + (b^{}_{\ell{m}})^2 \right) \, .
\label{variance1}
\end{equation}
The coefficients $V_{\ell{m}}$ are computed with the orthogonality relation of spherical harmonics:
\begin{eqnarray}
\left( V_{20}, V_{21}, V_{22} \right) &=& \left( \frac{1}{5}, \frac{3}{5}, \frac{12}{5} \right) \, ,
\nonumber \\
\left( V_{40}, V_{41}, V_{42}, V_{43}, V_{44} \right) &=& \left( \frac{1}{9}, \frac{10}{9}, 20, 280, 2240 \right) \, .
\label{variance2}
\end{eqnarray}
Table~\ref{TableVariance} gives the standard deviation (square root of the variance) of $\Psi^{}_{A,B,C}$ for synchronous rotation (eccentricity tides only or obliquity tides only).

\section*{Appendix F. Multi-layer incompressible body of uniform density}
\label{AppendixF}
\markright{APPENDIX F\hfill}

Consider an incompressible body of radius $R$ stratified into spherical homogeneous layers which can be solid or liquid, except the surface layer which is solid.
The $y_i$ functions (defined as in \citet{takeuchi1972}) required for the strains (Eqs.~(\ref{strainsRAD})-(\ref{strainsTAN})) have the following form in the solid layers \citep{sabadini2004}:
\begin{eqnarray}
y^{}_1 &=& \frac{h_2}{g} \left(  \frac{1}{7}  \, a \,\bar r^3 + b \, \bar r + \frac{1}{2} \, c\, \bar r^{-2} + d \, \bar r^{-4} \right) \, ,
\nonumber
 \\
y^{}_3 &=& \frac{h_2}{g} \left(  \frac{5}{42} \, a \, \bar r^3 + \frac{1}{2} \, b \, \bar r  - \frac{1}{3} \, d \, \bar r^{-4} \right) \, ,
\nonumber
 \\
\frac{r y^{}_4}{\tilde \mu} &=&  \frac{h_2}{g} \left(    \frac{8}{21} \, a \, \bar r^3 + b \, \bar r + \frac{1}{2} \, c \, \bar r^{-2} + \frac{8}{3} \, d \,\bar r^{-4} \right) \, .
\label{yiFun4}
\end{eqnarray}
where $\bar{r}=r/R$, $g$ is the surface gravity, $\tilde\mu$ is the viscoelastic shear modulus of the layer and $h_2=gy^{}_1(R)$ is the tidal Love number for radial displacement.
The dimensionless coefficients $(a,b,c,d)$ are constant in each layer.
The factorization of $h_2/g$ introduces one more unknown ($h_2$) in the problem but also one more constraint:
\begin{equation}
\frac{1}{7}  \, a_S + b_S + \frac{1}{2} \, c_S + d_S  = 1 \, ,
\label{bound1}
\end{equation}
where the index $S$ denotes the surface layer.
The boundary condition of no shear stress at the surface ($y^{}_4(R)=0$) yields another constraint:
\begin{equation}
\frac{8}{21} \, a_S + b_S + \frac{1}{2} \, c_S + \frac{8}{3} \, d_S = 0 \, .
\label{bound4}
\end{equation}
The two other boundary conditions of the problem involve the gravitational potential so that it is necessary to solve for strains, stresses and gravitational potential at the same time.

Suppose now that the density $\rho$ of the body is uniform.
In that case, it is possible to solve for the deformations independently of the gravitational potential.
First, it can be shown that the dimensionless constants $(a,b,c,d)$ depend only on the ratios between the parameters characterizing the different layers, that is their radii and shear moduli (the proof relies on the fact that the transition matrix between the layers $i$ and $i+1$ depends only on $R_i/R$ and $\tilde\mu_i/\tilde\mu_{i+1}$).
As a result the functions $(y^{}_1,y^{}_3,ry^{}_4/\tilde\mu)$ depend on a dimensional viscoelastic parameter only through the prefactor $h_2$.
Therefore the ratios of squared strains (determining the relative importance of the dissipation patterns) do not depend on the absolute scale of deformation (parameterized by $\tilde\mu_S/(\rho{g}R)$) but only on the ratios between the parameters characterizing the different layers (radii and shear moduli). 

Second, the boundary condition on the gravitational potential given by $Ry^{}_6(R)=5$ \citep{takeuchi1962,saito1974}) implies that
\begin{equation}
h_2 = \frac{5}{3} \, k_2 \, ,
\label{relationh2k2}
\end{equation}
where the tidal Love number for gravity  is defined by $k_2=y^{}_5(R)-1$ ($y^{}_5$ is the gravitational potential).
This relation was expected  since the radial deformation of the surface completely characterizes the gravity potential of the body when the density is uniform.

Third, the boundary condition on the radial stress ($y^{}_2(R)=0$) combined with Eqs.~(\ref{bound1})-(\ref{bound4}) implies that
\begin{equation}
h_2 = \frac{5}{2} \left( 1 + \left( \frac{19}{2} - 10 \, c_S  - \frac{35}{2} \, d_S \right) \frac{\tilde\mu_S}{\rho{g}R} \right)^{-1} \, .
\label{h2UniformDensity}
\end{equation}
Fourth, the tidal Love number $l_2=gy^{}_3(R)$ is related to $h_2$ by
\begin{equation}
l_2 = \left(  \frac{3}{10} - \frac{1}{4} \, c_S - \frac{7}{6} \, d_S \right) h_2 \, .
\label{l2UniformDensity}
\end{equation}
In Eqs.~(\ref{h2UniformDensity})-(\ref{l2UniformDensity}), the constants $(c_S,d_S)$ must be determined by relating them with propagation matrices to two unknown constants in the core (if it is solid) and applying the conditions (\ref{bound1})-(\ref{bound4}).
Each liquid layer introduces two additional unknowns but also two additional constraints \citep[e.g.][]{sabadini2004,jaraorue2011}.

For a one-layer body, the constants $(c,d)$ in Eqs.~(\ref{yiFun4}) vanish otherwise the solution diverges in $r=0$.
Eqs.~(\ref{h2UniformDensity})-(\ref{l2UniformDensity}) immediately yield the well-known formulas
\begin{eqnarray}
h_2 &=& \frac{5}{2} \left( 1 +  \frac{19}{2} \, \frac{\tilde\mu}{\rho{g}R} \right)^{-1} \, ,
\nonumber
\\
l_2 &=& \frac{3}{10} \, h_2 \, .
\label{h2l2OneLayer}
\end{eqnarray}
In that case, the solution of Eqs.~(\ref{bound1})-(\ref{bound4}) is
\begin{equation}
(a,b,c,d) = (-21/5,8/5,0,0) \, ,
\label{SolOneLayer}
\end{equation}
yielding the displacement functions
\begin{equation}
\left( y_1, y_3, \frac{r y_4}{\tilde \mu} \right) =   \frac{h_2}{g} \, \frac{\bar r}{5} \left( 8 - 3 \bar r^2, 4 - \frac{5}{2} \,\bar r^2 , 8 \left( 1 - \bar r^2 \right)  \right) \, .
\label{yiHomog}
\end{equation}

\section*{Appendix G. Two-layer  incompressible body of uniform density}
\label{AppendixG}
\markright{APPENDIX G\hfill}

Consider an incompressible body of uniform density $\rho$ composed of a viscoelastic core of radius $R_C$ and a viscoelastic mantle of radius $R$ with shear moduli $\tilde\mu_C$ and $\tilde\mu_M$, respectively.
The surface gravity is denoted $g$.
I define the dimensionless ratios
\begin{equation}
\left( \bar r \, , \, x \, , \, \xi \, , \, \mu_{red} \right) = \left( \frac{r}{R} \, , \, \frac{R_C}{R} \, , \, \frac{\tilde\mu_C}{\tilde\mu_M} \, , \,  \frac{\tilde\mu_M}{\rho{g}R} \right) \, .
\label{defxi}
\end{equation}

The coefficients $(a,b,c,d)$ in the displacement functions (\ref{yiFun4}) are ratios of polynomials in $x$ (dimensionless core radius) and $\xi$ (ratio of shear moduli):
\begin{equation}
\left( a , b ,c, d \right) = \left( p_a , p_b , p_c , p_d \right) / q \, .
\label{defa}
\end{equation}
The polynomial in the denominator is given by
\begin{eqnarray}
q &=&
4 \left( 24 + 40 \, x^3 - 45 \, x^7 - 19 \, x^{10} \right)
+ \, 2 \left( 89 + 15 \, x^3 - 5 \, x^7 + 76 \, x^{10} \right) \xi 
\nonumber \\
&&+ \,38  \left( 2 - 5 \, x^3 + 5 \, x^7 - 2 \, x^{10} \right) \xi^2  \, .
 \label{coeffq}
\end{eqnarray}
In the core, $p_c=p_d=0$ while $p_a$ and $p_b$ are given by
\begin{eqnarray}
p^C_a &=& - 294 \, \Big[
5 + \left( 2 - 5 \, x^3 + 8 \, x^5 \right) (\xi-1)
\Big] \, ,
\nonumber \\
p^C_b &=&  2 \, \Big[
280 + \left( 152 -21 \, x^2 -147 \, x^5 + 296 \, x^7 \right) (\xi-1)
\Big] \, .
\label{CoeffCore}
\end{eqnarray}
In the mantle, $p_c$ and $p_d$ are given by
\begin{eqnarray}
p^M_c &=&  - \frac{4}{5} \, x^3 \left(  \xi -1 \right) \Big[
16 \left( 40 - 21 \, x^2 - 19 \, x^7 \right)
+ 19 \left( 40 - 21 \, x^2 + 16 \, x^7 \right) \xi
\Big] \, ,
\nonumber \\
p^M_d &=& \frac{6}{5} \, x^5 \left(  \xi -1 \right) \Big[
2 \left( 64 - 45 \, x^2 - 19 \, x^5 \right)
+ 19 \left( 8 - 5 \, x^2 + 2 \, x^5 \right) \xi
\Big] \, ,
\label{CoeffMantlecd}
\end{eqnarray}
while $p_a$ and $p_b$ are related to them by Eqs.~(\ref{bound1})-(\ref{bound4}):
\begin{eqnarray}
p^M_a &=& - (21/5) \, q - 7 \, p^M_d \, ,
\nonumber \\
p^M_b &=& (8/5) \, q - (1/2) \, p^M_c \, .
\label{CoeffMantleab}
\end{eqnarray}
The substitution of Eqs.~(\ref{CoeffMantlecd}) into Eqs.~(\ref{h2UniformDensity})-(\ref{l2UniformDensity}) yields the displacement tidal Love numbers ($k_2$ is given by Eq.~(\ref{relationh2k2})):
\begin{eqnarray}
h_2 &=& \frac{5}{2
} \, \Big( 1+ \left( \frac{19}{2} - \frac{p_h}{q} \right) \mu_{red} \Big)^{-1} \, ,
\nonumber \\
l_2 &=& \left( \frac{3}{10} - \frac{p_l}{q} \right) h_2 \,,
\label{l2TwoLayers}
\end{eqnarray}
with
\begin{eqnarray}
p_h &=&  10 \, p^M_c  + (35/2) \, p^M_d \, ,
\nonumber \\
p_l &=&   (1/4) \, p^M_c + (7/6) \, p^M_d \, .
\label{coeffpk}
\end{eqnarray}

If $|\xi|=0$ (liquid core) or if $|\xi|\rightarrow\infty$ (infinitely rigid core), viscoelastic parameters appear in the functions $y_i$ through the prefactor $h_2$ so that tidal deformations depend depend only on the core radius $x$.
The Love number $h_2$ becomes
\begin{equation}
h_2(x) = \frac{5}{2} \, \Big( 1+ z(x) \, \mu_{red} \Big)^{-1} \, .
\label{h2x}
\end{equation}
If the core is liquid, $z(x)$ is given by
\begin{equation} 
z^{liq}(x) = 12 \, \frac{19 - 75 \, x^3 + 112 \, x^5 - 75 \, x^7 + 19 \, x^{10}}{24 + \, 40 \, x^3 - 45 \, x^7 - 19 \, x^{10}} \, .
\label{zliquid}
\end{equation}
If the core is infinitely rigid, $z(x)$ is given by
\begin{equation}
z^{rig}(x) = \frac{38 + \, 225 x^3 - \, 336 x^5 + \, 200 x^7 + 48 \, x^{10}}{ 2 \left( 2 - 5 \, x^3 + 5 \, x^7 - 2 \, x^{10} \right) } \, .
\label{zrigid}
\end{equation}
The thin shell limit is obtained first by setting $\xi=0$ (liquid core) and then by taking the limit $x\rightarrow1$.
In that case, $z^{liq}(x)\sim(60/11)(1-x)$ and $p_l/q\rightarrow3/110$ so that
\begin{eqnarray}
h_2 &\rightarrow& \frac{5}{2} \left(1+ \frac{60}{11} \, (1-x) \, \mu_{red} \right)^{-1}  \, ,
\nonumber \\
l_2 &\rightarrow& \frac{3}{11} \, h_2 \, .
\label{ThinShellLimit}
\end{eqnarray}

\subsection*{Acknowledgments}
This work was financially supported by the Belgian PRODEX program managed by the European Space Agency in collaboration with the Belgian Federal Science Policy Office.
I thank Attilio Rivoldini and Antony Trinh for discussions on tidal response functions.



\renewcommand{\baselinestretch}{0.5}


\scriptsize

\begin{thebibliography}{96}
\providecommand{\natexlab}[1]{#1}
\expandafter\ifx\csname urlstyle\endcsname\relax
  \providecommand{\doi}[1]{doi:\discretionary{}{}{}#1}\else
  \providecommand{\doi}{doi:\discretionary{}{}{}\begingroup
  \urlstyle{rm}\Url}\fi
  
  \markright{REFERENCES\hfill}

\bibitem[{\textit{{Alterman} et~al.}(1959)\textit{{Alterman}, {Jarosch}, and
  {Pekeris}}}]{alterman1959}
{Alterman}, Z., H.~{Jarosch}, and C.~L. {Pekeris} (1959), {Oscillations of the
  Earth}, \textit{Proc. R. Soc. London A}, \textit{252}, 80--95,
  \doi{10.1098/rspa.1959.0138}.

\bibitem[{\textit{{Anderson}}(1989)}]{anderson1989}
{Anderson}, D. (1989), \textit{{Theory of the Earth}}, Blackwell, Boston.

\bibitem[{\textit{{Baland} et~al.}(2012)\textit{{Baland}, {Yseboodt}, and {Van
  Hoolst}}}]{baland2012}
{Baland}, R.-M., M.~{Yseboodt}, and T.~{Van Hoolst} (2012), {Obliquity of the
  Galilean satellites: The influence of a global internal liquid layer},
  \textit{Icarus}, \textit{220}, 435--448, \doi{10.1016/j.icarus.2012.05.020}.

\bibitem[{\textit{{Balmino}}(1994)}]{balmino1994}
{Balmino}, G. (1994), {Gravitational potential harmonics from the shape of an
  homogeneous body}, \textit{Celest. Mech. Dyn. Astr.}, \textit{60}, 331--364,
  \doi{10.1007/BF00691901}.

\bibitem[{\textit{{Benjamin} et~al.}(2006)\textit{{Benjamin}, {Wahr}, {Ray},
  {Egbert}, and {Desai}}}]{benjamin2006}
{Benjamin}, D., J.~{Wahr}, R.~D. {Ray}, G.~D. {Egbert}, and S.~D. {Desai}
  (2006), {Constraints on mantle anelasticity from geodetic observations, and
  implications for the J$_{2}$ anomaly}, \textit{Geophys. J. Int.},
  \textit{165}, 3--16, \doi{10.1111/j.1365-246X.2006.02915.x}.

\bibitem[{\textit{{Beuthe}}(2008)}]{beuthe2008}
{Beuthe}, M. (2008), {Thin elastic shells with variable thickness for
  lithospheric flexure of one-plate planets}, \textit{Geophys.~J.~Int.},
  \textit{172}, 817--841, \doi{10.1111/j.1365-246X.2007.03671.x}.

\bibitem[{\textit{{Beuthe}}(2010)}]{beuthe2010a}
{Beuthe}, M. (2010), {East-west faults due to planetary contraction},
  \textit{Icarus}, \textit{209}, 795--817, \doi{10.1016/j.icarus.2010.04.019}.

\bibitem[{\textit{{Billings} and {Kattenhorn}}(2005)}]{billings2005}
{Billings}, S.~E., and S.~A. {Kattenhorn} (2005), {The great thickness debate:
  Ice shell thickness models for Europa and comparisons with estimates based on
  flexure at ridges}, \textit{Icarus}, \textit{177}, 397--412,
  \doi{10.1016/j.icarus.2005.03.013}.

\bibitem[{\textit{{Bills}}(2005)}]{bills2005a}
{Bills}, B.~G. (2005), {Free and forced obliquities of the Galilean satellites
  of Jupiter}, \textit{Icarus}, \textit{175}, 233--247,
  \doi{10.1016/j.icarus.2004.10.028}.

\bibitem[{\textit{{Bills} et~al.}(2005)\textit{{Bills}, {Neumann}, {Smith}, and
  {Zuber}}}]{bills2005b}
{Bills}, B.~G., G.~A. {Neumann}, D.~E. {Smith}, and M.~T. {Zuber} (2005),
  {Improved estimate of tidal dissipation within Mars from MOLA observations of
  the shadow of Phobos}, \textit{J. Geophys. Res.}, \textit{110}, E07004,
  \doi{10.1029/2004JE002376}.

\bibitem[{\textit{{B{\v e}hounkov{\'a}} et~al.}(2010)\textit{{B{\v
  e}hounkov{\'a}}, {Tobie}, {Choblet}, and {{\v C}adek}}}]{behounkova2010}
{B{\v e}hounkov{\'a}}, M., G.~{Tobie}, G.~{Choblet}, and O.~{{\v C}adek}
  (2010), {Coupling mantle convection and tidal dissipation: Applications to
  Enceladus and Earth-like planets}, \textit{J. Geophys. Res.}, \textit{115},
  E09011, \doi{10.1029/2009JE003564}.

\bibitem[{\textit{{Cassen} et~al.}(1980)\textit{{Cassen}, {Peale}, and
  {Reynolds}}}]{cassen1980}
{Cassen}, P., S.~J. {Peale}, and R.~T. {Reynolds} (1980), {Tidal dissipation in
  Europa - A correction}, \textit{Geophys. Res. Lett.}, \textit{7}, 987,
  \doi{10.1029/GL007i011p00987}.

\bibitem[{\textit{{Castillo-Rogez} et~al.}(2011)\textit{{Castillo-Rogez},
  {Efroimsky}, and {Lainey}}}]{castillo2011}
{Castillo-Rogez}, J.~C., M.~{Efroimsky}, and V.~{Lainey} (2011), {The tidal
  history of Iapetus: Spin dynamics in the light of a refined dissipation
  model}, \textit{J. Geophys. Res.}, \textit{116}, E09008,
  \doi{10.1029/2010JE003664}.

\bibitem[{\textit{{Chyba} et~al.}(1989)\textit{{Chyba}, {Jankowski}, and
  {Nicholson}}}]{chyba1989}
{Chyba}, C.~F., D.~G. {Jankowski}, and P.~D. {Nicholson} (1989), {Tidal
  evolution in the Neptune-Triton system}, \textit{Astron. Astrophys.},
  \textit{219}, L23--L26.

\bibitem[{\textit{{Dahlen}}(1976)}]{dahlen1976}
{Dahlen}, F.~A. (1976), {The passive influence of the oceans upon the rotation
  of the earth.}, \textit{Geophys. J. Int.}, \textit{46}, 363--406,
  \doi{10.1111/j.1365-246X.1976.tb04163.x}.

\bibitem[{\textit{{Durek} and {Ekstr{\"o}m}}(1995)}]{durek1995}
{Durek}, J.~J., and G.~{Ekstr{\"o}m} (1995), {Evidence of bulk attenuation in
  the asthenosphere from recordings of the Bolivia earthquake},
  \textit{Geophys. Res. Lett.}, \textit{22}, 2309--2312,
  \doi{10.1029/95GL01434}.

\bibitem[{\textit{{Efroimsky} and {Lainey}}(2007)}]{efroimsky2007}
{Efroimsky}, M., and V.~{Lainey} (2007), {Physics of bodily tides in
  terrestrial planets and the appropriate scales of dynamical evolution},
  \textit{J. Geophys. Res.}, \textit{112}, E12003, \doi{10.1029/2007JE002908}.

\bibitem[{\textit{{Fabrycky} et~al.}(2007)\textit{{Fabrycky}, {Johnson}, and
  {Goodman}}}]{fabrycky2007}
{Fabrycky}, D.~C., E.~T. {Johnson}, and J.~{Goodman} (2007), {Cassini States
  with Dissipation: Why Obliquity Tides Cannot Inflate Hot Jupiters},
  \textit{Astrophys. J.}, \textit{665}, 754--766, \doi{10.1086/519075}.

\bibitem[{\textit{{Fischer} and {Spohn}}(1990)}]{fischer1990}
{Fischer}, H.-J., and T.~{Spohn} (1990), {Thermal-orbital histories of
  viscoelastic models of Io (J1)}, \textit{Icarus}, \textit{83}, 39--65,
  \doi{10.1016/0019-1035(90)90005-T}.

\bibitem[{\textit{{Fortes}}(2012)}]{fortes2012}
{Fortes}, A.~D. (2012), {Titan's internal structure and the evolutionary
  consequences}, \textit{Planet. Space Sci.}, \textit{60}, 10--17,
  \doi{10.1016/j.pss.2011.04.010}.

\bibitem[{\textit{{Fung}}(1965)}]{fung1965}
{Fung}, Y.~C. (1965), \textit{{Foundations of solid mechanics}}, Prentice-Hall,
  Englewood Cliffs, New Jersey.

\bibitem[{\textit{{Gaeman} et~al.}(2012)\textit{{Gaeman}, {Hier-Majumder}, and
  {Roberts}}}]{gaeman2012}
{Gaeman}, J., S.~{Hier-Majumder}, and J.~H. {Roberts} (2012), {Sustainability
  of a subsurface ocean within Triton's interior}, \textit{Icarus},
  \textit{220}, 339--347, \doi{10.1016/j.icarus.2012.05.006}.

\bibitem[{\textit{{Gammon} et~al.}(1983)\textit{{Gammon}, {Kiefte}, {Clouter},
  and {Denner}}}]{gammon1983}
{Gammon}, P.~H., H.~{Kiefte}, M.~J. {Clouter}, and W.~W. {Denner} (1983),
  {Elastic constants of artificial and natural ice samples by Brillouin
  spectroscopy}, \textit{J. Glaciol.}, \textit{29}, 433--460.

\bibitem[{\textit{{Gladman} et~al.}(1996)\textit{{Gladman}, {Dane Quinn},
  {Nicholson}, and {Rand}}}]{gladman1996}
{Gladman}, B., D.~{Dane Quinn}, P.~{Nicholson}, and R.~{Rand} (1996),
  {Synchronous Locking of Tidally Evolving Satellites}, \textit{Icarus},
  \textit{122}, 166--192, \doi{10.1006/icar.1996.0117}.

\bibitem[{\textit{{Hamilton} et~al.}(2012)\textit{{Hamilton}, {Beggan},
  {Still}, {Beuthe}, {Lopes}, {Williams}, {Radebaugh}, and
  {Wright}}}]{hamilton2012}
{Hamilton}, C.~W., C.~D. {Beggan}, S.~{Still}, M.~{Beuthe}, R.~M.~C. {Lopes},
  D.~A. {Williams}, J.~{Radebaugh}, and W.~{Wright} (2012), {Spatial
  distribution of volcanoes on Io: Implications for tidal heating and magma
  ascent}, \textit{Earth~Planet.~Sci.~Lett.}, p. (in press),
  \doi{10.1016/j.epsl.2012.10.032}.

\bibitem[{\textit{{Han} and {Showman}}(2010)}]{han2010}
{Han}, L., and A.~P. {Showman} (2010), {Coupled convection and tidal
  dissipation in Europa's ice shell}, \textit{Icarus}, \textit{207}, 834--844,
  \doi{10.1016/j.icarus.2009.12.028}.

\bibitem[{\textit{{Hussmann} et~al.}(2006)\textit{{Hussmann}, {Sohl}, and
  {Spohn}}}]{hussmann2006}
{Hussmann}, H., F.~{Sohl}, and T.~{Spohn} (2006), {Subsurface oceans and deep
  interiors of medium-sized outer planet satellites and large trans-neptunian
  objects}, \textit{Icarus}, \textit{185}, 258--273,
  \doi{10.1016/j.icarus.2006.06.005}.

\bibitem[{\textit{{Iess} et~al.}(2010)\textit{{Iess}, {Rappaport}, {Jacobson},
  {Racioppa}, {Stevenson}, {Tortora}, {Armstrong}, and {Asmar}}}]{iess2010}
{Iess}, L., N.~J. {Rappaport}, R.~A. {Jacobson}, P.~{Racioppa}, D.~J.
  {Stevenson}, P.~{Tortora}, J.~W. {Armstrong}, and S.~W. {Asmar} (2010),
  {Gravity Field, Shape, and Moment of Inertia of Titan}, \textit{Science},
  \textit{327}, 1367--1369, \doi{10.1126/science.1182583}.

\bibitem[{\textit{{Iess} et~al.}(2012)\textit{{Iess}, {Jacobson}, {Ducci},
  {Stevenson}, {Lunine}, {Armstrong}, {Asmar}, {Racioppa}, {Rappaport}, and
  {Tortora}}}]{iess2012}
{Iess}, L., R.~A. {Jacobson}, M.~{Ducci}, D.~J. {Stevenson}, J.~I. {Lunine},
  J.~W. {Armstrong}, S.~W. {Asmar}, P.~{Racioppa}, N.~J. {Rappaport}, and
  P.~{Tortora} (2012), {The Tides of Titan}, \textit{Science}, \textit{337},
  457--459, \doi{10.1126/science.1219631}.

\bibitem[{\textit{{Jackson}}(2007)}]{jackson2007}
{Jackson}, I. (2007), {Properties of Rocks and Minerals - Physical Origins of
  Anelasticity and Attenuation in Rock}, in \textit{Treatise on Geophysics:
  Mineral physics}, edited by {{Spohn}, T.}, pp. 493--525, Cambridge University
  Press, Cambridge, \doi{10.1016/B978-044452748-6.00046-8}.

\bibitem[{\textit{{Jara-Oru{\'e}} and {Vermeersen}}(2011)}]{jaraorue2011}
{Jara-Oru{\'e}}, H.~M., and B.~L.~A. {Vermeersen} (2011), {Effects of
  low-viscous layers and a non-zero obliquity on surface stresses induced by
  diurnal tides and non-synchronous rotation: The case of Europa},
  \textit{Icarus}, \textit{215}, 417--438, \doi{10.1016/j.icarus.2011.05.034}.

\bibitem[{\textit{{Karato}}(2008)}]{karato2008}
{Karato}, S.-I. (2008), \textit{{Deformation of Earth materials}}, Cambridge
  University Press, Cambridge.

\bibitem[{\textit{{Karato}}(2010)}]{karato2010}
{Karato}, S.-I. (2010), {Rheology of the Earth's mantle: A historical review},
  \textit{Gondwana Res.}, \textit{18}, 17--45, \doi{10.1016/j.gr.2010.03.004}.

\bibitem[{\textit{{Karato} and {Wu}}(1993)}]{karato1993}
{Karato}, S.-I., and P.~{Wu} (1993), {Rheology of the upper mantle - A
  synthesis}, \textit{Science}, \textit{260}, 771--778,
  \doi{10.1126/science.260.5109.771}.

\bibitem[{\textit{{Kaula}}(1963)}]{kaula1963}
{Kaula}, W.~M. (1963), {Tidal Dissipation in the Moon},
  \textit{J.~Geophys.~Res.}, \textit{68}, 4959--4965.

\bibitem[{\textit{{Kaula}}(1964)}]{kaula1964}
{Kaula}, W.~M. (1964), {Tidal Dissipation by Solid Friction and the Resulting
  Orbital Evolution}, \textit{Rev. of Geophys.}, \textit{2}, 661--685,
  \doi{doi:10.1029/RG002i004p00661}.

\bibitem[{\textit{{Keszthelyi} et~al.}(2007)\textit{{Keszthelyi}, {Jaeger},
  {Milazzo}, {Radebaugh}, {Davies}, and {Mitchell}}}]{keszthelyi2007}
{Keszthelyi}, L., W.~{Jaeger}, M.~{Milazzo}, J.~{Radebaugh}, A.~G. {Davies},
  and K.~L. {Mitchell} (2007), {New estimates for Io eruption temperatures:
  Implications for the interior}, \textit{Icarus}, \textit{192}, 491--502,
  \doi{10.1016/j.icarus.2007.07.008}.

\bibitem[{\textit{{Khurana} et~al.}(2011)\textit{{Khurana}, {Jia}, {Kivelson},
  {Nimmo}, {Schubert}, and {Russell}}}]{khurana2011}
{Khurana}, K.~K., X.~{Jia}, M.~G. {Kivelson}, F.~{Nimmo}, G.~{Schubert}, and
  C.~T. {Russell} (2011), {Evidence of a global magma ocean in Io's interior},
  \textit{Science}, \textit{332}, 1186--1189, \doi{10.1126/science.1201425}.

\bibitem[{\textit{{Kirchoff} et~al.}(2011)\textit{{Kirchoff}, {McKinnon}, and
  {Schenk}}}]{kirchoff2011}
{Kirchoff}, M.~R., W.~B. {McKinnon}, and P.~M. {Schenk} (2011), {Global
  distribution of volcanic centers and mountains on Io: Control by
  asthenospheric heating and implications for mountain formation},
  \textit{Earth~Planet.~Sci.~Lett.}, \textit{301}, 22--30,
  \doi{10.1016/j.epsl.2010.11.018}.

\bibitem[{\textit{{Kohlstedt} and {Mackwell}}(2009)}]{kohlstedt2009}
{Kohlstedt}, D.~L., and S.~J. {Mackwell} (2009), {Strength and deformation of
  planetary lithospheres}, in \textit{Planetary Tectonics}, edited by
  {{Watters}, T.~R. and {Schultz}, R.~A.}, pp. 397--456, Cambridge University
  Press, Cambridge, \doi{10.1017/CBO9780511691645.010}.

\bibitem[{\textit{{Lainey} et~al.}(2009)\textit{{Lainey}, {Arlot}, {Karatekin},
  and {van Hoolst}}}]{lainey2009}
{Lainey}, V., J.-E. {Arlot}, {\"O}.~{Karatekin}, and T.~{van Hoolst} (2009),
  {Strong tidal dissipation in Io and Jupiter from astrometric observations},
  \textit{Nature}, \textit{459}, 957--959, \doi{10.1038/nature08108}.

\bibitem[{\textit{{Levrard}}(2008)}]{levrard2008}
{Levrard}, B. (2008), {A proof that tidal heating in a synchronous rotation is
  always larger than in an asymptotic nonsynchronous rotation state},
  \textit{Icarus}, \textit{193}, 641--643, \doi{10.1016/j.icarus.2007.10.003}.

\bibitem[{\textit{{Levrard} et~al.}(2007)\textit{{Levrard}, {Correia},
  {Chabrier}, {Baraffe}, {Selsis}, and {Laskar}}}]{levrard2007}
{Levrard}, B., A.~C.~M. {Correia}, G.~{Chabrier}, I.~{Baraffe}, F.~{Selsis},
  and J.~{Laskar} (2007), {Tidal dissipation within hot Jupiters: a new
  appraisal}, \textit{Astron. Astrophys.}, \textit{462}, L5--L8,
  \doi{10.1051/0004-6361:20066487}.

\bibitem[{\textit{{Lopes-Gautier} et~al.}(1999)\textit{{Lopes-Gautier},
  {McEwen}, {Smythe}, {Geissler}, {Kamp}, {Davies}, {Spencer}, {Keszthelyi},
  {Carlson}, {Leader}, {Mehlman}, {Soderblom}, and {Galileo NIMS And SSI
  Teams}}}]{lopes1999}
{Lopes-Gautier}, R., A.~S. {McEwen}, W.~B. {Smythe}, P.~E. {Geissler},
  L.~{Kamp}, A.~G. {Davies}, J.~R. {Spencer}, L.~{Keszthelyi}, R.~{Carlson},
  F.~E. {Leader}, R.~{Mehlman}, L.~{Soderblom}, and {Galileo NIMS And SSI
  Teams} (1999), {Active Volcanism on Io: Global Distribution and Variations in
  Activity}, \textit{Icarus}, \textit{140}, 243--264,
  \doi{10.1006/icar.1999.6129}.

\bibitem[{\textit{{Love}}(1911)}]{love1911}
{Love}, A.~E.~H. (1911), \textit{{Some problems of geodynamics}}, Cambridge
  University Press, Cambridge.

\bibitem[{\textit{{Malvern}}(1969)}]{malvern1969}
{Malvern}, L.~E. (1969), \textit{{Introduction to the mechanics of a continuous
  medium}}, Prentice-Hall, Upper Saddle River, New Jersey.

\bibitem[{\textit{{McKinnon}}(1999)}]{mckinnon1999}
{McKinnon}, W.~B. (1999), {Convective instability in Europa's floating ice
  shell}, \textit{Geophys. Res. Lett.}, \textit{26}, 951--954,
  \doi{10.1029/1999GL900125}.

\bibitem[{\textit{{Mitri} and {Showman}}(2008)}]{mitri2008}
{Mitri}, G., and A.~P. {Showman} (2008), {Thermal convection in ice-I shells of
  Titan and Enceladus}, \textit{Icarus}, \textit{193}, 387--396,
  \doi{10.1016/j.icarus.2007.07.016}.

\bibitem[{\textit{{Moore} et~al.}(2007)\textit{{Moore}, {Schubert}, {Anderson},
  and {Spencer}}}]{moore2007}
{Moore}, W.~B., G.~{Schubert}, J.~D. {Anderson}, and J.~R. {Spencer} (2007),
  {The interior of Io}, in \textit{Io After Galileo}, edited by R.~M.~C.
  {Lopes} and J.~R. {Spencer}, pp. 89--108, Springer-Praxis,
  \doi{10.1007/978-3-540-48841-5\_5}.

\bibitem[{\textit{{Munk} and {MacDonald}}(1960)}]{munk1960}
{Munk}, W.~H., and G.~J.~F. {MacDonald} (1960), \textit{{The rotation of the
  earth}}, Cambridge University Press, Cambridge.

\bibitem[{\textit{{Nakada} and {Karato}}(2012)}]{nakada2012}
{Nakada}, M., and S.-I. {Karato} (2012), {Low viscosity of the bottom of the
  Earth's mantle inferred from the analysis of Chandler wobble and tidal
  deformation}, \textit{Phys. Earth Planet. Int.}, \textit{192}, 68--80,
  \doi{10.1016/j.pepi.2011.10.001}.

\bibitem[{\textit{{Nimmo} and {Bills}}(2010)}]{nimmo2010}
{Nimmo}, F., and B.~G. {Bills} (2010), {Shell thickness variations and the
  long-wavelength topography of Titan}, \textit{Icarus}, \textit{208},
  896--904, \doi{10.1016/j.icarus.2010.02.020}.

\bibitem[{\textit{{Nimmo} and {Manga}}(2009)}]{nimmo2009}
{Nimmo}, F., and M.~{Manga} (2009), {Geodynamics of Europa's Icy Shell}, in
  \textit{Europa}, edited by R.~T. {Pappalardo}, W.~B. {McKinnon}, and K.~K.
  {Khurana}, pp. 381--404, University of Arizona Press, Tucson.

\bibitem[{\textit{{Nimmo} et~al.}(2007)\textit{{Nimmo}, {Thomas}, {Pappalardo},
  and {Moore}}}]{nimmo2007}
{Nimmo}, F., P.~C. {Thomas}, R.~T. {Pappalardo}, and W.~B. {Moore} (2007), {The
  global shape of Europa: Constraints on lateral shell thickness variations},
  \textit{Icarus}, \textit{191}, 183--192, \doi{10.1016/j.icarus.2007.04.021}.

\bibitem[{\textit{{Nimmo} et~al.}(2012)\textit{{Nimmo}, {Faul}, and
  {Garnero}}}]{nimmo2012}
{Nimmo}, F., U.~H. {Faul}, and E.~J. {Garnero} (2012), {Dissipation at tidal
  and seismic frequencies in a melt-free Moon}, \textit{J. Geophys. Res.},
  \textit{117}, E09005, \doi{10.1029/2012JE004160}.

\bibitem[{\textit{{Ojakangas} and
  {Stevenson}}(1989{\natexlab{a}})}]{ojakangas1989a}
{Ojakangas}, G.~W., and D.~J. {Stevenson} (1989{\natexlab{a}}), {Thermal state
  of an ice shell on Europa}, \textit{Icarus}, \textit{81}, 220--241,
  \doi{10.1016/0019-1035(89)90052-3}.

\bibitem[{\textit{{Ojakangas} and
  {Stevenson}}(1989{\natexlab{b}})}]{ojakangas1989b}
{Ojakangas}, G.~W., and D.~J. {Stevenson} (1989{\natexlab{b}}), {Polar wander
  of an ice shell on Europa}, \textit{Icarus}, \textit{81}, 242--270,
  \doi{10.1016/0019-1035(89)90053-5}.

\bibitem[{\textit{{Peale}}(1999)}]{peale1999}
{Peale}, S.~J. (1999), {Origin and Evolution of the Natural Satellites},
  \textit{Annu. Rev. Astron. Astrophys.}, \textit{37}, 533--602,
  \doi{10.1146/annurev.astro.37.1.533}.

\bibitem[{\textit{{Peale}}(2008)}]{peale2008}
{Peale}, S.~J. (2008), {Obliquity Tides in Hot Jupiters}, in \textit{Extreme
  Solar Systems}, \textit{ASP Conf. Ser.}, vol. 398, edited by {D.~Fischer,
  F.~A.~Rasio, S.~E.~Thorsett, \& A.~Wolszczan}, pp. 281--292.

\bibitem[{\textit{{Peale} and {Cassen}}(1978)}]{peale1978}
{Peale}, S.~J., and P.~{Cassen} (1978), {Contribution of tidal dissipation to
  lunar thermal history}, \textit{Icarus}, \textit{36}, 245--269,
  \doi{10.1016/0019-1035(78)90109-4}.

\bibitem[{\textit{{Peale} et~al.}(1979)\textit{{Peale}, {Cassen}, and
  {Reynolds}}}]{peale1979}
{Peale}, S.~J., P.~{Cassen}, and R.~T. {Reynolds} (1979), {Melting of Io by
  tidal dissipation}, \textit{Science}, \textit{203}, 892--894,
  \doi{10.1126/science.203.4383.892}.

\bibitem[{\textit{{Peltier}}(1982)}]{peltier1982}
{Peltier}, R. (1982), {Dynamics of the Ice Age Earth}, \textit{Adv. Geophys.},
  \textit{24}, 1--146, \doi{10.1016/S0065-2687(08)60519-1}.

\bibitem[{\textit{{Platzman}}(1984)}]{platzman1984}
{Platzman}, G.~W. (1984), {Planetary energy balance for tidal dissipation},
  \textit{Rev. Geophys. Space Phys.}, \textit{22}, 73--84,
  \doi{10.1029/RG022i001p00073}.

\bibitem[{\textit{{Ranalli}}(1995)}]{ranalli1995}
{Ranalli}, G. (1995), \textit{{Rheology of the earth (2nd ed.)}}, Chapman \&
  Hall, London.

\bibitem[{\textit{{Roberts} and {Nimmo}}(2008)}]{roberts2008}
{Roberts}, J.~H., and F.~{Nimmo} (2008), {Tidal heating and the long-term
  stability of a subsurface ocean on Enceladus}, \textit{Icarus}, \textit{194},
  675--689, \doi{10.1016/j.icarus.2007.11.010}.

\bibitem[{\textit{{Ross} and {Schubert}}(1986)}]{ross1986}
{Ross}, M., and G.~{Schubert} (1986), {Tidal dissipation in a viscoelastic
  planet}, \textit{J. Geophys. Res.}, \textit{91}, D447--D452,
  \doi{10.1029/JB091iB04p0D447}.

\bibitem[{\textit{{Ross} et~al.}(1990)\textit{{Ross}, {Schubert}, {Spohn}, and
  {Gaskell}}}]{ross1990}
{Ross}, M.~N., G.~{Schubert}, T.~{Spohn}, and R.~W. {Gaskell} (1990), {Internal
  structure of Io and the global distribution of its topography},
  \textit{Icarus}, \textit{85}, 309--325, \doi{10.1016/0019-1035(90)90119-T}.

\bibitem[{\textit{{Sabadini} and {Vermeersen}}(2004)}]{sabadini2004}
{Sabadini}, R., and B.~{Vermeersen} (2004), \textit{{Global dynamics of the
  Earth}}, Kluwer Academic Publishers, Dordrecht.

\bibitem[{\textit{{Saito}}(1974)}]{saito1974}
{Saito}, M. (1974), {Some problems of static deformation of the earth},
  \textit{J. Phys. Earth}, \textit{22}, 123--140.

\bibitem[{\textit{{Schmeling}}(1985)}]{schmeling1985}
{Schmeling}, H. (1985), {Numerical models on the influence of partial melt on
  elastic, anelastic and electric properties of rocks. Part I: elasticity and
  anelasticity}, \textit{Phys. Earth Planet. Int.}, \textit{41}, 34--57,
  \doi{10.1016/0031-9201(85)90100-1}.

\bibitem[{\textit{{Schubert} et~al.}(2004)\textit{{Schubert}, {Anderson},
  {Spohn}, and {McKinnon}}}]{schubert2004}
{Schubert}, G., J.~D. {Anderson}, T.~{Spohn}, and W.~B. {McKinnon} (2004),
  {Interior composition, structure and dynamics of the Galilean satellites}, in
  \textit{Jupiter}, edited by F.~{Bagenal}, T.~E. {Dowling}, and W.~B.
  {McKinnon}, pp. 281--306, Cambridge University Press, Cambridge.

\bibitem[{\textit{{Schubert} et~al.}(2009)\textit{{Schubert}, {Sohl}, and
  {Hussmann}}}]{schubert2009}
{Schubert}, G., F.~{Sohl}, and H.~{Hussmann} (2009), {Interior of Europa}, in
  \textit{Europa}, edited by R.~T. {Pappalardo}, W.~B. {McKinnon}, and K.~K.
  {Khurana}, pp. 353--367, University of Arizona Press, Tucson.

\bibitem[{\textit{{Segatz} et~al.}(1988)\textit{{Segatz}, {Spohn}, {Ross}, and
  {Schubert}}}]{segatz1988}
{Segatz}, M., T.~{Spohn}, M.~N. {Ross}, and G.~{Schubert} (1988), {Tidal
  dissipation, surface heat flow, and figure of viscoelastic models of Io},
  \textit{Icarus}, \textit{75}, 187--206, \doi{10.1016/0019-1035(88)90001-2}.

\bibitem[{\textit{{Sohl} et~al.}(2003)\textit{{Sohl}, {Hussmann}, {Schwentker},
  {Spohn}, and {Lorenz}}}]{sohl2003}
{Sohl}, F., H.~{Hussmann}, B.~{Schwentker}, T.~{Spohn}, and R.~D. {Lorenz}
  (2003), {Interior structure models and tidal Love numbers of Titan},
  \textit{J. Geophys. Res.}, \textit{108}, 5130, \doi{10.1029/2003JE002044}.

\bibitem[{\textit{{Sotin} et~al.}(2009)\textit{{Sotin}, {Tobie}, {Wahr}, and
  {McKinnon}}}]{sotin2009}
{Sotin}, C., G.~{Tobie}, J.~{Wahr}, and W.~B. {McKinnon} (2009), {Tides and
  Tidal Heating on Europa}, in \textit{Europa}, edited by R.~T. {Pappalardo},
  W.~B. {McKinnon}, and K.~K. {Khurana}, pp. 85--117, University of Arizona
  Press, Tucson.

\bibitem[{\textit{{Spada} et~al.}(2011)\textit{{Spada}, {Barletta}, {Klemann},
  {Riva}, {Martinec}, {Gasperini}, {Lund}, {Wolf}, {Vermeersen}, and
  {King}}}]{spada2011}
{Spada}, G., V.~R. {Barletta}, V.~{Klemann}, R.~E.~M. {Riva}, Z.~{Martinec},
  P.~{Gasperini}, B.~{Lund}, D.~{Wolf}, L.~L.~A. {Vermeersen}, and M.~A. {King}
  (2011), {A benchmark study for glacial isostatic adjustment codes},
  \textit{Geophys. J. Int.}, \textit{185}, 106--132,
  \doi{10.1111/j.1365-246X.2011.04952.x}.

\bibitem[{\textit{{Spohn}}(1997)}]{spohn1997}
{Spohn}, T. (1997), {Tides of Io}, \textit{Lecture Notes in Earth Sciences,
  Berlin Springer Verlag}, \textit{66}, 345--377, \doi{10.1007/BFb0011453}.

\bibitem[{\textit{{Spohn} and {Schubert}}(2003)}]{spohn2003}
{Spohn}, T., and G.~{Schubert} (2003), {Oceans in the icy Galilean satellites
  of Jupiter?}, \textit{Icarus}, \textit{161}, 456--467,
  \doi{10.1016/S0019-1035(02)00048-9}.

\bibitem[{\textit{{Stiles} et~al.}(2008)\textit{{Stiles}, {Kirk}, {Lorenz},
  {Hensley}, {Lee}, {Ostro}, {Allison}, {Callahan}, {Gim}, {Iess}, {Perci del
  Marmo}, {Hamilton}, {Johnson}, {West}, and {Cassini RADAR
  Team}}}]{stiles2008}
{Stiles}, B.~W., R.~L. {Kirk}, R.~D. {Lorenz}, S.~{Hensley}, E.~{Lee}, S.~J.
  {Ostro}, M.~D. {Allison}, P.~S. {Callahan}, Y.~{Gim}, L.~{Iess}, P.~{Perci
  del Marmo}, G.~{Hamilton}, W.~T.~K. {Johnson}, R.~D. {West}, and {Cassini
  RADAR Team} (2008), {Determining Titan's Spin State from Cassini RADAR
  Images}, \textit{Astron. J.}, \textit{135}, 1669--1680,
  \doi{10.1088/0004-6256/135/5/1669}.

\bibitem[{\textit{{Stiles} et~al.}(2010)\textit{{Stiles}, {Kirk}, {Lorenz},
  {Hensley}, {Lee}, {Ostro}, {Allison}, {Callahan}, {Gim}, {Iess}, {Perci del
  Marmo}, {Hamilton}, {Johnson}, {West}, and {Cassini RADAR
  Team}}}]{stiles2010}
{Stiles}, B.~W., R.~L. {Kirk}, R.~D. {Lorenz}, S.~{Hensley}, E.~{Lee}, S.~J.
  {Ostro}, M.~D. {Allison}, P.~S. {Callahan}, Y.~{Gim}, L.~{Iess}, P.~{Perci
  del Marmo}, G.~{Hamilton}, W.~T.~K. {Johnson}, R.~D. {West}, and {Cassini
  RADAR Team} (2010), {ERRATUM: ''Determining Titan's Spin State from Cassini
  Radar Images'' (2008, AJ, 135, 1669)}, \textit{Astron. J.}, \textit{139},
  311, \doi{10.1088/0004-6256/139/1/311}.

\bibitem[{\textit{{Tackley}}(2001)}]{tackley2001a}
{Tackley}, P.~J. (2001), {Convection in Io's asthenosphere: Redistribution of
  nonuniform tidal heating by mean flows}, \textit{J. Geophys. Res.},
  \textit{106}, 32,971--32,982, \doi{10.1029/2000JE001411}.

\bibitem[{\textit{{Tackley} et~al.}(2001)\textit{{Tackley}, {Schubert},
  {Glatzmaier}, {Schenk}, {Ratcliff}, and {Matas}}}]{tackley2001b}
{Tackley}, P.~J., G.~{Schubert}, G.~A. {Glatzmaier}, P.~{Schenk}, J.~T.
  {Ratcliff}, and J.-P. {Matas} (2001), {Three-Dimensional Simulations of
  Mantle Convection in Io}, \textit{Icarus}, \textit{149}, 79--93,
  \doi{10.1006/icar.2000.6536}.

\bibitem[{\textit{{Takeuchi} and {Saito}}(1972)}]{takeuchi1972}
{Takeuchi}, H., and M.~{Saito} (1972), {Seismic surface waves}, in
  \textit{Methods in computational physics, vol. 1}, edited by {{Bolt}, B.A.},
  pp. 217--295, Academic Press, New York.

\bibitem[{\textit{{Takeuchi} et~al.}(1962)\textit{{Takeuchi}, {Saito}, and
  {Kobayashi}}}]{takeuchi1962}
{Takeuchi}, H., M.~{Saito}, and N.~{Kobayashi} (1962), {Statical Deformations
  and Free Oscillations of a Model Earth}, \textit{J. Geophys. Res.},
  \textit{67}, 1141--1154, \doi{10.1029/JZ067i003p01141}.

\bibitem[{\textit{{Thomas} et~al.}(1998)\textit{{Thomas}, {Davies}, {Colvin},
  {Oberst}, {Schuster}, {Neukum}, {Carr}, {McEwen}, {Schubert}, and
  {Belton}}}]{thomas1998}
{Thomas}, P.~C., M.~E. {Davies}, T.~R. {Colvin}, J.~{Oberst}, P.~{Schuster},
  G.~{Neukum}, M.~H. {Carr}, A.~{McEwen}, G.~{Schubert}, and M.~J.~S. {Belton}
  (1998), {The Shape of Io from Galileo Limb Measurements}, \textit{Icarus},
  \textit{135}, 175--180, \doi{10.1006/icar.1998.5987}.

\bibitem[{\textit{{Tobie} et~al.}(2003)\textit{{Tobie}, {Choblet}, and
  {Sotin}}}]{tobie2003}
{Tobie}, G., G.~{Choblet}, and C.~{Sotin} (2003), {Tidally heated convection:
  Constraints on Europa's ice shell thickness}, \textit{J. Geophys. Res.},
  \textit{108}, 5124, \doi{10.1029/2003JE002099}.

\bibitem[{\textit{{Tobie} et~al.}(2005{\natexlab{a}})\textit{{Tobie},
  {Grasset}, {Lunine}, {Mocquet}, and {Sotin}}}]{tobie2005b}
{Tobie}, G., O.~{Grasset}, J.~I. {Lunine}, A.~{Mocquet}, and C.~{Sotin}
  (2005{\natexlab{a}}), {Titan's internal structure inferred from a coupled
  thermal-orbital model}, \textit{Icarus}, \textit{175}, 496--502,
  \doi{10.1016/j.icarus.2004.12.007}.

\bibitem[{\textit{{Tobie} et~al.}(2005{\natexlab{b}})\textit{{Tobie},
  {Mocquet}, and {Sotin}}}]{tobie2005}
{Tobie}, G., A.~{Mocquet}, and C.~{Sotin} (2005{\natexlab{b}}), {Tidal
  dissipation within large icy satellites: Applications to Europa and Titan},
  \textit{Icarus}, \textit{177}, 534--549, \doi{10.1016/j.icarus.2005.04.006}.

\bibitem[{\textit{{Veeder} et~al.}(2012)\textit{{Veeder}, {Davies}, {Matson},
  {Johnson}, {Williams}, and {Radebaugh}}}]{veeder2012}
{Veeder}, G.~J., A.~G. {Davies}, D.~L. {Matson}, T.~V. {Johnson}, D.~A.
  {Williams}, and J.~{Radebaugh} (2012), {Io: Volcanic thermal sources and
  global heat flow}, \textit{Icarus}, \textit{219}, 701--722,
  \doi{10.1016/j.icarus.2012.04.004}.

\bibitem[{\textit{{Wahr} et~al.}(2009)\textit{{Wahr}, {Selvans}, {Mullen},
  {Barr}, {Collins}, {Selvans}, and {Pappalardo}}}]{wahr2009}
{Wahr}, J., Z.~A. {Selvans}, M.~E. {Mullen}, A.~C. {Barr}, G.~C. {Collins},
  M.~M. {Selvans}, and R.~T. {Pappalardo} (2009), {Modeling stresses on
  satellites due to nonsynchronous rotation and orbital eccentricity using
  gravitational potential theory}, \textit{Icarus}, \textit{200}, 188--206,
  \doi{10.1016/j.icarus.2008.11.002}.

\bibitem[{\textit{{Wahr} et~al.}(2006)\textit{{Wahr}, {Zuber}, {Smith}, and
  {Lunine}}}]{wahr2006}
{Wahr}, J.~M., M.~T. {Zuber}, D.~E. {Smith}, and J.~I. {Lunine} (2006), {Tides
  on Europa, and the thickness of Europa's icy shell},
  \textit{J.~Geophys.~Res.}, \textit{111}, E12005, \doi{10.1029/2006JE002729}.

\bibitem[{\textit{{Winn} and {Holman}}(2005)}]{winn2005}
{Winn}, J.~N., and M.~J. {Holman} (2005), {Obliquity Tides on Hot Jupiters},
  \textit{Astrophys. J.}, \textit{628}, L159--L162, \doi{10.1086/432834}.

\bibitem[{\textit{{Wisdom}}(2004)}]{wisdom2004}
{Wisdom}, J. (2004), {Spin-Orbit Secondary Resonance Dynamics of Enceladus},
  \textit{Astrophys. J.}, \textit{128}, 484--491, \doi{10.1086/421360}.

\bibitem[{\textit{{Wisdom}}(2006)}]{wisdom2006}
{Wisdom}, J. (2006), {Dynamics of the Lunar Spin Axis}, \textit{Astrophys. J.},
  \textit{131}, 1864--1871, \doi{10.1086/499581}.

\bibitem[{\textit{{Wisdom}}(2008)}]{wisdom2008}
{Wisdom}, J. (2008), {Tidal dissipation at arbitrary eccentricity and
  obliquity}, \textit{Icarus}, \textit{193}, 637--640,
  \doi{10.1016/j.icarus.2007.09.002}.

\bibitem[{\textit{{Zschau}}(1978)}]{zschau1978}
{Zschau}, J. (1978), {Tidal friction in the solid earth: Loading tides versus
  body tides}, in \textit{Tidal Friction and the Earth's Rotation}, edited by
  {P.~Brosche \& J.~S\"undermann}, Springer-Verlag, pp. 62--94, New York.

\end{thebibliography}
\end{document}